\documentclass[twocolumn,pre,aps,showpacs,showkeys,amsmath]{revtex4}
\usepackage{graphicx}
\usepackage{adjustbox}
\usepackage{multirow}

\begin{document}

\title{Effect of Diverse Recoding of Granule Cells on Optokinetic Response in A Cerebellar Ring Network with Synaptic Plasticity}
\author{Sang-Yoon Kim}
\email{sykim@icn.re.kr}
\author{Woochang Lim}
\email{wclim@icn.re.kr}
\affiliation{Institute for Computational Neuroscience and Department of Science Education, Daegu National University of Education, Daegu 42411, Korea}

\begin{abstract}
We consider a cerebellar ring network for the optokinetic response (OKR), and investigate the effect of diverse recoding of granule (GR) cells on OKR by varying the connection probability $p_c$ from Golgi to GR cells.  For an optimal value of $p_c^*~(=0.06)$, individual GR cells exhibit diverse spiking patterns which are in-phase, anti-phase, or complex out-of-phase with respect to their population-averaged firing activity. Then, these diversely-recoded signals via parallel fibers (PFs) from GR cells are effectively depressed by the error-teaching signals via climbing fibers from the inferior olive which are also in-phase ones. Synaptic weights at in-phase PF-Purkinje cell (PC) synapses of active GR cells are strongly depressed via strong long-term depression (LTD), while those at anti-phase and complex out-of-phase PF-PC synapses are weakly depressed through weak LTD.  This kind of ``effective'' depression (i.e., strong/weak LTD) at the PF-PC synapses causes a big modulation in firings of PCs, which then exert effective inhibitory coordination on the vestibular nucleus (VN) neuron (which evokes OKR). For the firing of the VN neuron, the learning gain degree ${\cal{L}}_g$, corresponding to the modulation gain ratio, increases with increasing the learning cycle, and it saturates at about the 300th cycle. By varying $p_c$ from $p_c^*$, we find that a plot of saturated learning gain degree ${\cal L}_g^*$ versus $p_c$ forms a bell-shaped curve with a peak at $p_c^*$ (where the diversity degree in spiking patterns of GR cells is also maximum). Consequently, the more diverse in recoding of GR cells, the more effective in motor learning for the OKR adaptation.
\end{abstract}

\pacs{87.19.lw, 87.19.lu, 87.19.lv}
\keywords{Optokinetic response, Cerebellar ring network, Diverse recoding, Effective long-term depression, Effective motor learning}

\maketitle

\section{Introduction}
\label{sec:INT}
The cerebellum receives information from the sensory systems, the spinal cord and other parts of the brain and then regulates motor movements.
For a smoothly integrated body movement, the cerebellum activates a large set of spatially separated muscles in a precise order and timing.
Thus, the cerebellum plays an essential role in fine motor control (i.e., precise spatial and temporal motor control) for coordinating voluntary movements such as posture, balance, and locomotion, resulting in smooth and balanced muscular activity \cite{Ito1,Ito2,Ito3}.
Moreover, it is also involved in higher cognitive functions such as time perception and language processing \cite{Ito2,Ito3}.
Animals and humans with damaged cerebella are still able to initiate movements, but these movements become slow, inexact, and uncoordinated
\cite{Gil,Manto}.

The spatial information of movements (e.g., amplitude or velocity) is called ``gain,'' while the temporal information of movements (e.g., initiation or termination) is called ``timing'' \cite{Yama1}. The goal of cerebellar motor learning is to perform precise gain and temporal control for movements.
The cerebellar mechanisms for gain and timing control for eye movements have been studied in the two types of experimental paradigms; (1) gain control for the
optokinetic response (OKR) and the vestibulo-ocular reflex \cite{Ito1,VOR1} and (2) timing control for the eyeblink conditioning \cite{EB1,EB2}.
Here, we are concerned about gain adaptation of OKR. When the eye tracks a moving object with the stationary head, OKR may be seen. When the moving object is
out of the field of vision, the eye moves back rapidly to the original position where it first saw. In this way, OKR consists of two consecutive slow and fast phases. Experimental works on OKR in vertebrates such as rabbits, mice, and zebrafishes have been done in diverse aspects \cite{OKRExp1,OKRExp2,OKRExp3,OKRExp4,OKRExp5,OKRExp6,OKRExp7,OKRExp8}.

In the Marr-Albus-Ito theory for cerebellar computation \cite{Marr,Albus,Ito1}, the cerebellum is considered to act as a simple perceptron (i.e., pattern associator) which associates input [mossy fiber (MF)] patterns with output [Purkinje cell (PC)] patterns. The input patterns become more sparse and less similar to each other via recoding process in the granular layer, consisting of the granule (GR) and the Golgi (GO) cells.
Then, the recoded inputs are fed into the PCs via the parallel fibers (PFs) (i.e., the axons of GR cells).
In addition to the PF recoded signals, the PCs also receive the error-teaching signals through the climbing-fiber (CF) from the inferior olive (IO).
The PF-PC synapses are assumed to be the only synapses at which motor learning occurs.
Thus, synaptic plasticity may occur at the PF-PC synapses (i.e., their synaptic strengths may be potentiated or depressed).
Marr in \cite{Marr} assumes that a Hebbian type of long-term potentiation (LTP) (i.e., increase in synaptic strengths) occurs at the PF-PC synapses when both the PF and the CF signals are conjunctively excited \cite{Hebb,Br}. This Marr's theory (which directly relates the cerebellar function to its structure) represents a milestone in the history of cerebellum \cite{St}.
In contrast to Marr's learning via LTP, Albus in \cite{Albus} assumes that synaptic strengths at PF-PC synapses are depressed [i.e., an anti-Hebbian type of long-term depression (LTD) occurs] in the case of conjunctive excitations of both the PF and the CF signals.
In the case of Albus' learning via LTD, PCs learn when to stop their inhibition (i.e. when to disinhibit) rather than when to fire. In several later experimental works done by Ito et al., clear evidences for LTD were obtained \cite{Ito4,Ito5,Sakurai}. Thus, LTD became established as a unique type of synaptic plasticity for cerebellar motor learning \cite{Ito6,Ito7,Ito8,Ito9}.

In addition to experimental works on the OKR \cite{OKRExp1,OKRExp2,OKRExp3,OKRExp4,OKRExp5,OKRExp6,OKRExp7,OKRExp8}, computational works have also been performed \cite{OKRCom1,Yama1}. The Marr-Albus model of the cerebellum was also reformulated to incorporate dynamical responses in terms of the adaptive filter model (used in the field of engineering control) \cite{OKRCom2,OKRCom3}. The cerebellar structure may be mapped onto an adaptive filter structure. Through analysis-synthesis process of the adaptive filter model. the (time-varying) filter inputs (i.e., MF ``context'' signals for the post-eye-movement) are analyzed into diverse component signals (i.e., diversely recoded PF signals). Then, they are weighted (i.e., synaptic plasticity at PF-PC synapses) and recombined to generate the filter output (i.e., firing activity of PCs). The filter is adaptive because its weights are adjusted by an error-teaching signal (i.e., CF signal), employing the covariance learning rule \cite{CLR}. Using this adaptive filter model, gain adaptation of OKR was successfully simulated \cite{OKRCom1}.
Recently, Yamazaki and Nagao in \cite{Yama1} employed a spiking network model, which was originally introduced for Pavlovian delay eyeblink conditioning \cite{Yama2}. As elements in the spiking network, leaky integrate-and-fire neuron models were used, and parameter values for single cells and synaptic currents were
adopted from physiological data. Through a large-scale computer simulation, some features of OKR adaptation were successfully reproduced.

However, the effects of diverse recoding of GR cells on the OKR adaption in previous computational works are still needed to be more clarified in several dynamical aspects. First of all, dynamical classification of diverse PF signals (corresponding to the recoded outputs of GR cells) must be completely done for clear understanding their association with the error-teaching CF signals. Then, based on such dynamical classification of diverse spiking patterns of GR cells, synaptic plasticity at PF-PC synapses and subsequent learning progress could be more clearly understood. As a result, understanding on the learning gain and the learning progress for the OKR adaptation is expected to be so much improved.

To this end, we consider a cerebellar spiking ring network for the OKR adaptation, and first make a dynamical classification of
diverse spiking patterns of GR cells (i.e., diverse PF signals) by changing the connection probability $p_c$ from GO to GR cells in the granular layer.
An instantaneous whole-population spike rate $R_{\rm GR}(t)$ (which is obtained from the raster plot of spikes of individual neurons) may well describe collective firing activity in the whole population of GR cells \cite{Sparse2,Sparse1,Sparse3,Sparse4,Sparse5,Sparse6,W_Review,RM}. $R_{\rm GR}(t)$ is in-phase with respect to the sinusoidally-modulating MF input signal for the post-eye-movement, although it has a central flattened plateau due to inhibitory inputs from GO cells.

The whole population of GR cells is divided into GR clusters. These GR clusters show diverse spiking patterns which are in-phase, anti-phase, and complex out-of-phase relative to the instantaneous whole-population spike rate $R_{\rm GR}(t)$. Each spiking pattern is characterized in terms of the ``conjunction'' index, denoting the resemblance (or similarity) degree between the spiking pattern and the instantaneous whole-population spike rate $R_{\rm GR}(t)$ (corresponding to the population-averaged firing activity). To quantify the degree of diverse recoding of GR cells, we introduce the diversity degree $\cal{D}$, given by the relative standard deviation in the distribution of conjunction indices of all spiking patterns. We mainly consider an optimal case of $p_c^*~(=0.06)$ where the spiking patterns of GR clusters are the most diverse. In this case, ${\cal{D}^*} \simeq 1.613$ which is a quantitative measure for diverse recoding of GR cells in the granular layer. We also investigate dynamical origin of these diverse spiking patterns of GR cells. It is thus found that, diverse total synaptic inputs (including both the excitatory MF inputs and the inhibitory inputs from the pre-synaptic GO cells) into the GR clusters result in production of diverse spiking patterns (i.e. outputs) in the GR clusters.

Next, based on dynamical classification of diverse spiking patterns of GR clusters, we employ a refined rule for synaptic plasticity (constructed from the
experimental result in \cite{Safo}) and  make an intensive investigation on the effect of diverse recoding of GR cells on synaptic plasticity at PF-PC synapses and the subsequent learning process. PCs (corresponding to the output of the cerebellar cortex) receive both the diversely-recoded PF signals from GR cells and the error-teaching CF signals from the IO neuron. We also note that the CF signals are in-phase with respect to the instantaneous whole-population spike rate $R_{\rm GR}(t)$. In this case, CF signals may be regarded as ``instructors,'' while PF signals can be considered as ``students.''
Then, in-phase PF student signals are strongly depressed (i.e., their synaptic weights at PF-PC synapses are greatly decreased through strong LTD) by the in-phase CF
instructor signals. On the other hand, out-of-phase PF student signals are weakly depressed (i.e., their synaptic weights at PF-PC synapses are a little decreased
via weak LTD) due to the phase difference between the student PF and the instructor CF signals.
In this way, the student PF signals are effectively (i.e., strongly/weakly) depressed by the error-teaching instructor CF signals.

During learning cycles, the ``effective'' depression (i.e., strong/weak LTD) at PF-PC synapses may cause a big modulation in firing activities of PCs, which then exert effective inhibitory coordination on vestibular nucleus (VN) neuron (which evokes OKR eye-movement). For the firing activity of VN neuron, the learning gain degree ${\cal{L}}_g$, corresponding to the modulation gain ratio (i.e., normalized modulation divided by that at the 1st cycle), increases with learning cycle, and it eventually becomes saturated.

Saturation in the learning progress is clearly shown in the IO system. During the learning cycle, the IO neuron receives both the excitatory sensory signal for a desired eye-movement and the inhibitory signal from the VN neuron (representing a realized eye-movement). We introduce the learning progress degree ${\cal{L}}_p$, given by the ratio of the cycle-averaged inhibitory input from the VN neuron to the cycle-averaged excitatory input of the desired sensory signal. With increasing cycle, the cycle-averaged inhibition (from the VN neuron) increases (i.e., ${\cal{L}}_p$ increases), and converges to the constant cycle-averaged excitation (through the desired  signal). Thus, at about the 300th cycle, the learning progress degree becomes saturated at ${\cal{L}}_p =1$.
At this saturated stage, the cycle-averaged excitatory and inhibitory inputs to the IO neuron become balanced, and we get
the saturated learning gain degree ${\cal L}_g^*(\simeq 1.608)$ in the VN.

By changing $p_c$ from $p_c^*~(=0.06)$, we also investigate the effect of diverse recoding of GR cells on the OKR adaptation.
Thus, the plot of saturated learning gain degree ${\cal L}_g^*$ versus $p_c$ is found to form a bell-shaped curve with a peak (${\cal L}_g^* \simeq 1.608)$
at $p_c^*$. With increasing or decreasing $p_c$ from $p_c^*$, the diversity degree $\cal D$ in firing activities of GR cells also forms a bell-shaped curve
with a maximum value (${\cal{D}}^* \simeq 1.613$) at $p_c^*$. We note that both the saturated learning gain degree ${\cal L}_g^*$ and the diversity degree $\cal D$ have a strong correlation with the Pearson's correlation coefficient $r \simeq 0.9998$ \cite{Pearson}. Consequently, the more diverse in recoding of GR cells, the more effective in motor learning for the OKR adaptation.

This paper is organized as follows. In Sec.~\ref{sec:CRN}, we describe the cerebellar ring network for the OKR, composed of the granular layer, the Purkinje-molecular layer, and the VN-IO part. The governing equations for the population dynamics in the ring network are also presented, along with
a refined rule for the synaptic plasticity at the PF-PC synapses. Then, in the main Sec.~\ref{sec:MS}, we investigate the effect of diverse recoding of GR cells
on motor learning for the OKR adaptation by changing $p_c$. Finally, we give summary and discussion in Sec.~\ref{sec:SUM}.
In Appendix \ref{app:B}, glossary for various terms characterizing the cerebellar model is given to help readers keep track of them.

\begin{figure}
\includegraphics[width=0.8\columnwidth]{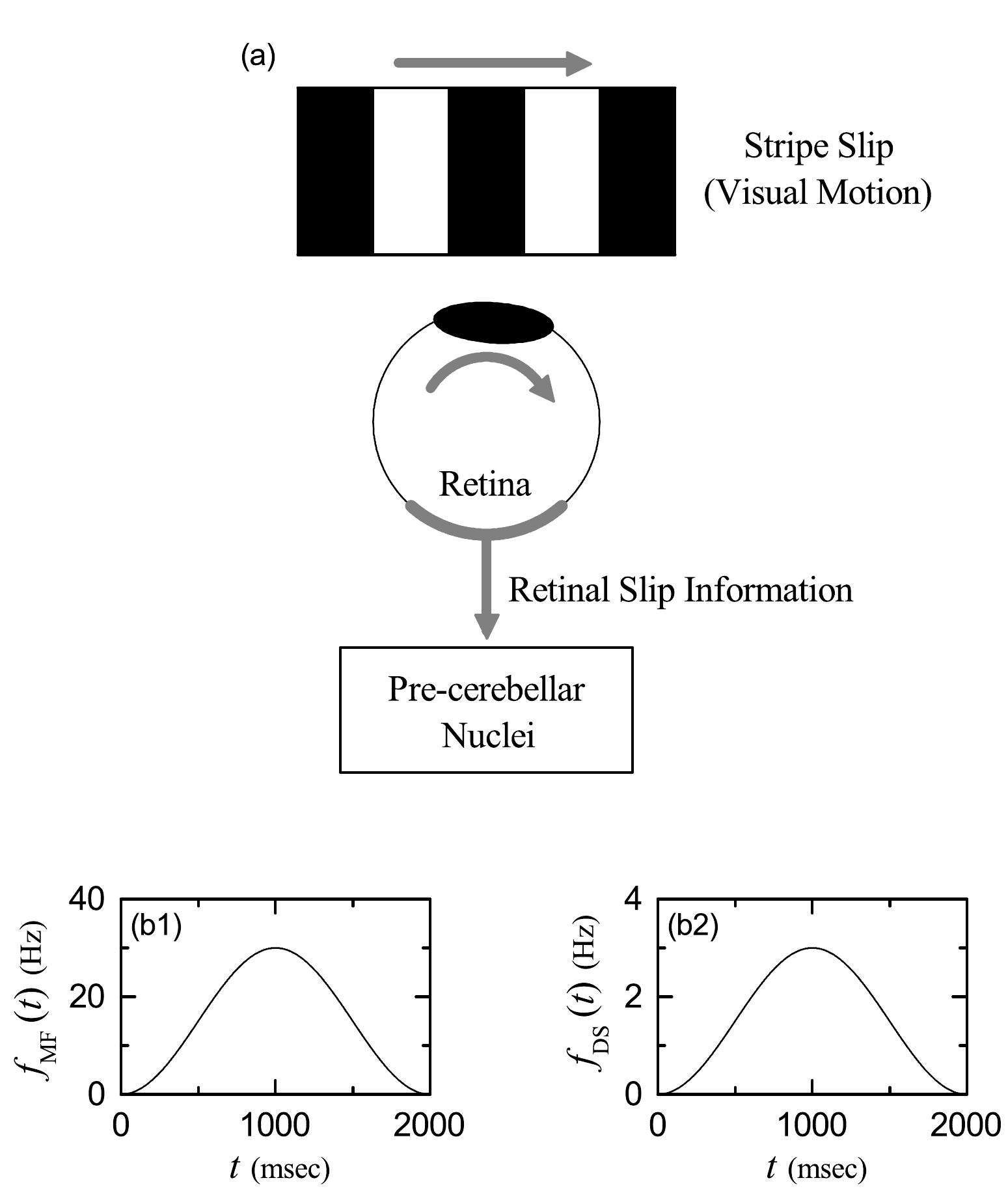}
\caption{(a) Optokinetic response when the eye tracks stripe slip. External sensory signals: (b1) firing rate $f_{\rm MF}(t)$ of the mossy-fiber (MF) context signal for the post-eye-movement and (b2) firing rate $f_{\rm DS}(t)$ of the inferior-olive (IO) desired signal (DS) for a desired eye-movement.
}
\label{fig:OKR}
\end{figure}

\section{Cerebellar Ring Network with Synaptic Plasticity}
\label{sec:CRN}
In this section, we describe our cerebellar ring network with synaptic plasticity for the OKR adaptation. Figure \ref{fig:OKR}(a) shows OKR which may be seen
when the eye tracks successive stripe slip with the stationary head. When each moving stripe is out of the field of vision, the eye moves back quickly to the original position where it first saw. Thus, OKR is composed of two consecutive slow and fast phases (i.e., slow tracking eye-movement and fast reset saccade).
It takes 2 sec (corresponding to 0.5 Hz) for one complete slip of each stripe. Slip of the visual image across large portions of the retina is the stimulus that stimulates optokinetic eye movements, and also the stimulus that produces the adaptation of the optokinetic system.

\subsection{MF Context Signal and IO Desired Signal}
\label{subsec:Signals}
There are two types of sensory signals which transfer the retinal slip information from the retina to their targets by passing intermediate pre-cerebellar nuclei
(PCN). In the 1st case, the retinal slip information first passes the pretectum in the midbrain, then passes the nucleus reticularis tegmentis pontis (NRTP) in the pons, and finally it is transferred to the granular layer (consisting of GR and GO cells) in the cerebellar cortex via MF sensory signal containing ``context'' for
the post-eye-movent. The MF context signals are modeled in terms of Poisson spike trains which modulate sinusoidally at the stripe-slip frequency $f_s=0.5$ Hz (i.e., one-cycle period: 2 sec) with the peak firing rate of 30 Hz (i.e., 30 spikes/sec) \cite{Yama1}. The firing frequency $f_{\rm MF}$ of Poisson spike trains for the MF context signal is
given by
\begin{equation}
f_{\rm MF}(t) = - {\overline f}_{\rm MF} \cos(2\pi f_s t) + {\overline f}_{\rm MF}; ~~~{\overline f}_{\rm MF}=15~{\rm Hz},
\label{eq:MF}
\end{equation}
which is shown in Fig.~\ref{fig:OKR}(b1).

In the 2nd case, the retinal slip information passes only the pretectum, and then (without passing NRTP) it is directly fed into to the IO via a sensory signal for a ``desired'' eye-movement. As in the MF context signals, the IO desired signals are also modeled in terms of the same kind of sinusoidally modulating Poisson spike trains at the stripe-slip frequency $f_s=0.5$ Hz.
The firing frequency $f_{\rm DS}$  of Poisson spike trains for the IO desired signal (DS) is given by:
\begin{equation}
f_{\rm DS}(t) = - {\overline f}_{\rm DS} \cos(2\pi f_s t) + {\overline f}_{\rm DS}; ~~~{\overline f}_{\rm DS}=1.5~{\rm Hz},
\label{eq:DS}
\end{equation}
which is shown in Fig.~\ref{fig:OKR}(b2).
In this case, the peak firing rate for the IO desired signal is reduced to 3 Hz to satisfy low mean firing rates ({\small $\sim$} 1.5 Hz) of individual IO neurons (i.e., corresponding to $1/10$ of the peak firing rate of the MF signal) \cite{IO1,IO2}.

\begin{figure}
\includegraphics[width=0.95\columnwidth]{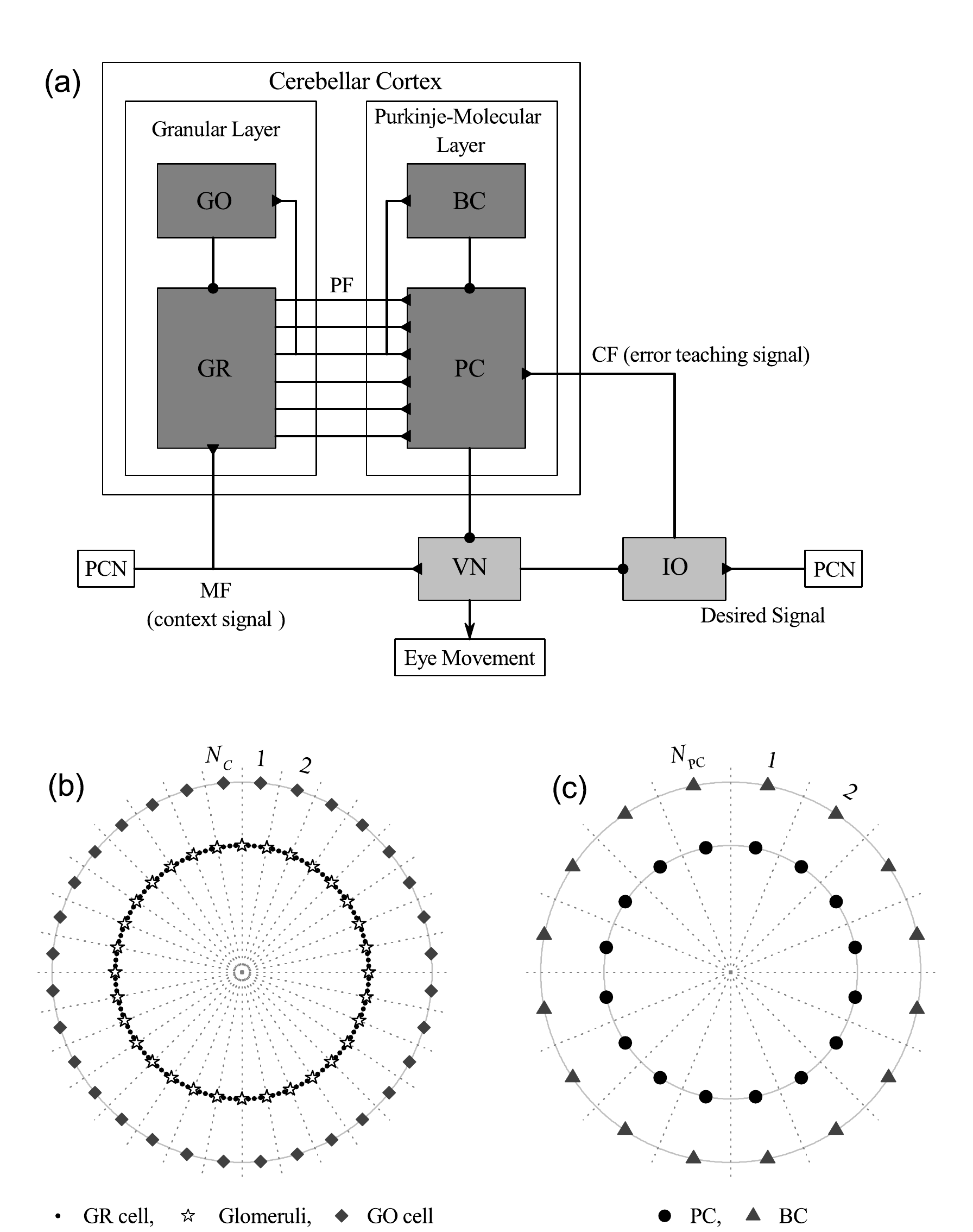}
\caption{Cerebellar Ring Network. (a) Box diagram for the cerebellar network. Lines with triangles and circles denote excitatory and inhibitory synapses, respectively. GR (granule cell), GO (Golgi cell), and PF (parallel fiber) in the granular layer, PC (Purkinje cell) and BC (basket cell) in the Purkinje-molecular layer, and other parts for VN (vestibular nuclei), IO(inferior olive), PCN(pre-cerebellar nuclei), MF (mossy fiber), and CF (climbing fiber).
(b) Schematic diagram for granular-layer ring network with concentric inner GR and outer GO rings. Numbers represent granular layer zones (bounded by dotted lines) for $N_C=32$. In each $I$th zone ($I=1, \cdots, N_C$), there exists the $I$th GR cluster on the inner GR ring. Each GR cluster consists of GR cells (solid circles), and it is bounded by 2 glomeruli (stars). On the outer GO ring in the $I$th zone, there exists the $I$th GO cell (diamonds).
(c) Schematic diagram for Purkinje-molecular-layer ring network with concentric inner PC and outer BC rings. Numbers represent the Purkinje-molecular-layer zones (bounded by dotted lines) for $N_{\rm PC}=16$. In each $J$th zone, there exist the $J$th PC (solid circle) on the inner PC ring and the $J$th BC (solid triangle)
on the outer BC ring.
}
\label{fig:RN}
\end{figure}

\subsection{Architecture for Cerebellar Ring Network}
\label{subsec:ACRN}
As in the famous small-world ring network \cite{SWN1,SWN2}, we develop a one-dimensional ring network with a simple architecture, which is in contrast to the
two-dimensional square-lattice network \cite{Yama1,Yama2}. This kind of ring network has advantage for computational and analytical efficiency, and its
visual representation may also be easily made.

Here, we employ such a cerebellar ring network for the OKR.
Figure \ref{fig:RN}(a) shows the box diagram for the cerebellar network. The granular layer, corresponding to the input layer of the cerebellar cortex, consists of the excitatory GR cells and the inhibitory GO cells. On the other hand, the Purkinje-molecular layer, corresponding to the output layer of the cerebellar cortex, is composed of the inhibitory PCs and the inhibitory BCs (basket cells). The MF context signal for the post-eye-movement is fed from the PCN (pre-cerebellar nuclei) to the GR cells. They are diversely recoded via inhibitory coordination of GO cells on GR cells in the granular layer. Then, these diversely-recoded outputs are fed via PFs to the PCs and the BCs in the Purkinje-molecular layer.

The PCs receive another excitatory error-teaching CF signals from the IO, along with the inhibitory inputs from BCs.
Then, depending on the type of PF signals (i.e., in-phase or out-of-phase PF signals), diverse PF (student) signals are effectively depressed by the (in-phase) error-teaching (instructor) CF signals. Such effective depression at PF-PC synapses causes a large modulation in firing activities of PCs (principal output cells in the cerebellar cortex). Then, the VN neuron generates the final output of the cerebellum (i.e., it evokes OKR eye-movement) through receiving both the inhibitory inputs from the PCs and the excitatory inputs via MFs. This VN neuron also provides inhibitory inputs for the realized eye-movement to the IO neuron which also receives the excitatory desired signals for a desired eye-movement from the PCN. Then, the IO neuron supplies excitatory error-teaching CF signals to the PCs.

Figure \ref{fig:RN}(b) shows a schematic diagram for the granular-layer ring network with concentric inner GR and outer GO rings.
Numbers represent granular-layer zones (bounded by dotted lines).
That is, the numbers 1, 2, $\cdots$, and $N_C$ denote the 1st, the 2nd, $\cdots$, and the $N_C$th granular-layer zones, respectively.
Hence, the total number of granular-layer zones is $N_C$; Fig.~2(b) shows an example for $N_C=32$.
In each $I$th zone ($I=1,\cdots, N_C$), there exists the $I$th GR cluster on the inner GR ring.
Each GR cluster consists of $N_{\rm GR}$ excitatory GR cells (solid circles). Then, location of each GR cell may be represented by the two indices $(I,i)$ which represent the $i$th GR cell in the $I$th GR cluster, where $i=1,\cdots,N_{\rm GR}$. Here, we consider the case of $N_C=2^{10}$ and $N_{\rm GR}=50$, and hence
the total number of GR cells is 51,200. In this case, the $I$th zone covers the angular range of $(I-1)~ \theta_{\rm GR}^* < \theta < I~ \theta_{\rm GR}^*$
($\theta_{\rm GR}^* = 0.35^{\circ}$). On the outer GO ring in each $I$th zone, there exists the $I$th inhibitory GO cell (diamond), and hence the total number of GO cells is $N_C$.

We note that each GR cluster is bounded by 2 glomeruli (corresponding to the axon terminals of the MFs) (stars).
GR cells within each GR cluster share the same inhibitory and excitatory synaptic inputs through their dendrites which contact the two glomeruli at both ends
of the GR cluster. Each glomerulus receives inhibitory inputs from nearby 81 (clockwise side: 41 and counter-clockwise side: 40) GO cells with a random connection
probability $p_c~(=0.06)$. Hence, on average, about 5 GO cell axons innervate each glomerulus. Thus, each GR cell receives about 10 inhibitory  inputs through 2 dendrites which synaptically contact the glomeruli at both boundaries. In this way, each GR cell in the GR cluster shares the same
inhibitory synaptic inputs from nearby GO cells through the intermediate glomeruli at both ends.

Also, each GR cell shares the same two excitatory inputs via the two glomeruli at both boundaries, because a glomerulus
receives an excitatory MF input. Here, we take into consideration stochastic variability of synaptic transmission from a glomerulus to GR cells,
and supply independent Poisson spike trains with the same firing rate to each GR cell for the excitatory MF signals.
In this GR-GO feedback system, each GO cell receives excitatory synaptic inputs via PFs from GR cells in the nearby 49 (central side: 1, clockwise side: 24 and counter-clockwise side: 24) GR clusters with a random connection probability $0.1$. Thus, 245 PFs (i.e. GR cell axons) innervate a GO cell.

Figure \ref{fig:RN}(c) shows a schematic diagram for the Purkinje-molecular-layer ring network with concentric inner PC and outer BC rings.
Numbers denote the Purkinje-molecular-layer zones (bounded by dotted lines). In each $J$th zone ($J=1,\cdots, N_{\rm PC}$), there exist the $J$th PC (solid circles) on the inner PC ring and the $J$th BC (solid triangles) on the outer BC ring. Here, we consider the case of $N_{\rm PC}=16,$ and hence the total numbers of PC and BC are 16, respectively. In this case, each $J$th ($J=1,\cdots,N_{\rm PC}$) zone covers the angular range of $(J-1)~ \theta_{\rm PC}^* < \theta < J~ \theta_{\rm PC}^*,$ where
$\theta_{\rm PC}^* \simeq 22.5^{\circ}$ (corresponding to about 64 zones in the granular-layer ring network).
We note that diversely-recoded PFs innervate PCs and BCs. Each PC (BC) in the $J$th Purkinje-molecular-layer zone receives excitatory synaptic inputs via PFs from
all the GR cells in the 288 GR clusters (clockwise side: 144 and counter-clockwise side: 144 when starting from the angle $\theta = (J-1)~ \theta_{\rm PC}^*$ in the granular-layer ring network). Thus, each PC (BC) is synaptically connected via PFs to the 14,400 GR cells (which corresponds to about 28 $\%$ of the total GR cells).
In addition to the PF signals, each PC also receives inhibitory inputs from nearby 3 BCs (central side: 1, clockwise side: 1 and counter-clockwise side: 1) and
excitatory error-teaching CF signal from the IO.

Outside the cerebellar cortex, for simplicity, we consider just one VN neuron and one IO neuron. Both excitatory inputs via 100 MFs and inhibitory inputs from all the 16 PCs are fed into the VN neuron. Then, the VN neuron evokes the OKR eye-movement and supplies inhibitory input for the realized eye-movement to the IO neuron.
One additional excitatory desired signal from the PCN is also fed into the IO neuron. Then, through integration of both excitatory and inhibitory inputs, the IO neuron provides excitatory error-teaching CF signals to the PCs.

\subsection{Leaky Integrate-And-Fire Neuron Model with Afterhyperpolarization Current}
\label{subsec:LIF}
As elements of the cerebellar ring network, we choose leaky integrate-and-fire (LIF) neuron models which incorporate additional afterhyperpolarization (AHP) currents that determine refractory periods \cite{LIF}. This LIF neuron model is one of the simplest spiking neuron models. Due to its simplicity, it can be easily
analyzed and simulated. Thus, it has been very popularly used as a neuron model.

The following equations govern dynamics of states of individual neurons in the $X$ population:
\begin{equation}
C_{X} \frac{dv_{i}^{(X)}}{dt} = -I_{L,i}^{(X)} - I_{AHP,i}^{(X)} + I_{ext}^{(X)} - I_{syn,i}^{(X)}, \;\;\; i=1, \cdots, N_{X},
\label{eq:GE}
\end{equation}
where $N_X$ is the total number of neurons in the $X$ population, $X=$ GR and GO in the granular layer, $X=$ PC and BC in the Purkinje-molecular layer, and in the other parts $X=$ VN and IO.
In Eq.~(1), $C_{X}$ (pF) represents the membrane capacitance of the cells in the $X$ population, and the state of the $i$th neuron in the $X$ population at a time $t$ (msec) is characterized by its membrane potential $v_i^{(X)}$ (mV). The time-evolution of $v_i^{(X)}(t)$ is governed by 4 types of currents (pA) into the
$i$th neuron in the $X$ population; the leakage current $I_{L,i}^{(X)}$, the AHP current $I_{AHP,i}^{(X)}$, the external constant current $I_{ext}^{(X)}$ (independent of $i$), and the synaptic current $I_{syn,i}^{(X)}$.

We note that the equation for a single LIF neuron model [without the AHP current and the synaptic current in Eq.~(\ref{eq:GE})] describes a simple
parallel resistor-capacitor (RC) circuit. Here, the leakage term is due to the resistor and the integration of the external current is due to the capacitor
which is in parallel to the resistor. Thus, in Eq.~(\ref{eq:GE}), the 1st type of leakage current $I_{L,i}^{(X)}$ for the $i$th neuron in the $X$ population
is given by:
\begin{equation}
I_{L,i}^{(X)} = g_{L}^{(X)} (v_{i}^{(X)} - V_{L}^{(X)}),
\label{eq:Leakage}
\end{equation}
where $g_L^{(X)}$ and $V_L^{(X)}$ are conductance (nS) and reversal potential for the leakage current, respectively.

When the membrane potential $v_i^{(X)}$ reaches a threshold $v_{th}^{(X)}$ at a time $t_{f,i}^{(X)}$, the $i$th neuron fires a spike.
After firing (i.e., $t \geq t_{f,i}^{(X)}$), the 2nd type of AHP current $I_{AHP,i}^{(X)}$ follows:
\begin{equation}
I_{AHP,i}^{(X)} = g_{AHP}^{(X)}(t) ~(v_{i}^{(X)} - V_{AHP}^{(X)})~~~{\rm ~for~} \; t \ge t_{f,i}^{(X)}.
\label{eq:AHP1}
\end{equation}
Here, $V_{AHP}^{(X)}$ is the reversal potential for the AHP current, and the conductance $g_{AHP}^{(X)}(t)$ is given by an exponential-decay
function:
\begin{equation}
g_{AHP}^{(X)}(t) = \bar{g}_{AHP}^{(X)}~  e^{-(t-t_{f,i}^{(X)})/\tau_{AHP}^{(X)}} ,
\label{eq:AHP2}
\end{equation}
where $\bar{g}_{AHP}^{(X)}$ and $\tau_{AHP}^{(X)}$ are the maximum conductance and the decay time constant for the AHP current.
As $\tau_{AHP}^{(X)}$ increases, the refractory period becomes longer.

The 3rd type of external constant current $I_{ext}^{(X)}$ for the cellular spontaneous discharge is supplied to only the PCs and the VN neuron
because of their high spontaneous firing rates \cite{PC1,PC2}. In Appendix \ref{app:A}, Table \ref{tab:SingleParm} shows the parameter values for the capacitance $C_X$, the leakage current $I_L^{(X)}$, the AHP current $I_{AHP}^{(X)}$, and the external constant current $I_{ext}^{(X)}$. These values are adopted from physiological data \cite{Yama2,Yama1}.

\subsection{Synaptic Currents}
\label{subsec:SC}
The 4th type of synaptic current $I_{syn,i}^{(X)}$ into the $i$th neuron in the $X$ population consists of the following 3 kinds of synaptic currents:
\begin{equation}
I_{syn,i}^{(X)} = I_{{\rm AMPA},i}^{(X,Y)} + I_{{\rm NMDA},i}^{(X,Y)} + I_{{\rm GABA},i}^{(X,Z)}.
\label{eq:ISyn1}
\end{equation}
Here, $I_{{\rm AMPA},i}^{(X,Y)}$ and $I_{{\rm NMDA},i}^{(X,Y)}$ are the excitatory AMPA ($\alpha$-amino-3-hydroxy-5-methyl-4-isoxazolepropionic acid) receptor-mediated and NMDA ($N$-methyl-$D$-aspartate) receptor-mediated currents from the pre-synaptic source $Y$ population to the post-synaptic $i$th neuron in the target $X$ population.
On the other hand, $I_{{\rm GABA},i}^{(X,Z)}$ is the inhibitory $\rm GABA_A$ ($\gamma$-aminobutyric acid type A) receptor-mediated current
from the pre-synaptic source $Z$ population to the post-synaptic $i$th neuron in the target $X$ population.

Similar to the case of the AHP current, the $R$ (= AMPA, NMDA, or GABA) receptor-mediated synaptic current $I_{R,i}^{(T,S)}$ from the pre-synaptic source $S$ population to the $i$th post-synaptic neuron in the target $T$ population is given by:
\begin{equation}
I_{R,i}^{(T,S)} = g_{R,i}^{(T,S)}(t)~(v_{i}^{(T)} - V_{R}^{(S)}),
\label{eq:ISyn2}
\end{equation}
where $g_{(R,i)}^{(T,S)}(t)$ and $V_R^{(S)}$ are synaptic conductance and synaptic reversal potential
(determined by the type of the pre-synaptic source $S$ population), respectively.
We get the synaptic conductance $g_{R,i}^{(T,S)}(t)$ from:
\begin{equation}
g_{R,i}^{(T,S)}(t) = \bar{g}_{R}^{(T)} \sum_{j=1}^{N_S} J_{ij}^{(T,S)}~ w_{ij}^{(T,S)} ~ s_{j}^{(T,S)}(t),
\label{eq:ISyn3}
\end{equation}
where $\bar{g}_{R}^{(T)}$  and $J_{ij}^{(T,S)}$  are the maximum conductance and the synaptic weight of the synapse
from the $j$th pre-synaptic neuron in the source $S$ population to the $i$th post-synaptic neuron in the target $T$ population, respectively.
The inter-population synaptic connection from the source $S$ population (with $N_s$ neurons) to the target $T$ population is given by the connection weight matrix
$W^{(T,S)}$ ($=\{ w_{ij}^{(T,S)} \}$) where $w_{ij}^{(T,S)}=1$ if the $j$th neuron in the source $S$ population is pre-synaptic to the $i$th neuron
in the target $T$ population; otherwise $w_{ij}^{(T,S)}=0$.

The post-synaptic ion channels are opened due to the binding of neurotransmitters (emitted from the source $S$ population) to receptors in the target
$T$ population. The fraction of open ion channels at time $t$ is represented by $s^{(T,S)}$. The time course of $s_j^{(T,S)}(t)$ of the $j$th neuron
in the source $S$ population is given by a sum of exponential-decay functions $E_{R}^{(T,S)} (t - t_{f}^{(j)})$:
\begin{equation}
s_{j}^{(T,S)}(t) = \sum_{f=1}^{F_{j}^{(s)}} E_{R}^{(T,S)} (t - t_{f}^{(j)}),
\label{eq:ISyn4}
\end{equation}
where $t_f^{(j)}$ and $F_j^{(s)}$ are the $f$th spike time and the total number of spikes of the $j$th neuron in the source $S$ population, respectively.
The exponential-decay function $E_{R}^{(T,S)} (t)$ (which corresponds to contribution of a pre-synaptic spike occurring at $t=0$ in the absence of synaptic delay)
is given by:
\begin{subequations}
\begin{eqnarray}
E_{R}^{(T,S)}(t) &=& e^{-t/\tau_{R}^{(T)}} \Theta(t)~~~{\rm or} \label{eq:ISyn5a} \\
 &=& (A_{1} e^{-t/\tau_{R,1}^{(T)}} + A_{2} e^{-t/\tau_{R,2}^{(T)}}) \Theta(t), \label{eq:ISyn5b}
\end{eqnarray}
\end{subequations}
where $\Theta(t)$ is the Heaviside step function: $\Theta(t)=1$ for $t \geq 0$ and 0 for $t <0$.
Depending on the source and the target populations, $E_{R}^{(T,S)} (t)$ may be a type-1 single exponential-decay function of
Eq.~(\ref{eq:ISyn5a}) or a type-2 dual exponential-decay function of Eq.~(\ref{eq:ISyn5b}). In the type-1 case, there exists one synaptic decay time constant $\tau_R^{(T)}$ (determined by the receptor on the post-synaptic target $T$ population), while in the type-2 case, two synaptic decay time constants, $\tau_{R,1}^{(T)}$ and $\tau_{R,2}^{(T)}$ exist.
In most cases, the type-1 single exponential-decay function of Eq.~(\ref{eq:ISyn5a}) appears, except for the two synaptic currents $I_{\rm GABA}^{\rm ( GR,GO)}$
and $I_{\rm NMDA}^{\rm (GO,GR)}$.

In Appendix \ref{app:A}, Table \ref{tab:SynParm} shows the parameter values for the maximum conductance $\bar{g}_{R}^{(T)}$, the synaptic weight $J_{ij}^{(T,S)}$, the synaptic reversal potential $V_{R}^{(S)}$, the synaptic decay time constant $\tau_{R}^{(T)}$, and the amplitudes $A_1$ and $A_2$ for the type-2 exponential-decay function in the granular layer, the Purkinje-molecular layer, and the other parts for the VN and IO, respectively. These values are adopted from physiological data \cite{Yama2,Yama1}.

\subsection{Synaptic Plasticity}
\label{subsec:SP}
We use a rule for synaptic plasticity, based on the experimental result in \cite{Safo}.
This rule is a refined one for the LTD in comparison to the rule used in \cite{Yama1,Yama2}, the details of which will be explained below.

The coupling strength of the synapse from the pre-synaptic neuron $j$ in the source $S$ population to the post-synaptic neuron $i$ in the target $T$ population is $J_{ij}^{(T,S)}$. Initial synaptic strengths are given in Table \ref{tab:SynParm}.
Here, we assume that learning occurs only at the PF-PC synapses. Hence, only the synaptic strengths $J_{ij}^{\rm (PC,PF)}$ of PF-PC synapses
may be modifiable, while synaptic strengths of all the other synapses are static.
[Here, the index $j$ for the PFs corresponds to the two indices $(M,m)$ for GR cells representing the $m$th ($1 \leq m \leq 50$) cell in the $M$th
($1 \leq M \leq 2^{10}$) GR cluster.] Synaptic plasticity at PF-PC synapses have been so much studied in diverse experimental \cite{Ito5,Ito6,Sakurai,Ito7,SPExp1,SPExp2,SPExp3,SPExp6,SPExp4,SPExp5,Safo,SPExp7,SPExp8,SPExp9} and computational \cite{Albus,SPCom1,SPCom2,SPCom3,SPCom4,SPCom5,Yama2,SPCom6,Yama1,SPCom7} works.

With increasing time $t$, synaptic strength for each PF-PC synapse is updated with the following multiplicative rule (depending on states) \cite{Safo}:
\begin{widetext}
\begin{equation}
J_{ij}^{\rm (PC,PF)}(t) \rightarrow J_{ij}^{\rm (PC,PF)}(t) + \Delta J_{ij}^{\rm (PC,PF)}(t),
\label{eq:SP}
\end{equation}
where
\begin{eqnarray}
\Delta J_{ij}^{\rm (PC,PF)}(t) &=& \Delta{\rm LTD}_{ij}^{(1)} + \Delta{\rm LTD}_{ij}^{(2)} + \Delta{\rm LTP}_{ij}, \label{eq:DeltaJ} \\
\Delta{\rm LTD}_{ij}^{(1)} &=& - \delta_{LTD} \cdot J_{ij}^{\rm (PC,PF)}(t) \cdot CF_{i}(t) \cdot \sum_{\Delta t=0}^{\Delta t_{r}^{*}} \Delta J_{LTD}(\Delta t),
\label{eq:LTD1} \\
\Delta{\rm LTD}_{ij}^{(2)} &=& - \delta_{LTD} \cdot J_{ij}^{\rm (PC,PF)}(t) \cdot [ 1- CF_{i}(t) ] \cdot PF_{ij}(t) \cdot D_{i}(t) \cdot \sum_{\Delta t=0}^{\Delta t_{l}^{*}} \Delta J_{LTD} (\Delta t),
\label{eq:LTD2} \\
\Delta{\rm LTP}_{ij} &=&  \delta_{LTP} \cdot [J_{0}^{\rm (PC,PF)} - J_{ij}^{\rm (PC,PF)}(t)] \cdot [1-CF_{i}(t)] \cdot PF_{ij}(t) \cdot [1-D_{i}(t)].
\label{eq:LTP}
\end{eqnarray}
\end{widetext}
Here, $J_{0}^{\rm (PC,PF)}$ is the initial value (=0.006) for the synaptic strength of PF-PC synapses.
Synaptic modification (LTD or LTP) occurs, depending on the relative time difference $\Delta t$ [= $t_{\rm CF}$ (CF activation time) - $t_{\rm PF}$ (PF activation time)] between the spiking times of the error-teaching instructor CF and the diversely-recoded student PFs.
In Eqs.~(\ref{eq:LTD1})-(\ref{eq:LTP}), $CF_i(t)$ represents a spike train of the CF signal coming into the $i$th PC.
When $CF_i(t)$ activates at a time $t$, $CF_i(t)=1$; otherwise, $CF_i(t)=0$. This instructor CF firing causes LTD at PF-PC synapses in conjunction with earlier ($\Delta t >0)$ student PF firings in the range of $t_{\rm CF} - \Delta t_r^* < t_{\rm PF} <t_{\rm CF}$ ($\Delta t_r^* \simeq 277.5$ msec), which corresponds to the
major LTD in Eq.~(\ref{eq:LTD1}).

We next consider the case of $CF_i(t)=0$, corresponding to Eqs.~(\ref{eq:LTD2}) and (\ref{eq:LTP}).
Here, $PF_{ij}(t)$ denotes a spike train of the PF signal from the $j$th pre-synaptic GR cell to the $i$th post-synaptic PC.
When $PF_{ij}(t)$ activates at time $t$, $PF_{ij}(t)=1$; otherwise, $PF_{ij}(t)=0$.
In the case of $PF_{ij}(t)=1$, PF firing may give rise to LTD or LTP, depending on the presence of earlier CF firings in an effective range.
If CF firings exist in the range of $t_{\rm PF} + \Delta t_l^* < t_{\rm CF} <t_{\rm PF}$ ($\Delta t_l^* \simeq -117.5$ msec), $D_i(t)=1$; otherwise $D_i(t)=0$.
When both $PF_{ij}(t)=1$ and $D_i(t)=1$, the PF firing causes another LTD at PF-PC synapses in association with earlier ($\Delta t <0$) CF firings
[see Eq.~(\ref{eq:LTD2})]. The likelihood for occurrence of earlier CF firings within the effective range is very low because mean firing rates of the CF signals
(corresponding to output firings of individual IO neurons) are {\small $\sim$} 1.5 Hz \cite{IO1,IO2}. Hence, this 2nd type of LTD is a minor one.
On the other hand, in the case of $D_i(t)=0$ (i.e., absence of earlier associated CF firings), LTP occurs due to the PF firing alone
[see Eq.~(\ref{eq:LTP})]. The update rate $\delta_{LTD}$ for LTD in Eqs.~(\ref{eq:LTD1}) and (\ref{eq:LTD2}) is 0.005, while the update rate $\delta_{LTP}$ for LTP
in Eqs.~(\ref{eq:LTP}) is 0.0005 (=$\delta_{LTD}/10$).

\begin{figure}
\includegraphics[width=0.65\columnwidth]{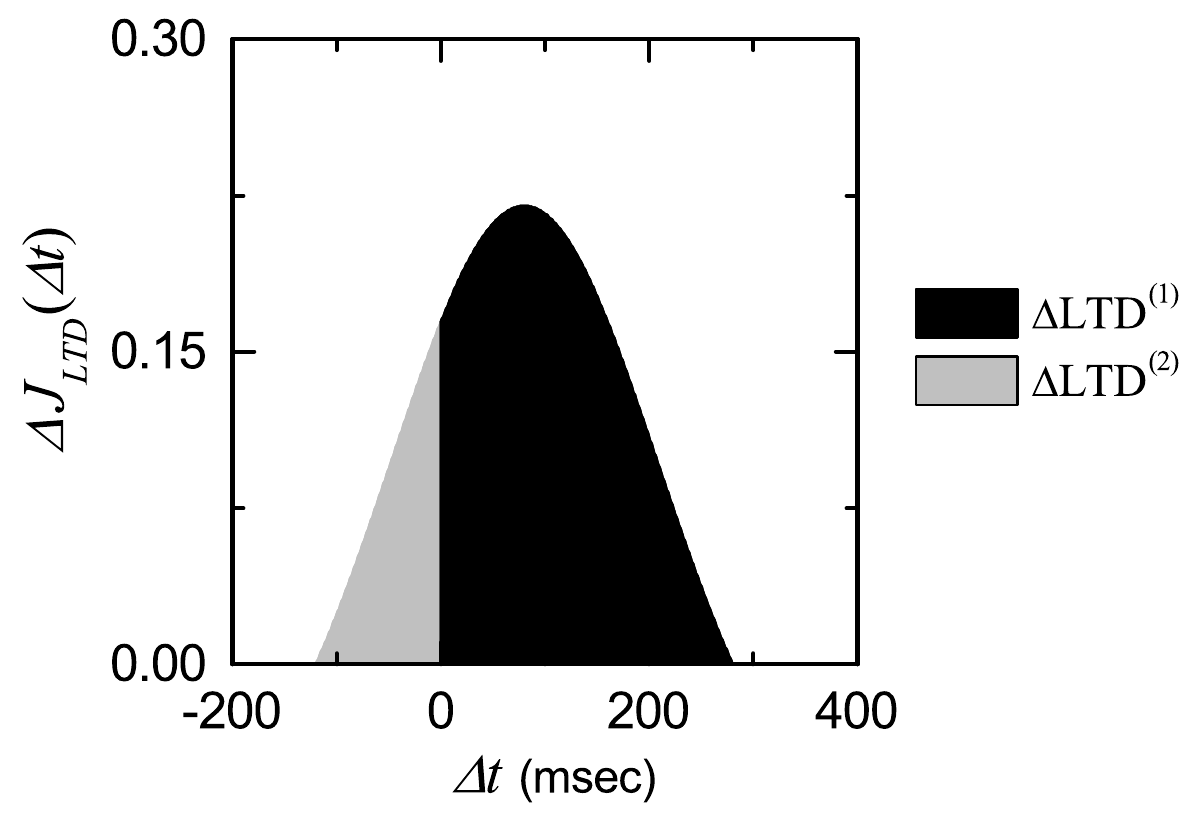}
\caption{Time window for the LTD at the PF-PC synapse. Plot of synaptic modification $\Delta J_{LTD}(\Delta t)$ for LTD versus $\Delta t$
[see Eq.~(\ref{eq:TW})].
}
\label{fig:TW}
\end{figure}

In the case of LTD in Eqs.~(\ref{eq:LTD1}) and (\ref{eq:LTD2}), the synaptic modification $\Delta J_{LTD} (\Delta t)$ varies depending on the relative time difference $\Delta t$ $(= t_{CF} - t_{PF}$).
We employ the following time window for the synaptic modification $\Delta J_{LTD} (\Delta t)$ \cite{Safo}:
\begin{equation}
\Delta J_{LTD}(\Delta t) = A  + B \cdot e^{-(\Delta t -t_0)^2/\sigma^2},
\label{eq:TW}
\end{equation}
where $A=-0.12$, $B=0.4$, $t_0 = 80$, and $\sigma=180$.
Figure \ref{fig:TW} shows the time window for $\Delta J_{LTD} (\Delta t)$.
As shown well in Fig.~\ref{fig:TW}, LTD occurs in an effective range of $\Delta t_l^* < \Delta t < \Delta t_r^*$.
We note that a peak exists at $t_0=80$ msec, and hence peak LTD occurs when PF firing precedes CF firing by 80 msec.
A CF firing causes LTD in conjunction with earlier PF firings in the black region ($0< \Delta t < \Delta t_r^*$), and
it also gives rise to another LTD in association with later PF firings in the gray region ($\Delta t_l^* < \Delta t <0$).
The effect of CF firing on earlier PF firings is much larger than that on later PF firings.
However, outside the effective range (i.e., $\Delta t > \Delta t_r^*$ or $< \Delta t_l^*$), PF firings alone leads to LTP,
due to absence of effectively associated CF firings.

Finally, we discuss the advantages of our refined rule for synaptic plasticity in comparison to the synaptic rule in \cite{Yama1,Yama2}.
Our rule is constructed from the experimental result in \cite{Safo}. In the presence of a CF firing, a major LTD  ($\Delta {\rm LTD}^{(1)}$) takes place in association with earlier PF firings in the range of $t_{\rm CF} - \Delta t_r^* < t_{\rm PF} <t_{\rm CF}$ ($\Delta t_r^* \simeq 277.5$ msec), while a minor LTD ($\Delta {\rm LTD}^{(2)}$) occurs in association with later PF firings in the range of $t_{\rm CF} < t_{\rm PF} <t_{\rm CF} - \Delta t_l^*$ ($\Delta t_l^* \simeq -117.5$ msec). The magnitude of LTD changes depending on $\Delta t$ (= $t_{\rm CF}$ - $t_{\rm PF}$); a peak LTD occurs for $\Delta t =80$ msec.
On the other hand, the rule in \cite{Yama1,Yama2} considers only the major LTD in conjunction with earlier PF firings in the range of $t_{\rm CF} - 50 < t_{\rm PF} <t_{\rm CF}$, the magnitude of major LTD is equal, independently of $\Delta t$, and minor LTD in association with later PF firings is not considered.
Outside the effective range of LTD, PF firings alone result in LTP in both rules.
However, we also note that some features of OKR were successfully reproduced by using the simple synaptic rule with only the major LTD in \citep{Yama1,Yama2}.

\subsection{Numerical Method for Integration}
\label{subsec:NM}
Numerical integration of the governing Eq.~(\ref{eq:GE}) for the time-evolution of states of individual neurons,
along with the update rule for synaptic plasticity of Eq.~(\ref{eq:SP}), is done by employing the 2nd-order Runge-Kutta method
with the time step 1 msec.
For each realization, we choose random initial points $v_i^{(X)}(0)$ for the $i$th neuron in the $X$ population
with uniform probability in the range of $v_i^{(X)}(0) \in (V_L^{(X)}-5.0, V_L^{(X)}+5.0)$; the values of $V_L^{(X)}$ are
given in Table \ref{tab:SingleParm}.

\section{Effect of Diverse Spiking Patterns of GR Clusters on Motor Learning for The OKR Adaption}
\label{sec:MS}
In this section, we study the effect of diverse recoding of GR cells on motor learning for the OKR adaptation
by varying the connection probability $p_c$ from the GO to the GR cells. We mainly consider an optimal case
of $p_c^*=0.06$ where the spiking patterns of GR clusters are the most diverse. In this case, we first make dynamical
classification of diverse spiking patterns of the GR clusters. Then, we make an intensive investigation on the effect of diverse recoding of GR cells on synaptic plasticity at PF-PC synapses and the subsequent learning process in the PC-VN-IO system. Finally, we vary $p_c$ from the optimal value $p_c^*$, and study dependence of the diversity degree $\cal{D}$ of spiking patterns and the saturated learning gain degree ${\cal L}_g^*$ on $p_c$. Both $\cal{D}$ and ${\cal L}_g^*$ are found to form bell-shaped curves with peaks at $p_c^*$, and they have strong correlation with the Pearson's coefficient $r\simeq 0.9998$. As a result, the more diverse in recoding of GR cells, the more effective in the motor learning for the OKR adaptation.

\subsection{Firing Activity in The Whole Population of GR Cells}
\label{subsec:WP}
As shown in Fig.~\ref{fig:RN}, recoding process is performed in the granular layer (corresponding to the input layer of the cerebellar cortex), consisting of GR and GO cells. In the GR-GO feedback system, GR cells (principal output cells in the granular layer) receive excitatory context signals
for the post-eye-movement via the sinusoidally-modulating MFs [see Fig.~\ref{fig:OKR}(b1)] and make recoding of context signals. In this recoding process, GO cells make effective inhibitory coordination for diverse recoding of GR cells. Thus, diversely recoded signals are fed into the PCs
(principal output cells in the cerebellar cortex) via PFs. Due to this type of diverse recoding of GR cells, the cerebellum was recently reinterpreted as a liquid state machine with powerful discriminating/separating capability (i.e., different input signals are transformed into more different ones via recoding process) rather than the simple perceptron in the Marr-Albus-Ito theory \cite{Yama3,Ma}.

\begin{figure}
\includegraphics[width=0.95\columnwidth]{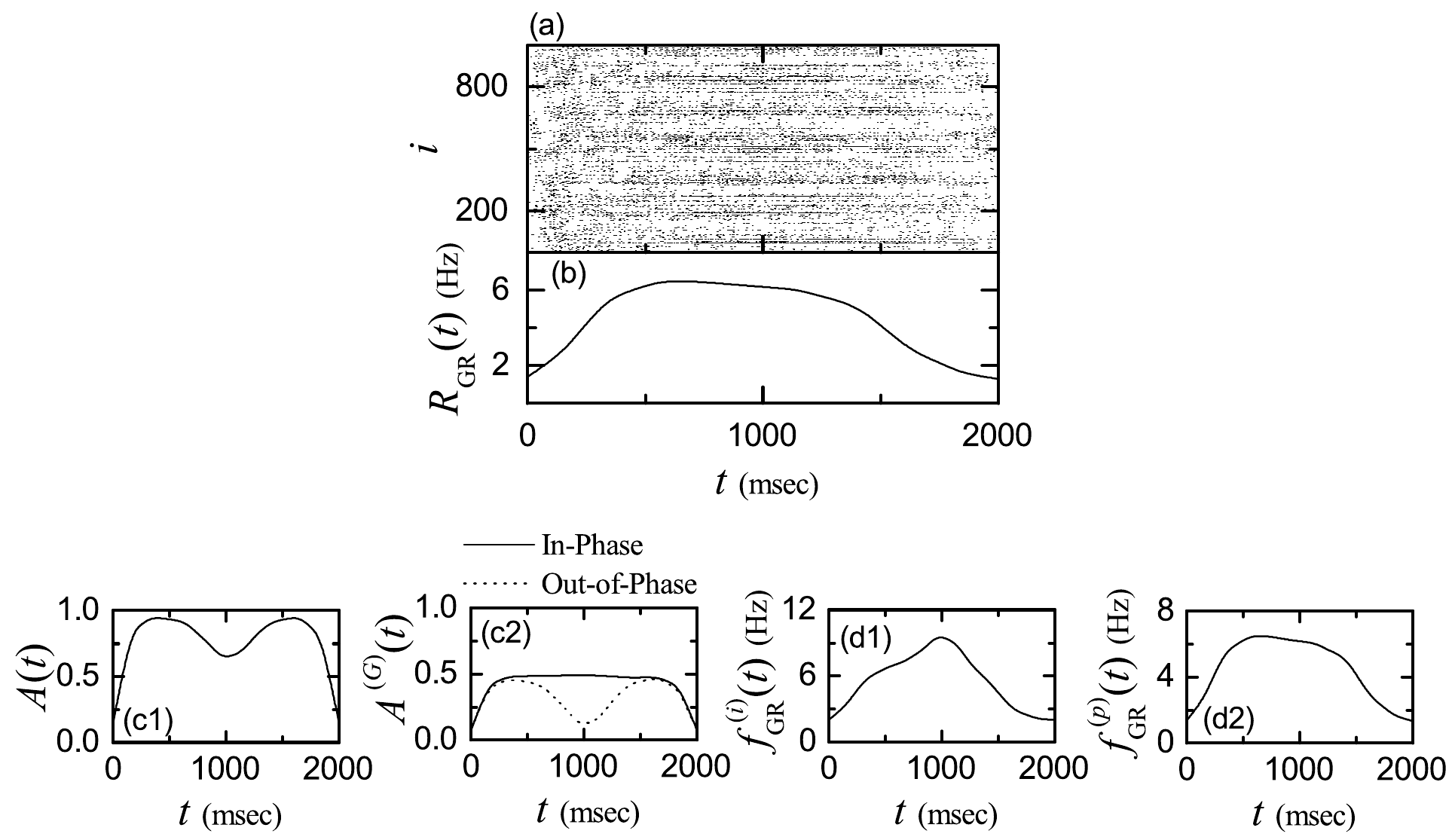}
\caption{Firing activity of GR cells in an optimal case of $p_c$ (connection probability from GO to GR cells) = 0.06. (a) Raster plots of spikes of $10^3$ randomly chosen GR cells. (b) Instantaneous whole-population spike rate $R_{\rm GR}(t)$ in the whole population of GR cells. Band width for
$R_{\rm GR}(t)$: $h=10$ msec. Plots of the activation degrees (c1) $A(t)$ in the whole population of GR cells and (c2) $A^{(G)}(t)$ in the $G$ spiking group
[$G:$ in-phase (solid curve) and out-of-phase (dotted curve)]. Plots of (d1) instantaneous individual firing rate $f_{\rm GR}^{(i)}(t)$ for the active GR cells and (d2) instantaneous population spike rate $f_{\rm GR}^{(p)}(t)$ in the whole population of GR cells. Bin size for (c1)-(d2): $\Delta t=$ 10 msec.
}
\label{fig:WP}
\end{figure}

We first consider the firing activity in the whole population of GR cells for $p_c^*=0.06$.
Collective firing activity may be well visualized in the raster plot of spikes which is a collection of spike trains of individual neurons.
Such raster plots of spikes are fundamental data in experimental neuroscience. As a population quantity showing collective firing behaviors,
we use an instantaneous whole-population spike rate $R_{\rm GR}(t)$ which may be obtained from the raster plots of spikes \cite{Sparse2,Sparse1,Sparse3,Sparse4,Sparse5,Sparse6,W_Review,RM}.
To obtain a smooth instantaneous whole-population spike rate, we employ the kernel density estimation (kernel smoother) \cite{Kernel}.
Each spike in the raster plot is convoluted (or blurred) with a kernel function $K_{h}(t)$ [such as a smooth Gaussian function in Eq.~(\ref{eq:Gaussian})],
and then a smooth estimate of instantaneous whole-population spike rate $R_{\rm GR}(t)$ is obtained by averaging the convoluted kernel function over all spikes of
GR cells in the whole population:
\begin{equation}
R_{\rm GR}(t) = \frac{1}{N} \sum_{i=1}^{N} \sum_{s=1}^{n_i} K_h (t-t_{s}^{(i)}),
\label{eq:IWPSR}
\end{equation}
where $t_{s}^{(i)}$ is the $s$th spiking time of the $i$th GR cell, $n_i$ is the total number of spikes for the $i$th GR cell, and $N$ is the total number of
GR cells (i.e., $N = N_c \cdot N_{\rm GR} = 51,200$).
Here, we use a Gaussian kernel function of band width $h$:
\begin{equation}
K_h (t) = \frac{1}{\sqrt{2\pi}h} e^{-t^2 / 2h^2}, ~~~~ -\infty < t < \infty.
\label{eq:Gaussian}
\end{equation}
Throughout the paper, the band width $h$ of $K_h(t)$ is 10 msec.

Figure \ref{fig:WP}(a) shows a raster plot of spikes of $10^3$ randomly chosen GR cells.
At the initial and the final stages of the cycle, GR cells fire sparse spikes, because the firing rates of Poisson spikes for the MF are low. On the other hand, at the middle stage, the firing rates for the MF are relatively high, and hence spikes of GR cells become relatively dense.
Figure \ref{fig:WP}(b) shows the instantaneous whole-population spike rate $R_{\rm GR}(t)$ in the whole population of GR cells.
$R_{\rm GR}(t)$ is basically in proportion to the sinusoidally-modulating inputs via MFs.  However, it has a different waveform with a central plateau.
At the initial stage, it rises rapidly, then a broad plateau appears at the middle stage, and at the final stage, it decreases slowly.
In comparison to the MF signal, the top part of $R_{\rm GR}(t)$ becomes lowered and flattened, due to the effect of inhibitory GO cells.
Thus, a central plateau emerges.

We next consider the activation degree of GR cells.
To examine it, we divide the whole learning cycle (2000 msec) into 200 bins (bin size: 10 msec).
Then, we get the activation degree $A_i$ for the active GR cells in the $i$th bin:
\begin{equation}
A_i = \frac {N_{a,i}} {N},
\label{eq:AD}
\end{equation}
where $N_{a,i}$ and $N$ are the number of active GR cells in the $i$th bin and the total number of GR cells, respectively.
Figure \ref{fig:WP}(c1) shows a plot of the activation degree $A(t)$ in the whole population of GR cells.
It is nearly symmetric, and has double peaks with a central valley at the middle stage; its values at both peaks are about 0.94
and the central minimum value is about 0.65.

\begin{figure}
\includegraphics[width=0.95\columnwidth]{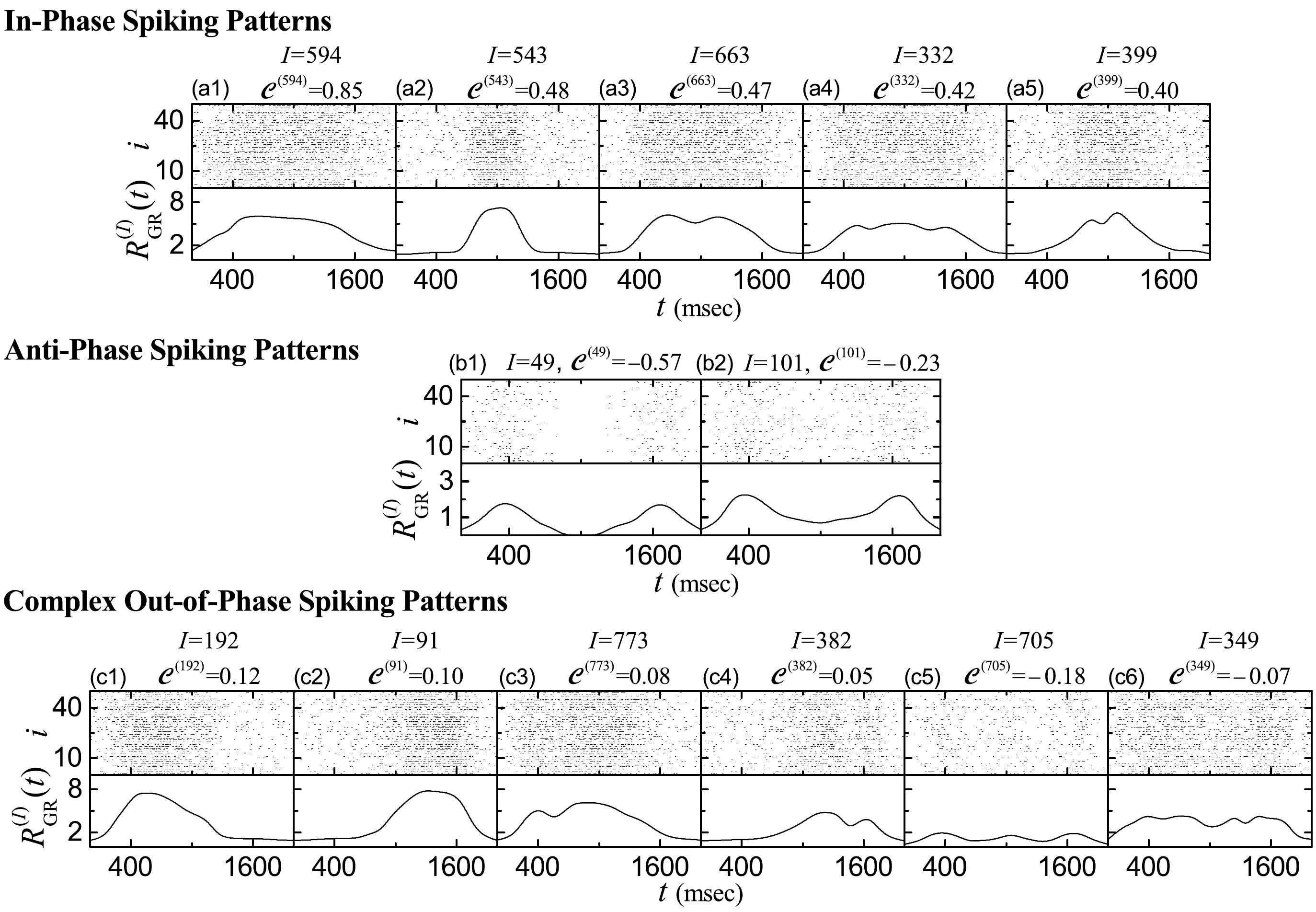}
\caption{Diverse spiking patterns in the GR clusters in the optimal case of $p_c^* = 0.06$. Raster plots of spikes and instantaneous cluster spike rates  $R_{\rm GR}^{(I)}(t)$ for diverse spiking patterns. Five in-phase spiking patterns in the $I$th GR clusters; $I=$ (a1) 594, (a2) 543, (a3) 663, (a4) 332, and
(a5) 399. Two anti-phase spiking patterns in the $I$th GR cluster; $I=$ (b1) 49 and (b2) 101.
Six complex out-of-phase spiking patterns in the $I$th GR clusters; $I=$ (c1) 192, (c2) 91, (c3) 773, (c4) 382, (c5) 705, and (c6) 349.
${\cal{C}}^{(I)}$  represents the conjunction index of the spiking pattern in the $I$th GR cluster.
}
\label{fig:DSP}
\end{figure}

Presence of the central valley in $A(t)$ is in contrast to the central plateau in $R_{\rm GR}(t)$.
Appearance of such a central valley may be understood as follows.
The whole population of GR cells can be decomposed into two types of in-phase and out-of-phase spiking groups.
Spiking patterns of in-phase (out-of-phase) GR cells are in-phase (out-of-phase) with respect to $R_{\rm GR}(t)$ (representing the
population-averaged firing activity in the whole population of GR cells); details will be given in Figs.~\ref{fig:DSP} and \ref{fig:Char}.
Then, the activation degree $A_i^{(G)}$ of active GR cells in the $G$ spiking group in the $i$th bin is given by:
\begin{equation}
A_i^{(G)} = \frac {N_{a,i}^{(G)}} {N},
\label{eq:SAD}
\end{equation}
where $N_{a,i}^{(G)}$ is the number of active GR cells in the $G$ spiking group ($G=i$ and $o$ for the in-phase and the out-of-phase spiking groups, respectively)
in the $i$th bin. The sum of $A_i^{(G)}(t)$ over the in-phase and the out-of-phase spiking groups is just the activation degree $A_i(t)$ in the whole population.
Figure \ref{fig:WP}(c2) shows plots of activation degree $A^{(G)}(t)$ in the in-phase (solid line) and the out-of-phase (dotted curve)
spiking groups. In the case of in-phase ($G=i)$ spiking group, $A^{(i)}(t)$ has a central plateau, while $A^{(o)}(t)$ has double peaks with a central
valley in the case of out-of-phase ($G=o$) spiking group. Hence, small contribution of out-of-phase spiking group at the middle stage leads to
emergence of the central valley in $A(t)$ in the whole population.

We note again that, in the whole population the activation degree $A(t)$ with a central valley is in contrast to $R_{\rm GR}(t)$ with a central plateau.
To understand this discrepancy, we consider the bin-averaged instantaneous individual firing rates $f_{\rm GR}^{(i)}$ of active GR cells:
\begin{equation}
   f_{\rm GR}^{(i)} = \frac {N_{s,i}} {N_{a,i}~\Delta t},
\label{eq:IFR}
\end{equation}
where $N_{s,i}$ is the number of spikes of GR cells in the $i$th bin, $N_{a,i}$ is the number of active GR cells in the $i$th bin, and
the bin size $\Delta t$ is 10 msec.
Figure \ref{fig:WP}(d1) shows a plot of $f_{\rm GR}^{(i)}(t)$ for the active GR cells. We note that active GR cells fire spikes at higher firing rates
at the middle stage because $f_{\rm GR}^{(i)}(t)$ has a central peak.
Then, the bin-averaged instantaneous population spike rate $f_{\rm GR}^{(p)}$ is given by the product of the activation degree $A_i$ of Eq.~(\ref{eq:AD})
and the instantaneous individual firing rate $f_{\rm GR}^{(i)}$ of Eq.~(\ref{eq:IFR}):
\begin{equation}
f_{\rm GR}^{(p)} =  A_i~f_{\rm GR}^{(i)} = \frac {N_{s,i}} {N~\Delta t}.
\label{eq:PSR}
\end{equation}

The instantaneous population spike rate $f_{\rm GR}^{(p)}(t)$ in Fig.~\ref{fig:WP}(d2) has a central plateau, as in the case of $R_{\rm GR}(t)$.
We note that both $f_{\rm GR}^{(p)}(t)$ and $R_{\rm GR}(t)$ correspond to bin-based estimate and kernel-based smooth estimate for the instantaneous
whole-population spike rate for the GR cells, respectively \cite{RM}.
In this way, although the activation degree $A(t)$ of GR cells are lower at the middle stage, their population spike rate becomes nearly the same as that
in the neighboring parts (i.e., central plateau is formed), due to the higher individual firing rates.

\subsection{Dynamical Classification of Spiking Patterns of GR Clusters}
\label{subsec:DS}
There are $N_C~(=2^{10})$ GR clusters. $N_{\rm GR}~(=50)$ GR cells in each GR cluster share the same inhibitory and excitatory inputs via their dendrites which
synaptically contact the two glomeruli (i.e., terminals of MFs) at both ends of the GR cluster [see Fig.~\ref{fig:RN}(b)]; nearby inhibitory GO cell axons innervate the two glomeruli. Hence, GR cells in each GR cluster show similar firing behaviors.
Similar to the case of $R_{\rm GR}(t)$ in Eq.~(\ref{eq:IWPSR}), the firing activity of the $I$th GR cluster is characterized in terms of its instantaneous cluster spike rate $R_{\rm GR}^{(I)}(t)$ ($I=1, \cdots, N_C$):
\begin{equation}
R_{\rm GR}^{(I)}(t) = \frac{1}{N_{\rm GR}} \sum_{i=1}^{N_{\rm GR}} \sum_{s=1}^{n_i^{(I)}} K_h (t-t_{s}^{(I,i)}),
\label{eq:ISPSR}
\end{equation}
where $t_{s}^{(I,i)}$ is the $s$th spiking time of the $i$th GR cell in the $I$th GR cluster and $n_i^{(I)}$ is the total number of spikes for the $i$th GR cell
in the $I$th GR cluster.

We introduce the conjunction index ${\cal{C}}^{(I)}$ of each GR cluster, representing the degree for the conjunction (association) of the spiking behavior [$R_{\rm GR}^{(I)}(t)$] of each $I$th GR cluster with that of the whole population [$R_{\rm GR}(t)$ in Fig.~\ref{fig:WP}(b)] [i.e., denoting the degree for the resemblance (similarity) between $R_{\rm GR}^{(I)}(t)$ and $R_{\rm GR}(t)$]. The conjunction index ${\cal{C}}^{(I)}$ is given by the cross-correlation at the zero-time lag
[i.e., $Corr_{\rm GR}^{(I)}(0)$] between $R_{\rm GR}^{(I)}(t)$ and $R_{\rm GR}(t)$:
\begin{equation}
Corr_{\rm GR}^{(I)} (\tau) = \frac{\overline{\Delta R_{\rm GR}(t+\tau) \Delta R_{\rm GR}^{(I)}(t)}}{\sqrt{\overline{\Delta R_{\rm GR}^{2}(t)}} \sqrt{\overline{{\Delta R_{\rm GR}^{(I)}}^2(t)}}},
\label{eq:CI}
\end{equation}
where $\Delta R_{\rm GR}(t) = R_{\rm GR}(t)-\overline{R_{\rm GR}(t)}$, $\Delta R_{\rm GR}^{(I)}(t) = R_{\rm GR}^{(I)}(t)-\overline{R_{\rm GR}^{(I)}(t)}$, and the overline denotes the time average over a cycle. We note that ${\cal{C}}^{(I)}$ represents well the phase difference (shift) between the spiking patterns [$R_{\rm GR}^{(I)}(t)$] of GR clusters and the firing behavior [$R_{\rm GR}(t)$] in the whole population.

In all the $2^{10}$ GR clusters, we obtain their conjunction indices ${\cal C}^{(I)}$, make intensive examination of the phase difference of $R_{\rm GR}^{(I)}(t)$
with respect to $R_{\rm GR}(t)$, and thus classify the whole GR clusters into the in-phase, the anti-phase, and the complex out-of-phase spiking groups.
Figure \ref{fig:DSP} shows examples for diverse spiking patterns of GR clusters. This type of diversity arises from inhibitory coordination of GO cells on the firing activity of GR cells in the GR-GO feedback system in the granular layer.

Five examples for ``in-phase'' spiking patterns in the $I$th ($I=$ 594, 543, 663, 332, and 399) GR clusters are given in Figs.~\ref{fig:DSP}(a1)-\ref{fig:DSP}(a5), respectively. Raster plot of spikes of $N_{\rm GR} (=50)$ GR cells and the corresponding instantaneous cluster spike rate $R_{\rm GR}^{(I)}(t)$ are shown, along with the value of ${\cal{C}}^{(I)}$ in each case of the $I$th GR cluster. In all these cases, the instantaneous cluster spike rates $R_{\rm GR}^{(I)}(t)$ are in-phase
relative to the instantaneous whole-population spike rate $R_{\rm GR}(t)$. Among them, in the case of $I=594$ with the maximum conjunction index
${\cal C}_{max}$ $(=0.85)$, $R_{\rm GR}^{(594)}(t)$ with a central plateau is the most similar (in-phase) to $R_{\rm GR}(t)$. In the next case of $I=543$ with
${\cal{C}}^{(I)}=0.48,$ $R_{\rm GR}^{(543)}(t)$ has a central sharp peak, and hence its similarity degree relative to $R_{\rm GR}(t)$ decreases. The remaining two cases of $I=663$ and 332 (with more than one central peaks) may be regarded as ones developed from the case of $I=594$. With increasing the number of peaks in the central part, the value of ${\cal{C}}^{(I)}$ decreases, and hence the resemblance degree relative to $R_{\rm GR}(t)$ is reduced. The final case of $I=399$ with double peaks can be considered as one evolved form the case of $I=543$. In this case, the value of ${\cal{C}}^{(I)}$ is reduced to 0.40.

Based on the examples in Figs.~\ref{fig:DSP}(a1)-\ref{fig:DSP}(a5), spiking patterns which have central plateau, central sharp peak, and two or more central peaks in the middle part of cycle are considered as in-phase spiking patterns relative to the instantaneous whole-population spike rate $R_{\rm GR}(t)$.
We make an intensive examination of the instantaneous cluster spike rates $R_{\rm GR}^{(I)}$ of the GR clusters with $C^{(I)}<0.40$, and determine the higher threshold ${\cal C}^*_h~(\simeq 0.39)$ between the in-phase and the complex out-of-phase spiking patterns.
For ${\cal C}^{(I)} > {\cal C}^*_h$ in-phase spiking patterns such as ones in Figs.~\ref{fig:DSP}(a1)-\ref{fig:DSP}(a5) appear.
On the other hand, when passing the higher threshold ${\cal C}^*_h$ from the above, complex out-of-phase spiking patterns
(with ${\cal C}^{(I)} < {\cal C}^*_h$) emerge. These complex out-of-phase spiking patterns have left-skewed (right-skewed) peaks near the 1st
(3rd) quartile of cycle [i.e., near $t=500$ (1500) msec], explicit examples of which will be given below in Figs.~\ref{fig:DSP}(c1)-\ref{fig:DSP}(c4).
Thus, all the GR clusters, exhibiting in-phase spiking patterns, constitute the in-phase spiking group where the range of ${\cal{C}}^{(I)}$ is
$({\cal C}^*_h,{\cal C}_{max})$; ${\cal C}_{max}=0.85$ and ${\cal C}^*_h \simeq 0.39$.

Next, we consider the ``anti-phase'' spiking patterns. Two examples for the anti-phase spiking patterns in the $I$th ($I=$ 49 and 101) GR clusters are given in Figs.~\ref{fig:DSP}(b1) and \ref{fig:DSP}(b2), respectively. We note that, in both cases, the instantaneous cluster spike rates $R_{\rm GR}^{(I)}(t)$ are anti-phase with respect to $R_{\rm GR}(t)$ in the whole population. In the case of $I=49$ with the minimum conjunction index ${\cal{C}}_{min}~(=-0.57)$, $R_{\rm GR}^{(49)}(t)$ is the most anti-phase relative to $R_{\rm GR}(t)$, and it has double peaks near the 1st and the 3rd quartiles and a central deep valley  at the middle of the cycle. The case of $I=101$ with ${\cal{C}}^{(I)}$ $(=-0.23)$ may be regarded as evolved from the case of $I=49$. It has an increased (but still negative) value of ${\cal{C}}^{(I)}$ $(=-0.23)$ due to the risen central shallow valley.

\begin{figure}
\includegraphics[width=0.9\columnwidth]{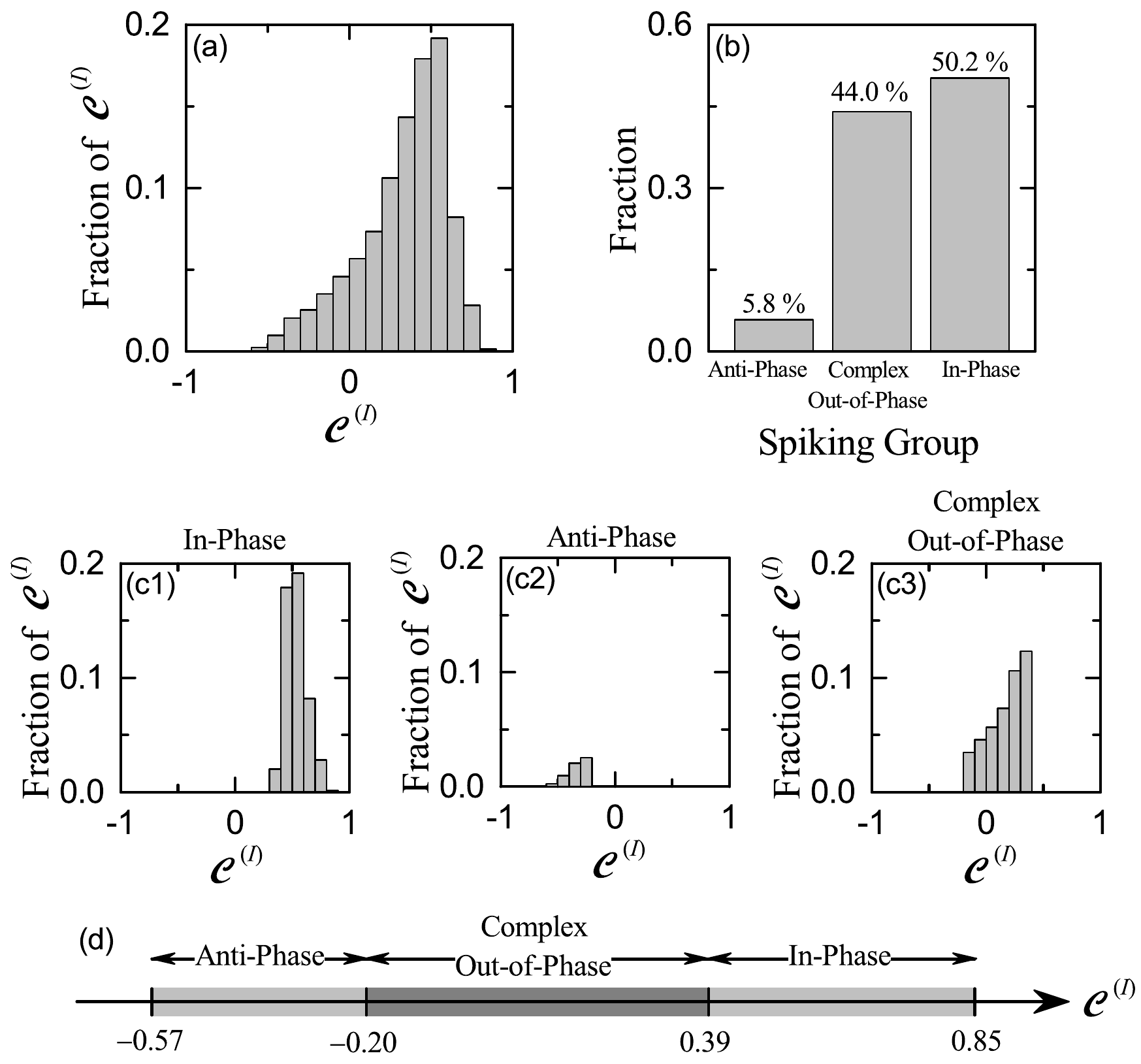}
\caption{Characterization of diverse spiking patterns in the GR clusters in the optimal case of $p_c^* = 0.06$.
(a) Distribution of conjunction indices $\{ {\cal{C}}^{(I)} \}$ for the GR clusters in the whole population.
(b) Fraction of spiking groups.  Distribution of conjunction indices $\{ {\cal{C}}^{(I)} \}$ for the (c1) in-phase, (c2) anti-phase, and (c3) complex out-of-phase spiking groups. Bin size for the histograms in (a) and in (c1)-(c3) is 0.1. (d) Bar diagram for the ranges of conjunction indices
$\{ {\cal{C}}^{(I)} \}$ for the in-phase, anti-phase, and complex out-of-phase spiking groups.
}
\label{fig:Char}
\end{figure}

Based on the examples in Figs.~\ref{fig:DSP}(b1)-\ref{fig:DSP}(b2), spiking patterns [$R_{\rm GR}^{(I)}(t)$] which have double peaks near the 1st and the 3rd quartiles and a central valley at the middle of cycle are regarded as anti-phase spiking patterns with respect to $R_{\rm GR}(t)$. Like the case of the above in-phase spiking patterns, through intensive examination of $R_{\rm GR}^{(I)}$ of the GR clusters with $C^{(I)} > -0.23,$ we determine the lower threshold ${\cal C}^*_l~(\simeq -0.20)$ between the anti-phase and the complex out-of-phase spiking patterns. For ${\cal C}^{(I)} < {\cal C}^*_l$ anti-phase spiking patterns such as ones in Figs.~\ref{fig:DSP}(b1)-\ref{fig:DSP}(b2) exist. In contrast, when passing the lower threshold ${\cal C}^*_l$ from the below, complex out-of-phase spiking patterns (with ${\cal C}^{(I)} > {\cal C}^*_l$) appear. These complex out-of-phase spiking patterns have a central peak which is transformed from the central valley at the middle of cycle, along with double peaks near the 1st and the 3rd quartiles. An explicit example will be given below in Fig.~\ref{fig:DSP}(c5).
Thus, all the GR clusters, showing anti-phase spiking patterns, form the anti-phase spiking group where the range of ${\cal{C}}^{(I)}$ is
$({\cal C}_{min},{\cal C}^*_l)$; ${\cal C}_{min}=-0.57$ and ${\cal C}^*_l \simeq -0.20$.

As discussed above, in the range of ${\cal C}^*_l~(\simeq -0.20) < {\cal{C}}^{(I)} < {\cal C}^*_h~(\simeq 0.39)$, a 3rd type of complex ``out-of-phase'' spiking patterns appear between the in-phase and the anti-phase spiking patterns.
Figure \ref{fig:DSP}(c1)-\ref{fig:DSP}(c6) show six examples for the complex out-of-phase spiking patterns in the $I$th ($I=$192, 91, 773, 382, 705, and 349)
GR clusters. The cases of $I=192$ and 91 seem to be developed from the in-phase spiking pattern in the $I$th ($I=594$ or 543) GR cluster.
In the case of $I=192$, $R_{\rm GR}^{(192)}$ has a left-skewed peak near the 1st quartile of the cycle, while in the case of $I=91$, $R_{\rm GR}^{(91)}$ has a right-skewed peak near the 3rd quartile. Hence, the values of ${\cal{C}}^{(I)}$ for $I=192$ and 91 are reduced to 0.12 and 0.10, respectively. In the next two cases of $I=773$ and 382, they seem to be developed from the cases of $I=192$ and 91, respectively. The left-skewed (right-skewed) peak in the case of $I=192$ (91)
is bifurcated into double peaks, which leads to more reduction of conjunction indices; ${\cal{C}}^{(773)}=0.08$ and ${\cal{C}}^{(382)}=0.05$.
In the remaining case of $I=705$, it seems to be evolved from the anti-phase spiking pattern in the $I=101$ case. The central valley for $I=101$
is transformed into a central peak. Thus, $R_{\rm GR}^{(705)}$ has three peaks, and its value of ${\cal{C}}^{(705)}$ is a little increased to -0.18.
As ${\cal{C}}^{(I)}$ is more increased toward the zero, $R_{\rm GR}^{(I)}$ becomes more complex, as shown in Fig.~\ref{fig:DSP}(c6) in the case of $I=349$
with ${\cal{C}}^{(349)}=-0.07$.

Results on characterization of the diverse in-phase, anti-phase, and complex out-of-phase spiking patterns are shown in Fig.~\ref{fig:Char}.
Figure \ref{fig:Char}(a) shows the plot of the fraction of conjunction indices ${\cal{C}}^{(I)}$ in the whole GR clusters.
${\cal{C}}^{(I)}$ increases slowly from the negative to the peak at 0.55, and then it decreases rapidly.
For this distribution $\{ {\cal{C}}^{(I)} \}$, the range is (-0.57, 0.85), the mean is 0.32, and the standard deviation is 0.516.
Then, the diversity degree $\cal{D}$ for the spiking patterns [$R_{\rm GR}^{(I)}(t)$] of all the GR clusters is given by:
\begin{eqnarray}
   {\cal{D}} &=& {\rm Relative~ Standard ~ Deviation}\nonumber \\
& & ~{\rm for ~the~Distribution~} \{ {\cal{C}}^{(I)} \},
\label{eq:DD}
\end{eqnarray}
where the relative standard deviation is just the standard deviation divided by the mean.
In the optimal case of $p_c^*=0.06$, ${\cal{D}}^* \simeq 1.613$, which is just a quantitative measure for the diverse recoding performed via feedback cooperation
between the GR and the GO cells in the granular layer. As will be seen later in Fig.~\ref{fig:Final}(b) for the plot of $\cal{D}$ versus $p_c$, ${\cal{D}}^*$ is
just the maximum, and hence spiking patterns of GR clusters at $p_c^*$ is the most diverse.

We decompose the whole GR clusters into the in-phase, anti-phase, and complex out-of-phase spiking groups.
Figure \ref{fig:Char}(b) shows the fraction of spiking groups. The in-phase spiking group with ${\cal C}^*_h~(\simeq 0.39) < {\cal C}^{(I)} < {\cal C}_{max}~(=0.85)$
is a major one with fraction 50.2$\%$, while the anti-phase spiking group with ${\cal C}_{min}~(=-0.57) < {\cal C}^{(I)} < {\cal C}^*_l~(\simeq -0.20)$
is a minor one with fraction 5.8$\%$. Between them (${\cal C}^*_l < {\cal C}^{(I)} < {\cal C}^*_h$), the complex out-of-phase spiking group with fraction 44$\%$ exists. In this case, the spiking-group ratio, given by the ratio of the fraction of the in-phase spiking group to that
of the out-of-phase spiking group (consisting of both the anti-phase and  complex out-of-phase spiking groups), is ${\cal{R}}^* \simeq 1.008.$
Thus, in the optimal case of $p_c^*=0.06$, the fractions between the in-phase and the out-of-phase spiking groups are well balanced.
Under such good balance between the in-phase and the out-of-phase spiking groups, spiking patterns of the GR clusters are the most diverse.

Figures \ref{fig:Char}(c1)-\ref{fig:Char}(c3) also show the plots of the fractions of conjunction indices ${\cal{C}}^{(I)}$
of the GR clusters in the in-phase, anti-phase, and complex out-of-phase spiking groups, respectively. The ranges for the distributions
$\{ {\cal{C}}^{(I)} \}$ in the three spiking groups are also given in the bar diagram in Fig.~\ref{fig:Char}(d).
In the case of in-phase spiking group, the distribution $\{ {\cal{C}}^{(I)} \}$ with a peak at 0.55 has only positive values in the range of
$({\cal C}^*_h,{\cal C}_{max})$ (${\cal C}_{max}=0.85$ and ${\cal C}^*_h \simeq 0.39$), and its mean and standard deviations are 0.538 and 0.181, respectively.
On the other hand, in the case of the anti-phase spiking group,
the distribution $\{ {\cal{C}}^{(I)} \}$ with a peak at -0.25 has only negative values in the range of
$({\cal C}_{min},{\cal C}^*_l)$ (${\cal C}_{min}=-0.57$ and ${\cal C}^*_l \simeq -0.20$),
and its mean and standard deviations are -0.331 and 0.135, respectively. Between the in-phase and the anti-phase spiking groups, there exists an
intermediate complex out-of-phase spiking group. In this case, the range for the distribution $\{ {\cal{C}}^{(I)} \}$ with a peak at 0.35 is
$({\cal C}^*_l,{\cal C}^*_h$), and the mean and the standard deviation are 0.174 and 0.242, respectively. As will be seen in the next subsection, these in-phase, anti-phase, and complex out-of-phase spiking groups play their own roles in the synaptic plasticity at PF-PC synapses, respectively.

\begin{figure}
\includegraphics[width=0.9\columnwidth]{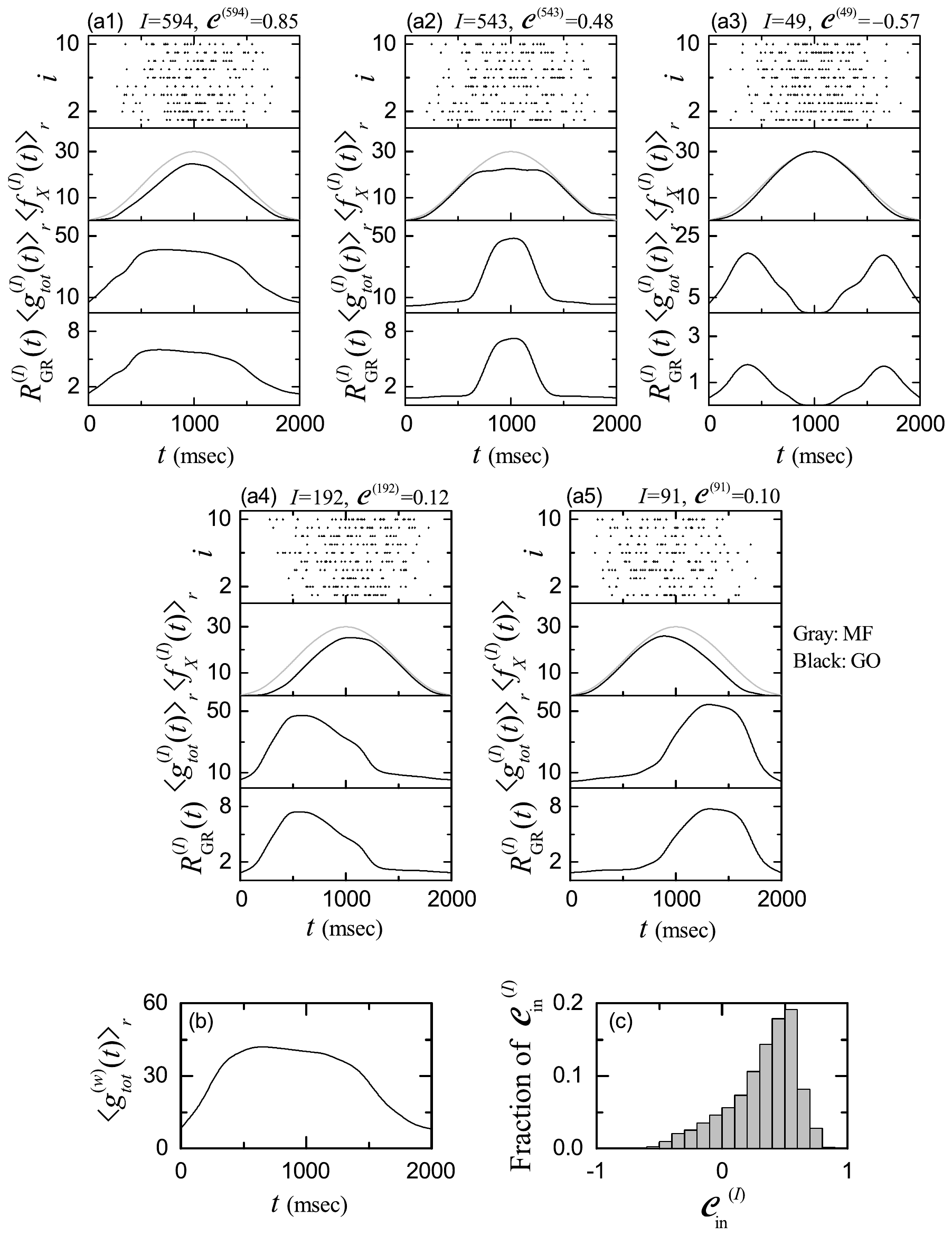}
\caption{Dynamical origin for the diverse spiking patterns in the GR clusters in the optimal case of $p_c^* = 0.06$.
In-phase spiking patterns for $I$= (a1) 594 and (a2) 543, anti-phase spiking pattern for (a3) $I=49$, and complex out-of-phase spiking patterns
for $I$= (a4) 192 and (a5) 91. In (a1)-(a5), top panel: raster plots of spikes in the sub-population of pre-synaptic GO cells innervating the $I$th GR cluster, 2nd panel: plots of $f_X^{(I)}(t):$ bin-averaged instantaneous spike rates of the MF signals ($X= {\rm MF}$) into the $I$th GR cluster (gray line) and bin-averaged instantaneous sub-population of pre-synaptic GO cells ($X={\rm GO}$) innervating the $I$th GR cluster (black line); $\langle \cdots \rangle_r$ represents the realization average (number of realizations is 100), 3rd panel: time course of $\langle g_{tot}^{(I)}(t) \rangle_r:$ conductance of total synaptic inputs (including both the excitatory and inhibitory inputs) into the $I$th GR cluster, and bottom panel: plots of $R_{\rm GR}^{(I)}(t):$ instantaneous cluster spike rate in the $I$th GR cluster. (b) Plot of cluster-averaged conductance $\langle g_{tot}^{(w)}(t) \rangle_r$ of total synaptic inputs into the GR clusters versus $t$.
(c) Distribution of conjunction indices $\{ {\cal C}_{\rm in}^{(I)} \}$ for the conductances of total synaptic inputs into the GR clusters.
}
\label{fig:Origin}
\end{figure}

Finally, we study the dynamical origin of diverse spiking patterns in the $I$th GR clusters.
As examples, we consider two in-phase spiking patterns for $I=594$ and 543 [see the spiking patterns in Figs.~\ref{fig:DSP}(a1) and \ref{fig:DSP}(a2)],
one anti-phase spiking pattern for $I=49$ [see the spiking pattern Fig.~\ref{fig:DSP}(b1)], and two complex out-of-phase spiking patterns for
$I=192$ and 91 [see the spiking patterns in Figs.~\ref{fig:DSP}(c1) and \ref{fig:DSP}(c2)].
In Fig.~\ref{fig:Origin}, (a1)-(a5) correspond to the cases of $I=$594, 543, 49, 192, and 91, respectively.

Diverse recodings for the MF signals are made in the GR layer, composed of excitatory GR and inhibitory GO cells (i.e., in the GR-GO cell feedback loop).
In this case, spiking activities of GR cells are controlled by two types of synaptic input currents (i.e., excitatory synaptic inputs through MF signals and inhibitory synaptic inputs from randomly connected GO cells).
Then, we make investigations on the dynamical origin of diverse spiking patterns of the GR clusters (shown in Fig.~\ref{fig:DSP}) through analysis of total synaptic inputs into the GR clusters.
Synaptic current is given by the product of synaptic conductance $g$ and potential difference [see Eq.~(\ref{eq:ISyn2})].
Here, synaptic conductance determines the time-course of synaptic current. Hence, it is enough to consider the time-course of synaptic conductance.
The synaptic conductance $g$ is given by the product of synaptic strength per synapse, the number of synapses $M_{\rm syn}$, and the fraction $s$ of open (post-synaptic) ion channels [see Eq.~(\ref{eq:ISyn3})]. Here, the synaptic strength per synapse is given by the product of maximum synaptic conductance
$\bar{g}$ and synaptic weight $J$, and the time-course of $s(t)$ is given by a summation for exponential-decay functions over pre-synaptic spikes, as shown
in Eqs.~(\ref{eq:ISyn3}) and (\ref{eq:ISyn4}).

We make an approximation of the fraction $s(t)$ of open ion channels (i.e., contributions of summed effects of pre-synaptic spikes) by the bin-averaged spike rate $f_X^{(I)}(t)$ of pre-synaptic neurons ($X=$ MF and GO); $f_{\rm MF}^{(I)}(t)$ is the bin-averaged spike rate of the MF signals into the $I$th GR cluster
and $f_{\rm GO}^{(I)}(t)$ is the bin-averaged spike rate of the pre-synaptic GO cells innervating the $I$th GR cluster.
Then, the conductance $g_X^{(I)}(t)$ of synaptic input from $X$ (=MF or GO) into the $I$th GR cluster ($I=1, \cdots, N_C$) is given by:
\begin{equation}
 g_X^{(I)}(t) \simeq {\rm M}_f^{(R)} \cdot f_X^{(I)}(t).
\label{gX}
\end{equation}
Here, the multiplication factor ${\rm M}_f^{(R)}$ [= maximum synaptic conductance ${\bar g}_R$ $\times$ synaptic weight
$J^{({\rm GR},X)}$ $\times$ number of synapses $M_{\rm syn}^{({\rm GR},X)}$] varies depending on $X$ and the receptor $R$ on the post-synaptic GR cells.
In the case of excitatory synaptic currents into the $I$th GR cluster with AMPA receptors via the MF signal, ${\rm M}_f^{\rm (AMPA)} = 2.88;$
${\bar g}_{\rm AMPA}=0.18,$ $J^{\rm (GR,MF)}=8.0,$ and $M_{\rm syn}^{({\rm GR})}=2$.
In contrast, in the case of the $I$th GR cluster with NMDA receptors, ${\bar g}_{\rm NMDA}=0.025,$ and hence ${\rm M}_f^{(\rm NMDA)} = 0.4,$ which is much less than ${\rm M}_f^{(\rm AMPA)}$.
For the inhibitory synaptic current from pre-synaptic GO cells to the $I$th GR cluster with GABA receptors,
${\rm M}_f^{(\rm GABA)} = 2.72$; $\bar{g}_{\rm GABA}=0.028,$ $J^{\rm (GR,GO)}=10,$ and $M_{\rm syn}^{\rm (GR,GO)} = 9.7.$
Then, the conductance $g_{tot}^{(I)}$ of total synaptic inputs (including both the excitatory and the inhibitory inputs) into the $I$th GR cluster is given by:
\begin{eqnarray}
g_{tot}^{(I)}(t) &=& g_{\rm MF}^{(I)} - g_{\rm GO}^{(I)} = g_{\rm AMPA}^{(I)} + g_{\rm NMDA}^{(I)} - g_{\rm GO}^{(I)} \nonumber \\
             & = & 3.28~ f_{\rm MF}^{(I)}(t) - 2.72~ f_{\rm GO}^{(I)}(t).
\label{eq:gTOT}
\end{eqnarray}
Total synaptic input with conductance $g_{tot}^{(I)}(t)$ is fed into GR cells in the $I$th GR cluster, and then the corresponding output,
given by the instantaneous cluster spike rate $R_{\rm GR}^{(I)}(t),$ emerges.
Through averaging $g_{tot}^{(I)}(t)$ over all the GR clusters, we obtain the cluster-averaged conductance $g_{tot}^{(w)}(t)$ of total synaptic inputs into the GR clusters:
\begin{equation}
g_{tot}^{(w)}(t) = {\frac {1} {N_C}} \sum_{I=1}^{N_C} g_{tot}^{(I)}(t).
\label{eq:gWTOT}
\end{equation}
The cluster-averaged total synaptic input with $g_{tot}^{(w)}(t)$ gives rise to the cluster-averaged output, given by the instantaneous whole-population spike rate
$R_w(t)$ [$= {\frac {1} {N_C}} \sum_{I=1}^{N_C} R_{\rm GR}^{(I)}(t)$].

In Figs.~\ref{fig:Origin}(a1)-\ref{fig:Origin}(a5), the top panels show the raster plots of spikes in the sub-populations of pre-synaptic
GO cells innervating the $I$th GR clusters. We obtain bin-averaged (sub-population) spike rates $f_{\rm GO}^{(I)}(t)$ from the raster plots.
The bin-averaged spike rate of pre-synaptic GO cells in the $i$th bin is given by $\frac {n_i^{(s)}} {N_{pre}~\Delta t}$, where $n_i^{(s)}$ is the
number of spikes in the $i$th bin, $\Delta t$ (=10 msec) is the bin size, and $N_{pre}$ (=10) is the number of pre-synaptic GO cells.
Via an average over 100 realizations, we obtain the realization-averaged (bin-averaged) spike rate of pre-synaptic GO cells $\langle f_{\rm GO}^{(I)}(t) \rangle_r$ because $N_{pre}~(=10)$ is small; $\langle \cdots \rangle_r$ represent a realization-average.
The 2nd panels show $\langle f_{\rm GO}^{(I)}(t) \rangle_r$ (black line) and $\langle f_{\rm MF}^{(I)}(t) \rangle_r$ (gray line).
We note that $\langle f_{\rm GO}^{(I)}(t) \rangle$ changes depending on $I$, while $\langle f_{\rm MF}^{(I)}(t) \rangle$ is independent of $I$.
In contrast to the spiking activity of GR cells [which exhibit random repetition of transitions between active (bursting) and inactive (silent) states
(see Fig.~\ref{fig:WP}(a))], GO cells exhibit relatively regular spikings, which may be well seen in slightly-modified sinusoidal-like bin-averaged spike rate $\langle f_{\rm GO}^{(I)}(t) \rangle_r$ \cite{Yama2,Heine}.
Then, we may get the realization-averaged conductance $\langle g_{tot}^{(I)}(t) \rangle_r$ of total synaptic inputs in Eq.~(\ref{eq:gTOT}), which is shown in
the 3rd panels. These conductances $\langle g_{tot}^{(I)}(t) \rangle_r$ of total synaptic inputs show diverse patterns depending on $I$, although
$\langle f_{\rm GO}^{(I)}(t) \rangle$, related to inhibitory synaptic input, exhibits relatively regular patterns.

We note that the shapes of $\langle g_{tot}^{(I)}(t) \rangle_r$ (corresponding to the total input into the $I$th GR cluster) in the 3rd panels are nearly the same as those of $R_{\rm GR}^{(I)}(t)$ (corresponding to the output of the $I$th GR cluster) in the bottom panels.
Hence, we expect that in-phase (out-of-phase) inputs into the GR clusters may result in generation of in-phase (out-of-phase) outputs (i.e., responses) in the GR clusters. To confirm this point clearly, similar to case of the spiking patterns [$R_{\rm GR}^{(I)}(t)$] (i.e., the outputs) in the GR clusters, we introduce the
conjunction index for the total synaptic input into the $I$th GR cluster between $\langle g_{tot}^{(I)}(t) \rangle_r$ (conductance of total synaptic input into the $I$th GR cluster) and the cluster-averaged conductance of total synaptic inputs $\langle g_{tot}^{(w)}(t) \rangle_r$.
Figure \ref{fig:Origin}(b) shows the plot of $\langle g_{tot}^{(w)}(t) \rangle_r$ versus $t$. We also note that the shape of $\langle g_{tot}^{(w)}(t) \rangle_r$
is similar to the instantaneous whole-population spike rate $R_{\rm GR}(t)$ in Fig.~\ref{fig:WP}(b).

As in the case of the conjunction index ${\cal C}^{(I)}$ for the spiking patterns (i.e. outputs) in the $I$th GR cluster [see Eq.~(\ref{eq:CI})],
the conjunction index ${\cal{C}}^{(I)}_{\rm in}$ for the total synaptic input is given by the cross-correlation at the zero-time lag (i.e., $Corr_{\rm in}^{(I)}(0)$) between $\langle g_{tot}^{(I)}(t) \rangle_r$ and $\langle g_{tot}^{(w)}(t) \rangle_r$:
\begin{equation}
Corr_{\rm in}^{(I)} (\tau) = \frac{\overline{ \Delta \langle g_{tot}^{(w)}(t+ \tau) \rangle_r \Delta \langle g_{tot}^{(I)}(t) \rangle_r}}
{ \sqrt{\overline{ {\Delta \langle g_{tot}^{(w)}(t) \rangle_r}^2}} \sqrt{\overline{ {\Delta \langle g_{tot}^{(I)}(t) \rangle_r}^2}}},
\label{eq:INCI}
\end{equation}
where $\Delta \langle g_{tot}^{(w)}(t) \rangle_r = \langle g_{tot}^{(w)}(t) \rangle_r-\overline{ \langle g_{tot}^{(w)}(t) \rangle_r}$, $\Delta \langle g_{tot}^{(I)}(t) \rangle_r = \langle g_{tot}^{(I)}(t) \rangle_r-\overline{ \langle g_{tot}^{(I)}(t) \rangle_r}$, and the overline represents the time average.
Thus, we have two types of conjunction indices, ${\cal C}^{(I)}$ [output conjunction index: given by $Corr_{\rm GR}^{(I)}(0)$] and ${\cal C}^{(I)}_{\rm in}$
[input conjunction index: given by $Corr_{\rm in}^{(I)}(0)$] for the output and the input in the $I$th GR cluster, respectively.

Figure \ref{fig:Origin}(c) shows the plot of fraction of input conjunction indices $\{ {\cal C}^{(I)}_{\rm in} \}$ in the whole GR clusters.
We note that the distribution of input conjunction indices in Fig.~\ref{fig:Origin}(c) is nearly the same as that of
output conjunction indices in Fig.~\ref{fig:Char}(a). ${\cal{C}}^{(I)}_{\rm in}$ increases slowly from the negative value to the peak at 0.55, and then it decreases rapidly. In this distribution of $\{ {\cal{C}}^{(I)}_{\rm in} \}$, the range is (-0.57, 0.85), the mean is 0.321, and the standard deviation is 0.516.
Then, we obtain the diversity degree ${\cal{D}}_{\rm in}$ for the total synaptic inputs $\{ \langle g_{tot}^{(I)}(t) \rangle_r \}$ of all the GR clusters:
\begin{eqnarray}
   {\cal{D}_{\rm in}} &=& {\rm Relative~ Standard ~ Deviation}\nonumber \\
& & ~{\rm for ~the~Distribution~of~} \{ {\cal{C}}^{(I)}_{\rm in} \}.
\label{eq:INDD}
\end{eqnarray}
Hence, ${\cal{D}}_{\rm in} \simeq 1.607$ for the synaptic inputs, which is nearly the same as ${\cal D}^*~(\simeq 1.613)$ for the spiking patterns of GR cells.
Consequently, diverse total synaptic inputs into the GR clusters lead to generation of diverse outputs (i.e., spiking patterns) of the GR cells.

\begin{figure}
\includegraphics[width=0.95\columnwidth]{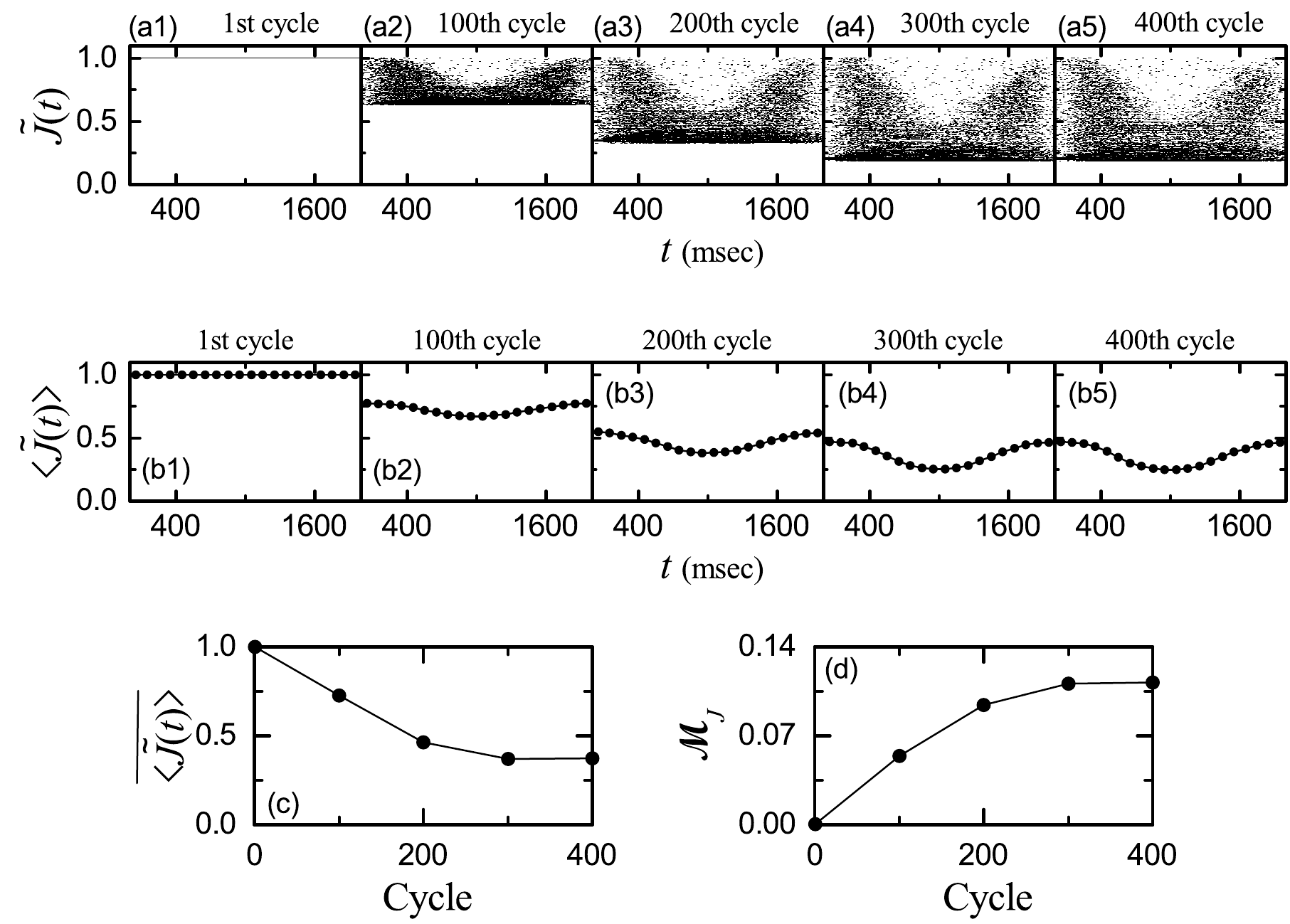}
\caption{Change in synaptic weights of active PF-PC synapses during learning in the optimal case of $p_c^* = 0.06$. (a1)-(a5) Cycle-evolution of distribution of normalized synaptic weights $\tilde{J}$ of active PF signals. (b1)-(b5) Cycle-evolution of bin-averaged (normalized) synaptic weights $\langle {\tilde J} \rangle$ of active PF signals. Bin size: $\Delta t =100$ msec. Plots of (c) cycle-averaged mean $\overline{ \langle {\tilde J} \rangle }$ and (d) modulation ${\cal{M}}_J$ for
$\langle {\tilde J} \rangle$ versus cycle.
}
\label{fig:SW}
\end{figure}

\subsection{Effect of Diverse Recoding in GR Clusters on Synaptic Plasticity at PF-PC Synapses}
\label{subsec:SW}
Based on dynamical classification of spiking patterns of GR clusters, we investigate the effect of diverse recoding in the GR clusters on
synaptic plasticity at PF-PC synapses. As shown in the above subsection, MF context input signals for the post-eye-movement are diversely recoded in the granular layer (corresponding to the input layer of the cerebellar cortex). The diversely-recoded in-phase and out-of-phase PF (student) signals (corresponding to the
outputs from the GR cells) are fed into the PCs (i.e., principal cells of the cerebellar cortex) and the BCs in the Purkinje-molecular layer
(corresponding to the output layer of the cerebellar cortex). The PCs also receive in-phase error-teaching CF (instructor) signals from the IO, along with the
inhibitory inputs from the BCs. Then, the synaptic weights at the PF-PC synapses vary depending on the relative phase difference between the PF (student)
signals and the CF (instructor) signals.

We first consider the change in normalized synaptic weights $\tilde J$ of active PF-PC synapses during learning in the optimal case of $p_c^* = 0.06$;
\begin{equation}
{\tilde {J}}_{ij}(t) = \frac {J_{ij}^{\rm (PC,PF)}(t)} {J_{0}^{\rm (PC,PF)}}.
\label{NSW}
\end{equation}
where the initial synaptic strength ($J_{0}^{\rm (PC,PF)}=0.006$) is the same for all PF-PC synapses.
Figures \ref{fig:SW}(a1)-\ref{fig:SW}(a5) show cycle-evolution of distribution of ${\tilde J}(t)$ of active PF-PC synapses.
With increasing the learning cycle, normalized synaptic weights ${\tilde J}(t)$ are decreased due to LTD at PF-PC synapses,
and eventually their distribution seems to be saturated at about the 300th cycle.
We note that in-phase PF signals are strongly depressed (i.e., strong LTD) by the in-phase CF signals, while out-of-phase PF signals are weakly
depressed (i.e., weak LTD) due to the phase difference between the PF and the CF signals. As shown in Fig.~\ref{fig:WP}(c2), the activation degree
$A^{(G)}$ of the in-phase spiking group ($G=i$) is dominant at the middle stage of the cycle, while at the other parts of the cycle,
the activation degrees of the in-phase ($G=i$) and the out-of-phase ($G=o$) spiking groups are comparable. Consequently, strong LTD occurs at
the middle stage, while at the initial and final stages somewhat less LTD takes place due to contribution of both the out-of-phase
spiking group (with weak LTD) and the in-phase spiking group.

To more clearly examine the above cycle evolutions, we get the bin-averaged (normalized) synaptic weight in each $i$th bin (bin size: $\Delta t=$ 100 msec):
\begin{equation}
{\langle {\tilde J}(t) \rangle_i} = {\frac {1} {N_{s,i}}} \sum_{f=1}^{N_{s,i}} {\tilde J}_{i,f}(t),
\label{eq:BSW}
\end{equation}
where ${\tilde J}_{i,f}$ is the normalized synaptic weight of the $f$th active PF signal in the $i$th bin, and $N_{s,i}$ is the total number of
active PF signals in the $i$th bin.
Figures \ref{fig:SW}(b1)-\ref{fig:SW}(b5) show cycle-evolution of bin-averaged (normalized) synaptic weights $\langle {\tilde J}(t) \rangle$
of active PF signals. In each cycle,  $\langle {\tilde J}(t) \rangle$ forms a well-shaped curve.
With increasing the cycle, the well curve comes down, its modulation [=(maximum - minimum)/2] increases, and saturation seems to occur at about the 300th cycle.

\begin{figure}
\includegraphics[width=0.95\columnwidth]{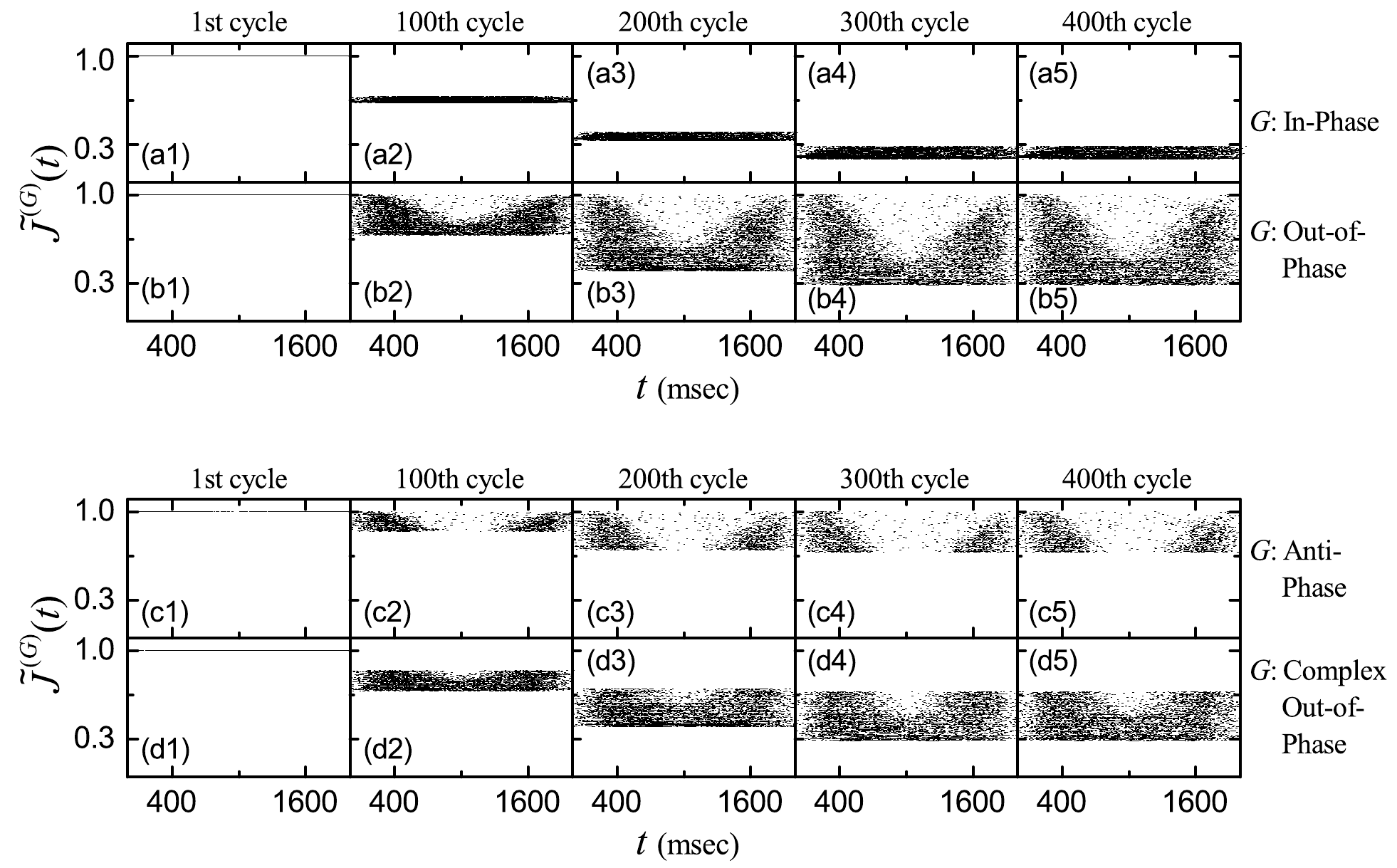}
\caption{Change in synaptic weights of active PF-PC synapses in each spiking group during learning in the optimal case of $p_c^* = 0.06$. Cycle-evolution of distributions of normalized synaptic weights ${\tilde{J}}^{(G)}$ of active PF signals in the $G$ spiking group [G: (a1)-(a5) in-phase and (b1)-(b5)  out-of-phase]. Cycle-evolution of distributions of normalized synaptic weights ${\tilde{J}}^{(G)}$ of active out-of-phase PF signals in the $G$ spiking group [G: (c1)-(c5) anti-phase and (d1)-(d5) complex out-of-phase].
}
\label{fig:SGSW}
\end{figure}

We also get the cycle-averaged mean $\overline{ \langle {\tilde J}(t) \rangle }$ via time average of $\langle {\tilde J}(t) \rangle$ over a cycle:
\begin{equation}
\overline{ \langle {\tilde J} \rangle } = {\frac {1} {N_b}} \sum_{i=1}^{N_b} \langle {\tilde J}(t) \rangle_i,
\label{eq:MSW}
\end{equation}
where $N_b$ is the number of bins for cycle averaging, and the overbar represents the time average over a cycle.
Figures \ref{fig:SW}(c) and \ref{fig:SW}(d) show plots of the cycle-averaged mean $\overline{ \langle {\tilde J}(t) \rangle }$
and the modulation ${\cal {M}}_J$ for $\langle {\tilde J}(t) \rangle$  versus cycle.
The cycle-averaged mean $\overline{ \langle {\tilde J}(t) \rangle }$ decreases from 1 to 0.372 due to LTD at PF-PC synapses.
However, strength of the LTD varies depending on the stages of the cycle.
At the middle stage, strong LTD occurs, due to dominant contribution of in-phase active PF signals.
On the other hand, at the initial and the final stages, somewhat less LTD takes place, because both the out-of-phase spiking group
(with weak LTD) and the in-phase spiking group make contributions together. As a result, with increasing cycle, the middle-stage part
comes down more rapidly than the initial and final parts, and hence the modulation ${\cal {M}}_J$ increases from 0 to 0.112.

We now decompose the whole active PF signals into the in-phase and the out-of-phase active PF signals, and make an intensive investigation
on their effect on synaptic plasticity at PF-PC synapses. Figure \ref{fig:SGSW} shows cycle-evolution of distributions of normalized synaptic weights ${\tilde J}^{(G)}$ of active PF signals in the $G$ spiking group [(a1)-(a5) in-phase ($G=i$) and (b1)-(b5) out-of-phase ($G=o$)] in the optimal case of $p_c^* = 0.06$; the out-of-phase spiking group consists of the anti-phase and the complex out-of-phase spiking groups. With increasing learning cycle, normalized synaptic weights ${\tilde J}^{(G)}$ for the in-phase and the out-of-phase PF signals are decreased, and saturated at about the 300th cycle.

\begin{figure*}
\includegraphics[width=1.45\columnwidth]{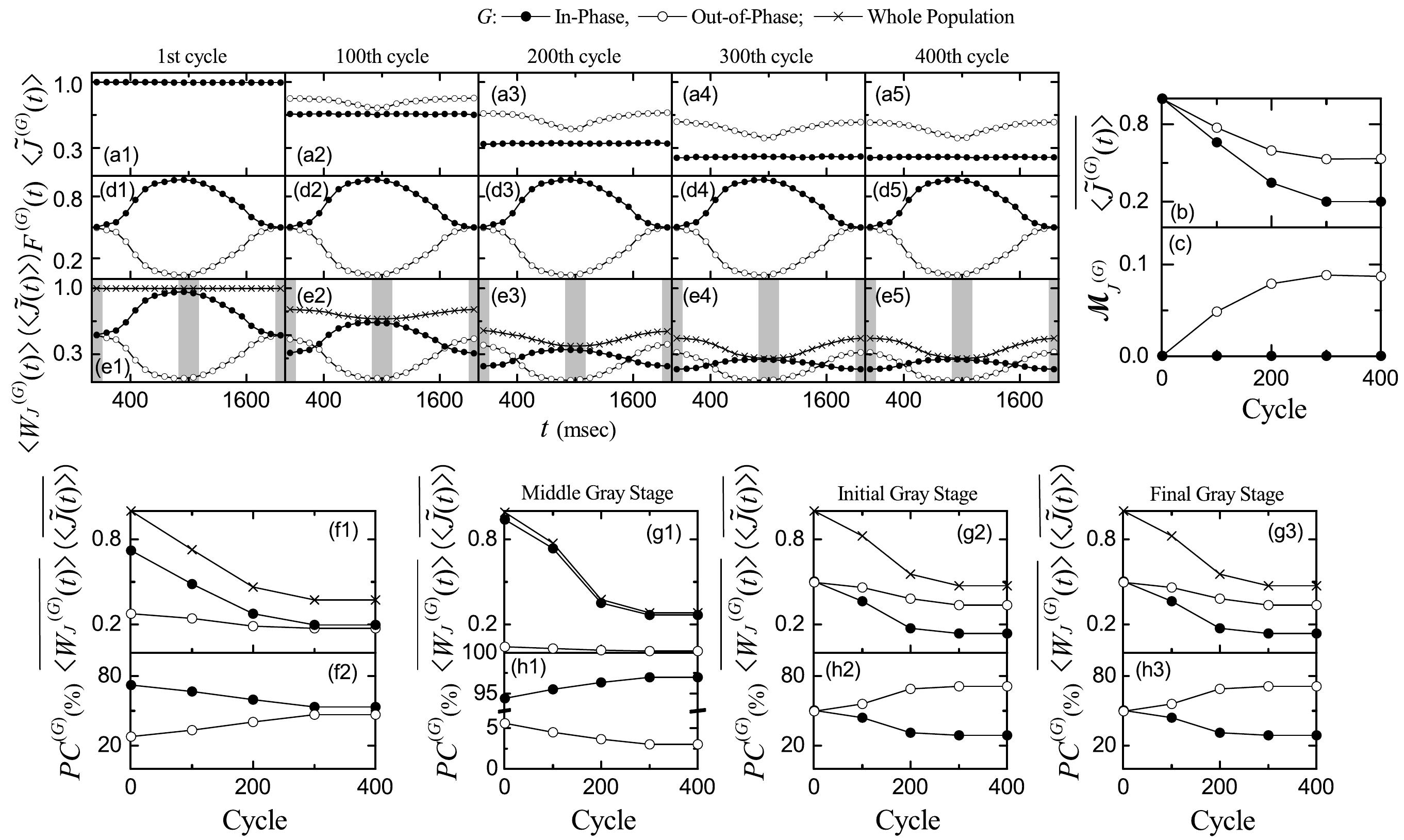}
\caption{Optimal case of $p_c^* = 0.06$. (a1)-(a5) Cycle-evolution of bin-averaged (normalized) synaptic weights $\langle {\tilde J}^{(G)} \rangle$ of active PF signals in the $G$ spiking-group. Plots of (b) cycle-averaged mean $\overline{ \langle {\tilde J}^{(G)} \rangle }$ and (c) modulation ${\cal{M}}_J^{(G)}$ for
$\langle {\tilde J}^{(G)} \rangle$ versus cycle. (d1)-(d5) Cycle-evolution of firing fraction $F^{(G)}$ of active PF signals in the $G$ spiking group.
(e1)-(e5) Cycle-evolution of weighted bin-averaged synaptic weights $\langle {W_J}^{(G)} \rangle$ of active PF signals in the $G$ spiking group; for comparison, bin-averaged synaptic weights $\langle {\tilde J} \rangle$ in the whole population of GR cells are also given. (f1) Plots of cycle-averaged means $\overline {\langle {W_J}^{(G)} \rangle}$ and $\overline {\langle {\tilde J} \rangle}$ for $\langle {W_J}^{(G)} \rangle$ and $\langle {\tilde J} \rangle$ versus cycle. (f2) Plots of percentage contributions $PC^{(G)}$ of the $G$ spiking group (i.e., ${\overline {\langle {W_J}^{(G)} \rangle}} /  {\overline {\langle {\tilde J} \rangle}}$) versus cycle. Left, right, and middle vertical gray bands in each cycle in (e1)-(e5) denote the initial, final, and middle stages, respectively.
Plots of stage-averaged values $\overline {\langle {W_J}^{(G)} \rangle}$ and $\overline {\langle {\tilde J} \rangle}$  versus cycle at the (g1) middle,(g2) initial, and (g3) final stages  of a cycle. Plots of percentage contributions $PC^{(G)}$  of the $G$ spiking group versus cycle at the (h1) middle, (h2) initial, and (h3) final stages of a cycle.
}
\label{fig:SWFF}
\end{figure*}

We note that the strength of LTD varies distinctly depending on the type of spiking group. The student PF signals (corresponding to the axons of the GR cells)
are classified as in-phase or out-of-phase PF signals with respect to the ``reference'' signal $R_{\rm GR}(t)$. Here, $R_{\rm GR}(t)$
is the instantaneous whole-population spike rate of Eq.~(\ref{eq:IWPSR})(denoting the population-averaged firing activity in the whole population of the GR cells). It is basically in proportion to the sinusoidal MF context signal $f_{\rm MF}(t)$ of Eq.~(\ref{eq:MF}), although its top part becomes lowered and flattened due to
inhibitory coordination of GO cells. The PF student signals (coming from the GR clusters) are characterized in terms of their conjunction indices ${\cal C}^{(I)}$
of Eq.~(\ref{eq:CI}) between the instantaneous cluster spike rate $R_{\rm GR}^{(I)}(t)$ and the reference signal $R_{\rm GR}(t)$.
The ranges of $\{ {\cal C}^{(I)} \}$ for the in-phase and the out-of-phase spiking groups are (0.39, 0.85) and (-0.57, 0.39), respectively, as shown
in Fig.~\ref{fig:Char}(d). The range for the out-of-phase spiking group is broader than that for the in-phase spiking group. These student PF signals are depressed
by the instructor CF signals (coming from the IO neuron). We also note that the instructor CF signals are in-phase with the reference signal $R_{\rm GR}(t)$, because
the CF signals are also basically in proportion to the sinusoidal IO desired signal $f_{\rm DS}(t)$ of Eq.~(\ref{eq:DS}) which is in-phase with $f_{\rm MF}(t)$.
Hence, in-phase student PF signals are strongly depressed by the instructor CF signals, because they are ``well-matched'' with the in-phase CF signals.
On the other hand, out-of-phase student PF signals are weakly depressed by the instructor CF signals, because they are ``ill-matched'' with the in-phase
CF signals.

In the case of active in-phase PF signals, the distribution of their normalized synaptic weights ${\tilde J}^{(G)}(t)$ ($G=i$) forms the bottom dense
bands in Figs.~\ref{fig:SW}(a1)-\ref{fig:SW}(a5), due to strong LTD at the in-phase PF-PC synapses. The (vertical) widths of these bottom bands are narrow
because of the narrow range of $\{ C^{(I)} \}$. On the other hand, in the case of active out-of-phase PF signals, the distribution of
$\{ {\tilde J}^{(G)}(t) \}$ ($G=o$) constitutes the upper sparse well-shaped parts in Figs.~\ref{fig:SW}(a1)-\ref{fig:SW}(a5), because of weak
LTD at the out-of-phase PF-PC synapses. Due to the broad range of $\{ C^{(I)} \}$, the heights of the well-shaped parts for the out-of-phase PF signals
are higher than those of the bottom bands for the in-phase PF signals with the narrow range. Moreover, the shapes for the distributions of
$\{ {\tilde J}^{(G)}(t) \}$ are consistent with the activation degrees $A^{(G)}$ of Eq.(\ref{eq:SAD}) of the active PF signals in the $G$ spiking group
which is also shown in Fig.~\ref{fig:WP}(c2). The activation degree $A^{(o)}$ ($G=o$) for the out-of-phase spiking group has a small minimum
in the middle stage of cycle, which leads to the well-shaped distributions of $\{ {\tilde J}^{(G)}(t) \}$ ($G=o$).
Hence, the in-phase PF signals (with strong LTD) make a large contribution in the middle stage of cycle, while at the initial and the final stages,
contributions of both the out-of-phase PF signals (with weak LTD) and the in-phase PF signals are comparable.

In the above way, effective depression (i.e., strong/weak LTD) at PF-PC synapses occurs, depending on the spiking type
(in-phase or out-of-phase) of the active PF signals; strong (weak) LTD takes place for the in-phase (out-of-phase) PF signals.
However, contributions of these in-phase and out-of-phase spiking groups vary depending on the stages of cycle.
Figure \ref{fig:WP}(c2) shows the activation degree $A^{(G)}$ of the in-phase ($G=i$) and the out-of-phase ($G=o$) spiking groups.
In the middle stage of cycle, the in-phase spiking group has a larger activation degree $A^{(i)}$. On the other hand, at the
initial and the final stages, the activation degrees of the in-phase and the out-of-phase spiking groups are comparable.
Hence, strong LTD takes place in the middle stage, due to a large contribution of the in-phase spiking group.
Thus, the in-phase spiking group makes a big contribution to formation of the minimum of bin-averaged (normalized) synaptic
weights $\langle {\tilde J}(t) \rangle$ in Figs.~\ref{fig:SW}(b1)-\ref{fig:SW}(b5).
In contrast, at the initial and the final stages of cycle, less LTD occurs because of comparable contributions of both
the out-of-phase spiking group with weak LTD and the in-phase spiking group with strong LTD.
Hence, maxima of $\langle {\tilde J}(t) \rangle$  appear at the initial and the final stages of cycle because both the (weakly-depressed) out-of-phase
and the (strongly-depressed) in-phase spiking groups contribute together.

In this way, the in-phase and the out-of-phase spiking groups play their own roles in formation of modulation of $\langle {\tilde J}(t) \rangle$.
The minimum of $\langle {\tilde J}(t) \rangle$ in the middle stage of cycle is formed via a large contribution of the in-phase spiking group with strong LTD, while formation of the maxima of $\langle {\tilde J}(t) \rangle$ in the initial and the final stages is made via comparable contributions of the out-of-phase spiking group
with weak LTD and the in-phase spiking group with strong LTD. Consequently, this kind of constructive interplay between the in-phase (strong LTD) and the out-of-phase (weak LTD) spiking groups leads to a big modulation of $\langle {\tilde J}(t) \rangle$, as shown in Fig.~\ref{fig:SW}(d).
(More detailed discussion on this point will be given below in Fig.~\ref{fig:SWFF}.)

We also make further decomposition of the out-of-phase PF signals into the anti-phase and the complex out-of-phase ones.
Figure \ref{fig:SGSW} also shows cycle-evolution of distributions of normalized synaptic weights ${\tilde J}^{(G)}$ of active out-of-phase PF signals in the $G$ spiking group [$G$: (c1)-(c5) anti-phase and (d1)-(d5) complex out-of-phase].
As the learning cycle is increased, normalized synaptic weights ${\tilde J}^{(G)}$ for the anti-phase and the complex out-of-phase PF signals are decreased and saturated at about the 300th cycle. In the case of anti-phase PF signals, weak depression occurs, and they constitute the top part for the out-of-phase PF signals in
Figs.~\ref{fig:SGSW}(b1)-\ref{fig:SGSW}(b5). On the other hand, in the case of complex out-of-phase PF signals, intermediate LTD takes place, and they form the bottom part for the out-of-phase PF signals in Figs.~\ref{fig:SGSW}(b1)-\ref{fig:SGSW}(b5).
These anti-phase (weak LTD) and complex out-of-phase (intermediate LTD) spiking groups make contribution to formation of maxima of
$\langle {\tilde J}(t) \rangle$ at the initial and the final stages of cycle, together with the in-phase spiking group with strong LTD.

Figures \ref{fig:SWFF}(a1)-\ref{fig:SWFF}(a5) show  cycle-evolutions of bin-averaged (normalized) synaptic weights $\langle {\tilde J}^{(G)}(t) \rangle$ of
active PF signals in the $G$ spiking-group [i.e., corresponding to the bin-averages for the distributions of ${\tilde J}^{(G)}(t)$ in Figs.~\ref{fig:SGSW}(a1)-\ref{fig:SGSW}(a5)
and Figs.~\ref{fig:SGSW}(b1)-\ref{fig:SGSW}(b5)]; $G:$ in-phase (solid circles) and out-of-phase (open circles).
In the case of in-phase PF signals, they are strongly depressed without modulation. On the other hand, in the case of out-of-phase PF signals, they are weakly depressed with modulation; at the initial and the final stages, they are more weakly depressed in comparison with the case at the middle stage.

The cycle-averaged means $\overline{ \langle {\tilde J}^{(G)}(t) \rangle }$ and modulations ${\cal{M}}_J^{(G)}$  for
$\langle {\tilde J}^{(G)}(t) \rangle$ are given in Figs.~\ref{fig:SWFF}(b) and \ref{fig:SWFF}(c), respectively. Both cycle-averaged means in the cases of the
in-phase and the out-of-phase PF signals decrease, and saturations occur at about the 300th cycle. In comparison with the case of out-of-phase PF signals (open circles), the cycle-averaged means in the case of in-phase PF signals (solid circles) are more reduced; the saturated limit value in the case of in-phase (out-of-phase) PF signals is 0.199 (0.529). In contrast, modulation occurs only for the out-of-phase PF signals (open circles), it increases with cycle, and becomes saturated at about the 300th cycle where the saturated value is 0.087.

In addition to the above bin-averaged (normalized) synaptic weights $\langle {\tilde J}^{(G)}(t) \rangle$, we need another information on firing fraction
$F^{(G)}(t)$ of active PF signals in the $G$ (in-phase or out-of-phase) spiking group to obtain the contribution of each spiking group to the bin-averaged synaptic weights $\langle {\tilde J}(t) \rangle$ of active PF signals in the whole population. The firing fraction $F_i^{(G)}$ of active PF signals in the $G$ spiking group in the $i$th bin is given by:
\begin{equation}
F_i^{(G)} = \frac {N_{s,i}^{(G)}} {N_{s,i}},
\label{eq:FF}
\end{equation}
where $N_{s,i}$ is the total number of active PF signals in the $i$th bin and $N_{s,i}^{(G)}$ is the number of active PF signals in the $G$ spiking group in the $i$th bin. We note that $F^{(G)}(t)$ is the same, independently of the learning cycle, because firing activity of PF signals depends only on the GR and GO cells in the granular layer.

Figures \ref{fig:SWFF}(d1)-\ref{fig:SWFF}(d5) show the firing fraction $F^{(G)}(t)$ of active PF signals in the $G$ spiking group.
The firing fraction $F^{(G)}(t)$ for the in-phase ($G=i$) active PF signals (solid circles) forms a bell-shaped curve, while $F^{(G)}(t)$ for the
out-of-phase ($G=o$) active PF signals (open circles) forms a well-shaped curve. The bell-shaped curve for the in-phase PF signal is higher than the well-shaped curve for the out-of-phase PF signal. For the in-phase PF signals, the firing fraction $F^{(i)}(t)$  is about 0.94 (i.e., $94\%$) at the middle stage, and about 0.51 (i.e., $51\%$) at the initial and the final stages. On the other hand, for the out-of-phase PF signals, the firing fraction $F^{(o)}(t)$ is about 0.49 (i.e., $49\%$) at the initial and the final stages, and about 0.06 (i.e., $6\%$) at the middle stage. Consequently, the fraction of in-phase active PF signals is dominant at the middle stage, while at the initial and the final stages, the fractions of both in-phase and out-of-phase active PF signals are nearly the same.

The weighted bin-averaged synaptic weight $\langle W_J^{(G)} \rangle_i$ for each $G$ spiking group in the $i$th bin is given by the product of
the firing fraction $F_i^{(G)}$ and the bin-averaged (normalized) synaptic weight $\langle {\tilde J}^{(G)} \rangle_i$:
\begin{equation}
\langle W_J^{(G)} \rangle_i = F_i^{(G)}~\langle {\tilde J}^{(G)} \rangle_i,
\label{eq:WSW}
\end{equation}
where the firing fraction $F_i^{(G)}$ plays a role of a weighting function for $\langle {\tilde J}^{(G)} \rangle_i$.
Then, the bin-averaged (normalized) synaptic weight $\langle {\tilde J} \rangle_i$ of active PF signals in the whole population in the $i$th bin
[see Eq.~(\ref{eq:BSW})] is given by the sum of weighted bin-averaged synaptic weights $\langle W_J^{(G)} \rangle_i$ of all spiking groups:
\begin{equation}
{\langle {\tilde J} \rangle_i} = \sum_G^{\rm all~spiking~groups} \langle W_J^{(G)} \rangle_i.
\label{eq:SWSW}
\end{equation}
Hence, $\langle W_J^{(G)}(t) \rangle$ represents contribution of the $G$ spiking group to $\langle {\tilde J}(t) \rangle$ of active PF signals in the whole population.

Figures \ref{fig:SWFF}(e1)-\ref{fig:SWFF}(e5) show cycle-evolution of weighted bin-averaged synaptic weights $\langle W_J^{(G)}(t) \rangle$ of active PF signals
in the $G$ spiking group [$G$: in-phase (solid circles) and out-of-phase (open circles)]. In the case of in-phase PF signals, the bin-averaged (normalized) synaptic weights $\langle {\tilde J}^{(G)}(t) \rangle$ are straight horizontal lines, the firing fraction $F^{(G)}(t)$ is a bell-shaped curve, and hence their product
leads to a bell-shaped curve for the weighted bin-averaged synaptic weight $\langle W_J^{(G)}(t) \rangle$. With increasing the cycle, the horizontal
straight lines for $\langle {\tilde J}^{(G)}(t) \rangle$ come down rapidly, while there is no change with cycle in $F^{(G)}(t)$.
Hence, the bell-shaped curves for $\langle W_J^{(G)}(t) \rangle$ also come down quickly, their modulations also are reduced in a fast way, and they become
saturated at about the 300th cycle.

On the other hand, in the case of out-of-phase PF signals, the bin-averaged (normalized) synaptic weights $\langle {\tilde J}^{(G)}(t) \rangle$ lie on a well-shaped curve, the firing fraction $F^{(G)}(t)$ also is a well-shaped curve, and then their product
results in a well-shaped curve for the weighted bin-averaged synaptic weight $\langle W_J^{(G)}(t) \rangle$. With increasing the cycle, the well-shaped curves for $\langle {\tilde J}^{(G)}(t) \rangle_i$ come down slowly, while there is no change with cycle in $F^{(G)}(t)$. Hence, the well-shaped curves for
$\langle W_J^{(G)}(t) \rangle$ also come down gradually, their modulations also are reduced little by little, and eventually they get saturated at about the 300th cycle.

For comparison, bin-averaged (normalized) synaptic weights $\langle {\tilde J}(t) \rangle$ of active PF signals in the whole population (crosses) are also given
in Figs.~\ref{fig:SWFF}(e1)-\ref{fig:SWFF}(e5), and they form a well-shaped curve. According to Eq.~(\ref{eq:SWSW}), the sum of the values at the solid circle (in-phase) and the open circle (out-of-phase) at a time $t$ in each cycle is just the value at the cross (whole population). At the middle stage of each cycle, contributions of in-phase PF signals (solid circles) are dominant
[i.e., contributions of out-of-phase PF signals (open circles) are negligible], while at the initial and the final stages, contributions of out-of-phase PF signals
are larger than those of in-phase PF signals (both contributions must be considered).
Consequently, $\langle {\tilde J}(t) \rangle$ of active PF signals in the whole population becomes more reduced at the middle stage than at the initial and the final stages, due to the dominant effect (i.e., strong LTD) of in-phase active PF signals at the middle stage, which results in a well-shaped curve for
$\langle {\tilde J}(t) \rangle$ in the whole population.

We make a quantitative analysis for contribution of $\langle W_J^{(G)}(t) \rangle$ in each $G$ spiking group to $\langle {\tilde J}(t) \rangle$ in the whole population. Figure \ref{fig:SWFF}(f1) shows plots of cycle-averaged weighted synaptic weight $\overline{ \langle W_J^{(G)}(t) \rangle }$
(i.e., time average of $\langle W_J^{(G)}(t) \rangle$ over a cycle) in the $G$ spiking-group [$G$: in-phase (solid circles) and out-of-phase (open circles)] and cycle-averaged synaptic weight $\overline{ \langle {\tilde J}(t) \rangle }$ of Eq.~(\ref{eq:MSW}) in the whole population (crosses) versus cycle.
The cycle-averaged weighted synaptic weights $\overline{ \langle W_J^{(G)}(t) \rangle }$ in the in-phase spiking group (solid circles) are larger than those in the out-of-phase spiking group (open circles), and their sums correspond to the cycle-averaged synaptic weight $\overline{ \langle {\tilde J}(t) \rangle }$ in the whole population (crosses). With increasing cycle, both $\overline{ \langle W_J^{(G)}(t) \rangle }$ and $\overline{ \langle {\tilde J}(t) \rangle }$ become saturated at about the 300th cycle. In the in-phase spiking group $\overline{ \langle W_J^{(G)}(t) \rangle }$ decreases rapidly from 0.722 to 0.198, while
$\overline{ \langle W_J^{(G)}(t) \rangle }$ in the out-of-phase spiking group decreases slowly from 0.273 to 0.174.
Thus, the saturated values of $\overline{ \langle W_J^{(G)}(t) \rangle }$  in both the in-phase and the out-of-phase spiking groups become close.

The percentage contribution $PC^{(G)}$ of $\overline{ \langle W_J^{(G)}(t) \rangle }$ in the $G$ spiking group to $\overline{ \langle {\tilde J}(t) \rangle }$
in the whole population is given by:
\begin{equation}
PC^{(G)}(\%) = \frac { \overline{ \langle W_J^{(G)}(t) \rangle } } { \overline{\langle {\tilde J}(t) \rangle} } \times 100.
\label{eq:PC}
\end{equation}
Figure \ref{fig:SWFF}(f2) shows a plot of $PC^{(G)}$ versus cycle [$G$: in-phase (solid circles) and out-of-phase (open circles)].
$PC^{(G)}$ of the in-phase spiking group decreases from 72.2 $\%$ to 53.2 $\%$, while $PC^{(G)}$ of the out-of-phase spiking group increases from 27.3 $\%$ to 46.8 $\%$. Thus, the saturated values of $PC^{(G)}$ of both the in-phase and the out-of-phase spiking groups get close.

We are particularly interested in the left, the right, and the middle vertical gray bands in each cycle in Figs.~\ref{fig:SWFF}(e1)-\ref{fig:SWFF}(e5) which denote the initial ($0 < t < 100$ msec), the final ($1900 < t < 2000$ msec), and the middle ($900 < t < 1100$ msec) stages, respectively.
In the case of in-phase (out-of-phase) spiking group, the maximum (minimum) of $\langle W_J^{(G)}(t) \rangle$ appears at the middle stage, while
the minimum (maximum) occurs at the initial and the final stages.
Figures \ref{fig:SWFF}(g1)-\ref{fig:SWFF}(g3) show plots of stage-averaged weighted synaptic weight $\overline{ \langle W_J^{(G)}(t) \rangle }$
[i.e., time average of $\langle W_J^{(G)}(t) \rangle$ over a stage] in the $G$ spiking-group [$G$: in-phase (solid circles) and out-of-phase (open circles)] and stage-averaged synaptic weight $\overline{ \langle {\tilde J}(t) \rangle }$ [i.e., time average of $\langle {\tilde J}(t) \rangle $ over the stage] in the whole population (crosses) versus cycle in the middle, the initial, and the final stages, respectively.
The sum of the values of $\overline{ \langle W_J^{(G)}(t) \rangle }$  at a time $t$ in the in-phase and the out-of-phase spiking groups
corresponds to the value of $\overline{ \langle {\tilde J}(t) \rangle }$ in the whole population.
As the cycle is increased, both $\overline{ \langle W_J^{(G)}(t) \rangle }$ and $\overline{ \langle {\tilde J}(t) \rangle }$ become saturated at about the
300th cycle. Figures \ref{fig:SWFF}(h1)-\ref{fig:SWFF}(h3) also show plots of percentage contribution $PC^{(G)}$ of the $G$ spiking group (i.e., ratio of the stage-averaged weighted synaptic weight $\overline{ \langle W_J^{(G)}(t) \rangle }$ in the $G$ spiking group to the stage-averaged synaptic weight $\overline{ \langle {\tilde J}(t) \rangle }$ in the whole population) in the middle, the initial, and the final stages, respectively [$G$: in-phase (solid circles) and out-of-phase (open circles)].

In the case of in-phase spiking group, $\overline{ \langle W_J^{(G)}(t) \rangle }$ decreases rapidly with cycle in all the 3 stages, while
in the case of out-of-phase spiking group, it also decreases in a relatively slow way with cycle.
At the middle stage, $\overline{ \langle W_J^{(G)}(t) \rangle }$ in the in-phase spiking group (solid circles) is much higher than that in the out-of-phase spiking group (open circles), and it decreases rapidly from 0.944 to 0.266. In this case, the percentage contribution $PC^{(G)}$ of the in-phase spiking group increases from 94.4 \% to 97.0 \%. Consequently, the contribution of in-phase spiking group is dominant at the middle stage, which leads to strong LTD at the PF-PC synapses.
On the other hand, at the initial and the final stages, with increasing cycle $\overline{ \langle W_J^{(G)}(t) \rangle }$ in the out-of-phase spiking group
becomes larger than that in the in-phase spiking group. The percentage contribution $PC^{(G)}$ of the out-of-phase spiking group increases from 49.2 \% to 70.2 \%, while that of the in-phase spiking group decreases from 50.8 \% to 29.8 \%. As a result, the contribution of out-of-phase spiking group at the initial and the final stages is larger than that of in-phase spiking group, which results in weak LTD at the PF-PC synapses.

In the above way, under good balance between the in-phase and the out-of-phase spiking groups (i.e., spiking-group ratio ${\cal{R}}^* \simeq 1.008$) in the optimal case of $p_c^*=0.06$, effective depression (i.e., strong/weak LTD) at the PF-PC synapses causes a big modulation in synaptic plasticity, which also leads to large modulations in firing activities of the PCs and the VN neuron (i.e., emergence of effective learning process). Hence, diverse recoding in the granular layer (i.e.,
appearance of diverse spiking patterns in the GR clusters) which results in effective (strong/weak) LTD at the PF-PC synapses is essential for effective motor learning for the OKR adaptation, which will be discussed in the following subsection.

\begin{figure}
\includegraphics[width=0.9\columnwidth]{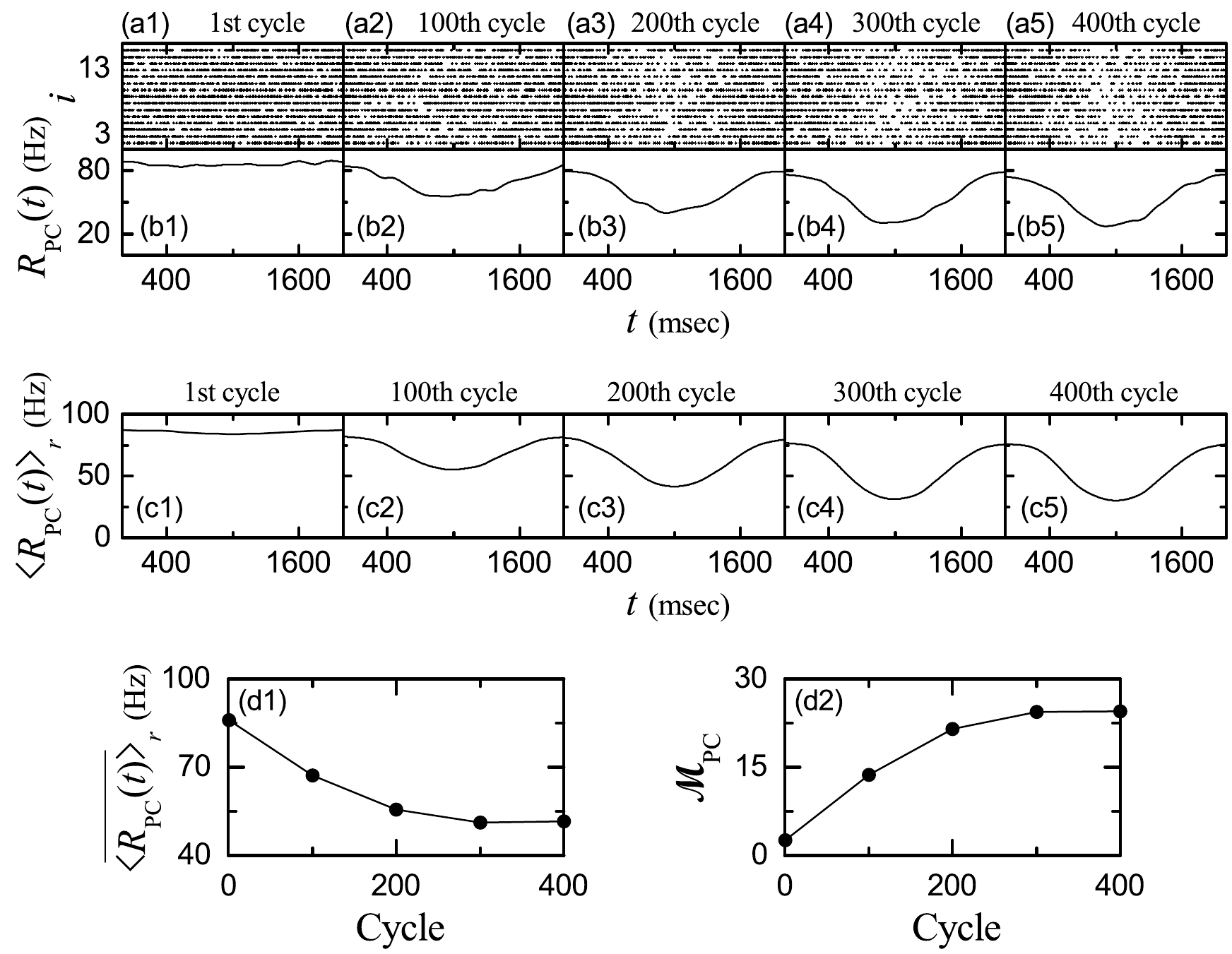}
\caption{Change in firing activity of PCs during learning in the optimal case of $p_c^* = 0.06$. (a1)-(a5) Raster plots of spikes of PCs and (b1)-(b5) instantaneous
population spike rates $R_{\rm PC}(t)$. (c1)-(c5) Realization-averaged instantaneous population spike rates $\langle R_{\rm PC} (t) \rangle_r$; number of realizations is 100. Plots of (d1) cycle-averaged mean $\overline{ \langle R_{\rm PC} (t) \rangle_r }$ and (d2) modulations ${\cal M}_{\rm PC}$ for $\langle R_{\rm PC} (t) \rangle_r$ versus cycle.
}
\label{fig:PC}
\end{figure}

\subsection{Effect of PF-PC Synaptic Plasticity on Subsequent Learning Process in The PC-VN-IO System}
\label{subsec:LP}
As a result of diverse recoding in the GR clusters, strong LTD occurs at the middle stage of a cycle due to dominant contribution of the in-phase spiking group.
On the other hand, at the initial and the final stages, somewhat less LTD takes place due to contribution of both the out-of-phase spiking
group (with weak LTD) and the in-phase spiking group.
Due to this kind of effective (strong/weak) LTD at the PF-PC synapses, a big modulation in synaptic plasticity at the PF-PC synapses (i.e., big modulation in bin-averaged normalized synaptic weight $\langle {\tilde J}(t) \rangle$) emerges. In this subsection, we investigate the effect of PF-PC synaptic plasticity with a big modulation on the subsequent learning process in the PC-VN-IO system.

Figure \ref{fig:PC} shows change in firing activity of PCs during learning in the optimal case of $p_c^* = 0.06$.
Cycle-evolutions of raster plots of spikes of 16 PCs and the corresponding instantaneous population spike rates $R_{\rm PC}(t)$ are shown
in Figs.~\ref{fig:PC}(a1)-\ref{fig:PC}(a5) and Figs.~\ref{fig:PC}(b1)-\ref{fig:PC}(b5), respectively.
Since the number of PCs is small, $R_{\rm PC}(t)$ seems to be a little rough. To get a smooth estimate for $R_{\rm PC}(t)$, we make 100 realizations.
Realization-averaged smooth instantaneous population spike rates $\langle R_{\rm PC}(t) \rangle_r$ are given in
Figs.~\ref{fig:PC}(c1)-\ref{fig:PC}(c5); $\langle \cdots \rangle_r$ denotes realization average and the number of realizations is 100. $\langle R_{\rm PC}(t) \rangle_r$ seems to be saturated at about the 300th cycle.

With increasing the learning cycle, raster plots of spikes of all the 16 PCs become more and more sparse at the middle stage, which may be clearly
seen in the instantaneous population spike rate $\langle R_{\rm PC}(t) \rangle_r$. Due to the effect of synaptic plasticity at the PF-PC synapses, the minimum of $\langle R_{\rm PC}(t) \rangle_r$ appears at the middle stage, while the maximum occurs at the initial and the final stages.
Thus, the modulation of $\langle R_{\rm PC}(t) \rangle_r$ increases with increasing the cycle.

\begin{figure}
\includegraphics[width=0.9\columnwidth]{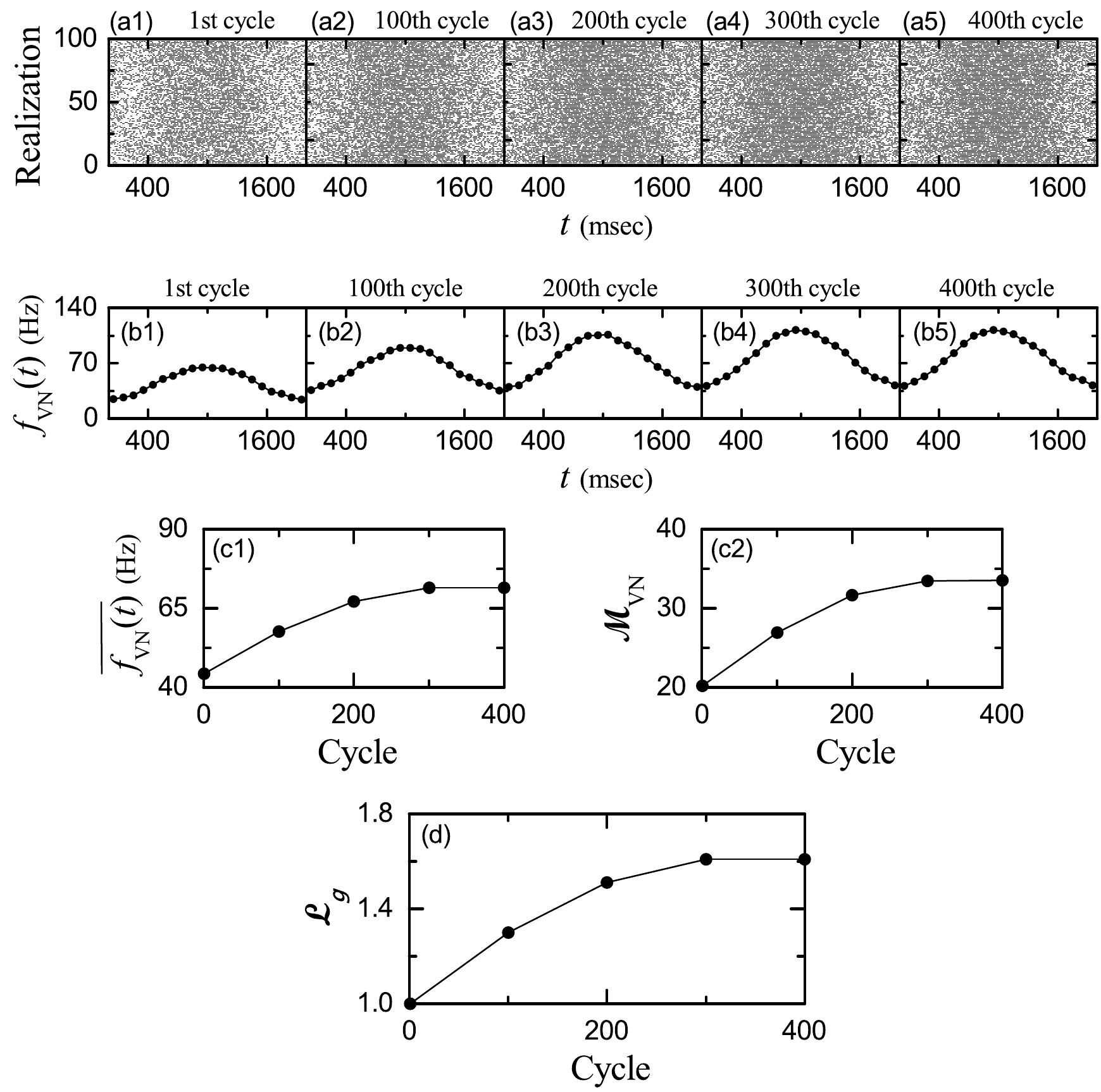}
\caption{Change in firing activity of the VN neuron during learning in the optimal case of $p_c^* = 0.06$.
(a1)-(a5) Raster plots of spikes of the VN neuron (i.e., collection of spike trains for all the realizations; number of realizations is 100) and (b1)-(b5) bin-averaged instantaneous individual firing rate $f_{\rm VN}(t)$; the bin size is $\Delta t=$ 100 msec. Plots of (c1) cycle-averaged mean $\overline{f_{\rm VN}(t)}$ and (c2) modulation $M_{\rm VN}$ for $f_{\rm VN}(t)$ versus cycle. (d) Plot of learning gain degree ${\cal L}_g$ versus cycle.
}
\label{fig:VN}
\end{figure}

In-phase PF (student) signals are strongly depressed by the in-phase CF (instructor) signals, while out-of-phase PF signals are weakly pressed due to the
phase difference between the out-of-phase PF signals and the in-phase CF signals. Fraction of in-phase PF signals at the middle stage of a cycle is dominant.
On the other hand, at the initial and the final stages, fraction of out-of-phase PF signals are larger than that of in-phase PF signals.
Thus, bin-averaged normalized synaptic weights $\langle {\tilde J}(t) \rangle$ form a well-shaped curve, as shown in Figs.~\ref{fig:SW}(b1)-\ref{fig:SW}(b5).
That is, strong LTD occurs at the middle stage, while at the initial and the final stages, weak LTD takes place.
As a result of this kind of effective depression (strong/weak LTD) at the (excitatory) PF-PC synapses, $\langle R_{\rm PC}(t) \rangle_r$ becomes lower at the middle stage (strong LTD) than at the initial and the final stages (weak LTD). Thus, $\langle R_{\rm PC}(t) \rangle_r$ forms a well-shaped curve with a minimum at the middle stage.

Figures \ref{fig:PC}(d1) and \ref{fig:PC}(d2) show plots of cycle-averaged mean $\overline{ \langle R_{\rm PC}(t) \rangle_r }$
(i.e., time average of $\langle R_{\rm PC}(t) \rangle_r$ over a cycle) and modulation ${\cal M}_{\rm PC}$ of $\langle R_{\rm PC}(t) \rangle_r$ versus
cycle, respectively. Due to LTD at the PF-PC synapses, the cycle-averaged mean $\overline{ \langle R_{\rm PC}(t) \rangle_r }$ decreases monotonically from 86.1 Hz, and it becomes saturated at 51.7 Hz at about the 300th cycle. On the other hand, the modulation ${\cal M}_{\rm PC}$ increases monotonically from 2.6 Hz, and it get saturated at 24.1 Hz at about the 300th cycle. Consequently, a big modulation occurs in ${\cal M}_{\rm PC}$
due to the effective depression (strong/weak LTD) at the PF-PC synapses.
These PCs (principal cells of the cerebellar cortex) exert effective inhibitory coordination on the VN neuron which evokes OKR eye-movement.

Figure \ref{fig:VN} shows change in firing activity of the VN neuron which produces the final output of the cerebellum during learning in the optimal case of $p_c^* = 0.06$. Cycle-evolutions of raster plots of spikes of the VN neuron (i.e., collection of spike trains for all the realizations; number of realizations is 100)
and  the bin-averaged instantaneous individual firing rates $f_{\rm VN}(t)$ [i.e., the number of spikes of the VN neuron in a bin divided by the bin width
($\Delta t = 100$ msec)] are shown in Figs.~\ref{fig:VN}(a1)-\ref{fig:VN}(a5) and Figs.~\ref{fig:VN}(b1)-\ref{fig:VN}(b5), respectively.
$f_{\rm VN}(t)$ seems to be saturated at about the 300th cycle.

In contrast to the case of PCs, as the cycle is increased, raster plots of spikes of the VN neuron become more and more dense at the middle stage, which may be clearly seen in the instantaneous individual firing rates $f_{\rm VN}(t)$.
Due to the effective inhibitory coordinations of PCs on the VN neuron, the maximum of $f_{\rm VN}(t)$ appears at the middle stage, while the minimum occurs at the initial and the final stages. Thus, $f_{\rm VN}(t)$ forms a bell-shaped curve.

\begin{figure}
\includegraphics[width=0.9\columnwidth]{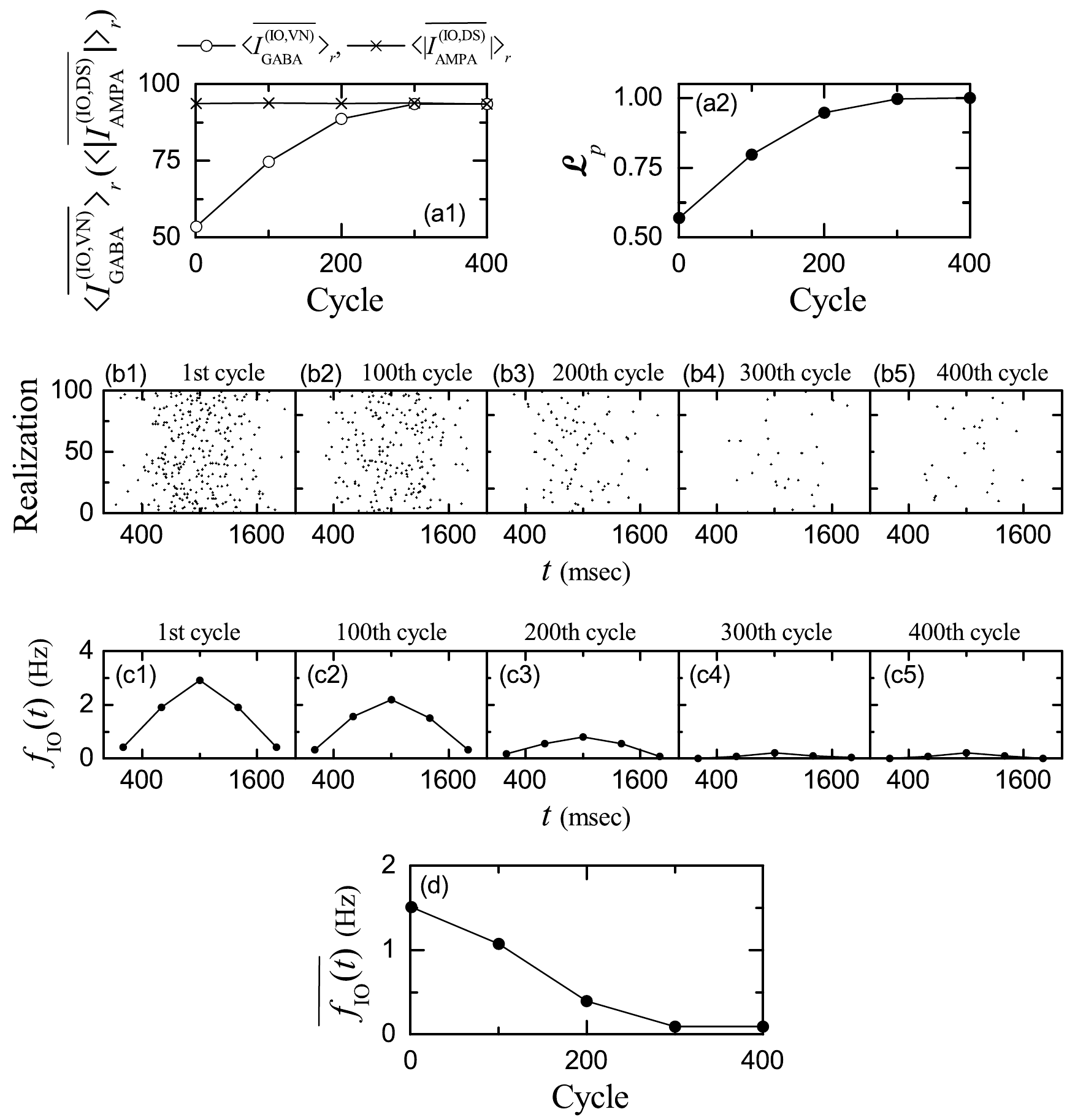}
\caption{Change in firing activity of the IO neuron during learning in the optimal case of $p_c^* = 0.06$.
Plots of (a1) realization-average for the cycle-averaged inhibitory synaptic current from the VN neuron ($\langle \overline {I_{\rm GABA}^{\rm (IO,VN)}} \rangle_r$) (open circles) and realization-average for the magnitude of the cycle-averaged excitatory synaptic current through the IO desired signal
($\langle |\overline {I_{\rm AMPA}^{\rm (IO,DS)}}| \rangle_r$) (crosses) versus cycle; number of realizations $\langle \cdots \rangle_r$ is 100.
(a2) Plot of learning progress degree ${\cal L}_p$ versus cycle. (b1)-(b5) Raster plots of spikes of the IO neuron (i.e., collection of spike trains for all the realizations; number of realizations is 100) and (c1)-(c5) bin-averaged instantaneous individual firing rate $f_{\rm IO}(t)$; the bin size is $\Delta t=$ 400 msec.
(d) Plot of cycle-averaged individual firing rate $\overline{ f_{\rm IO}(t) }$ versus cycle.
}
\label{fig:IO}
\end{figure}

Figures \ref{fig:VN}(c1) and \ref{fig:VN}(c2) show plots of cycle-averaged mean $\overline{ f_{\rm VN}(t) }$
of $f_{\rm VN}(t)$ [i.e., time average of $f_{\rm VN}(t)$ over a cycle] and modulation ${\cal M}_{\rm VN}$ of $f_{\rm VN}(t)$ versus
cycle, respectively. Due to the decreased inhibitory inputs from the PCs, the cycle-averaged mean $\overline{ f_{\rm VN}(t) }$ increases monotonically
from 44.3 Hz, and it becomes saturated at 71.5 Hz at about the 300th cycle. Also, the modulation of $f_{\rm VN}(t)$ increases from 20.2 Hz, and it
gets saturated at 32.5 Hz at about the 300th cycle. As a result of effective inhibitory coordination of PCs,
a big modulation occurs in ${\cal M}_{\rm VN}$.

The learning gain degree ${\cal L}_g$, corresponding to the modulation gain ratio, is given by the normalized modulation of $f_{\rm VN}(t)$ divided by that at the 1st cycle:
\begin{equation}
  {\cal L}_g = \frac {{\cal M}_{\rm VN}} {{\cal M}_{\rm VN}~{\rm at~the~1st~cycle}},
\label{LGD}
\end{equation}
where ${\cal M}_{\rm VN}$ at the 1st cycle is 20.2 Hz.
Figure \ref{fig:VN}(d) shows a plot of ${\cal L}_g$ versus cycle.
${\cal L}_g$ increases monotonically from 1, and it becomes saturated at about the 300th cycle.
Thus, we get the saturated learning gain degree ${\cal L}_g^*$ $(\simeq 1.608)$.
As will be seen in the next subsection, ${\cal L}_g^*~(\simeq 1.608)$ is the largest one among the available ones.
Hence, in the optimal case of $p_c^*=0.06$ where spiking patterns of GR clusters with the diversity degree ${\cal{D}}^*~(\simeq 1.613)$
are the most diverse, motor learning for the OKR adaptation with the saturated learning gain degree
${\cal L}_g^*~(\simeq 1.608)$ is the most effective.

Learning progress may be seen clearly in the IO system. During the learning cycle, the IO neuron receives both the excitatory desired
signal for a desired eye-movement and the inhibitory signal from the VN neuron (denoting a realized eye-movement).
We introduce the learning progress degree ${\cal{L}}_p$, given by the ratio of the cycle-averaged inhibitory input from the VN neuron to
the magnitude of the cycle-averaged excitatory input via the desired signal:
\begin{equation}
{\cal L}_p = \frac {\overline{ I_{\rm GABA}^{\rm (IO,VN)} }} {|\overline{ I_{\rm AMPA}^{\rm (IO,DS)}}|},
\label{eq:IER}
\end{equation}
where $\overline{ I_{\rm GABA}^{\rm (IO,VN)} }$ is the cycle-averaged inhibitory GABA receptor-mediated current from the VN neuron into the IO neuron,
and $\overline{ I_{\rm AMPA}^{\rm (IO,DS)} }$ is the cycle-averaged excitatory AMPA receptor-mediated current into the IO neuron via the desired signal;
no (excitatory) NMDA receptors exist on the IO neuron. [Note that the 4th term in Eq.~(\ref{eq:GE}) is given by $- I_{syn,i}^{(X)}(t)$, because
$I_{\rm GABA}^{\rm (IO,CN)} >0$ and $I_{\rm AMPA}^{\rm (IO,US)} <0.$]

Figure \ref{fig:IO}(a1) shows plots of $\overline{ I_{\rm GABA}^{\rm (IO,VN)} }$ (open circles)  and $|\overline{ I_{\rm AMPA}^{\rm (IO,DS)}}|$ (crosses) versus cycle in the optimal case of $p_c^* = 0.06$. With increasing the cycle, the cycle-averaged inhibitory input from the VN neuron increases, and converges to the constant magnitude of the cycle-averaged excitatory input through the IO desired signal. Thus, as shown in Fig.~\ref{fig:IO}(a2), ${\cal{L}}_p$ increases with cycle, and at about the 300th cycle, it becomes saturated at ${\cal{L}}_p =1$. In this saturated case, the cycle-averaged excitatory and inhibitory inputs into the IO
neuron are balanced.

We also study the firing activity of IO neuron during learning process. Figures~\ref{fig:IO}(b1)-\ref{fig:IO}(b5) and Figures~\ref{fig:IO}(c1)-\ref{fig:IO}(c5)
show cycle-evolutions of raster plots of spikes of the IO neuron (i.e., collection of spike trains for all the realizations; number of realizations is 100)
and  the bin-averaged instantaneous individual firing rates $f_{\rm IO}$ [i.e., the number of spikes of the IO neuron in a bin divided by the bin width
($\Delta t = 400$ msec)], respectively. In the 1st cycle, relatively dense spikes appear at the middle stage of the cycle in the raster plot of spikes, due to the effect of excitatory IO desired signal. However, with increasing the cycle, spikes at the middle stage become sparse, because of increased inhibitory input from the VN neuron. In this case, the bin-averaged instantaneous individual firing rate $f_{\rm IO}(t)$ of the IO neuron forms a bell-shaped curve due to the sinusoidally-modulating desired input signal into the IO neuron. With increasing the cycle, the amplitude of $f_{\rm IO}(t)$ decreases due to the inhibitory input from the VN neuron, and it becomes saturated at about the 300th cycle. Thus, the cycle-averaged individual firing rate $\overline{ f_{\rm IO}(t) }$ is decreased from 1.51 Hz to 0.09 Hz, as shown in Fig.~\ref{fig:IO}(d). The firing output of the IO neuron is fed into the PCs via the CFs. Hence, with increasing the cycle, the error-teaching CF instructor signals become weaker and saturated at about the 300th cycle.

While the saturated CF signals are fed into the PCs, saturation for the cycle-averaged bin-averaged synaptic weights $\overline{ \langle {\tilde J}(t) \rangle }$ appears [see Fig.~\ref{fig:SW}(c)]. Then, the subsequent learning process in the PC-VN system also becomes saturated, and we get the saturated learning gain degree ${\cal{L}}_g^*(\simeq 1.608)$, which is shown in Fig.~\ref{fig:VN}(d).

\begin{figure}
\includegraphics[width=0.9\columnwidth]{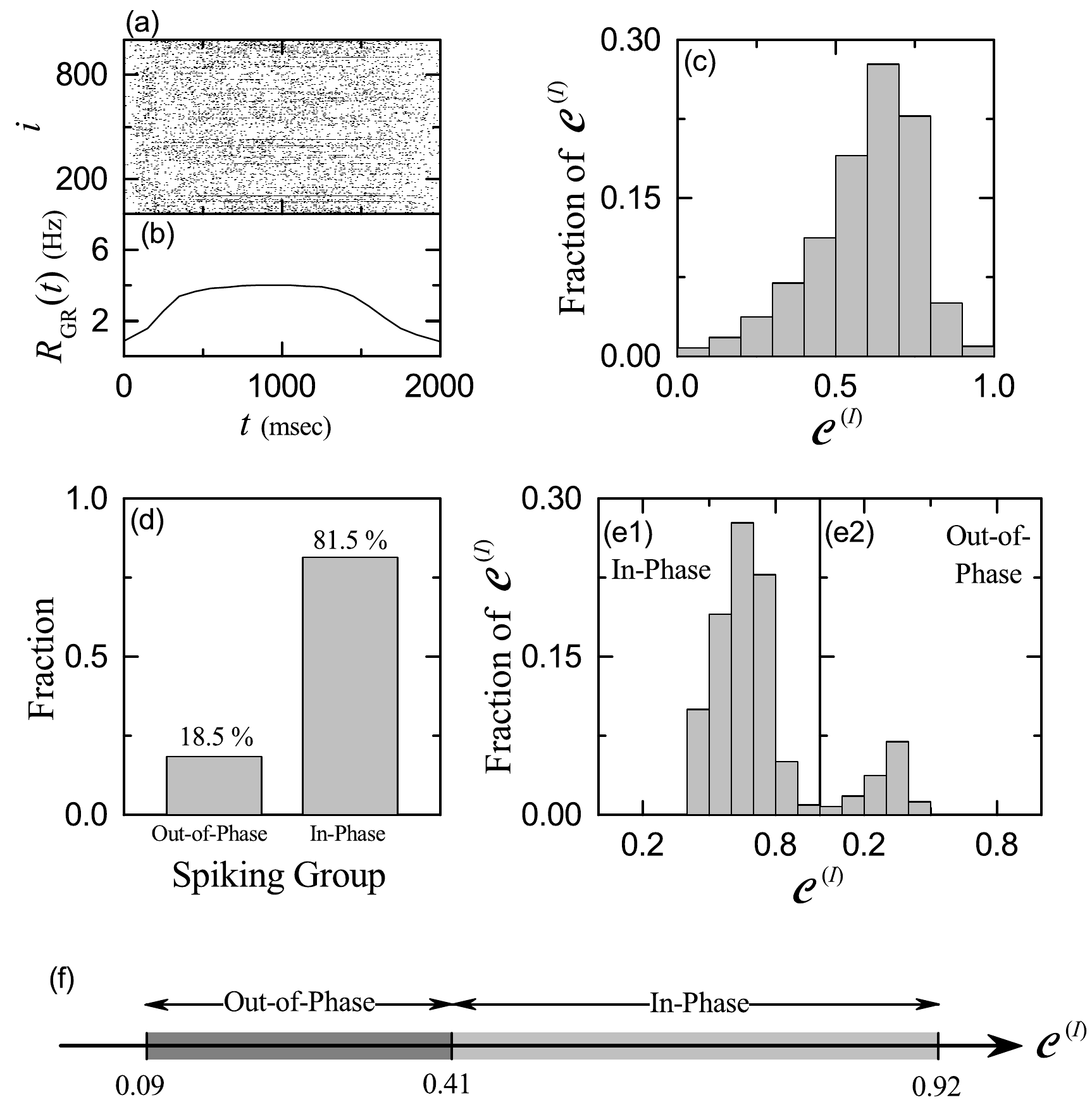}
\caption{Highly-connected case of $p_c=0.6$.
(a) Raster plot of spikes of $10^3$ randomly chosen GR cells. (b) Instantaneous whole-population spike rate $R_{\rm GR}(t)$ in the whole population of GR cells.
Band width for $R_{\rm GR}(t)$: $h=10$ msec. (c) Distribution of conjunction indices $\{ {\cal{C}}^{(I)} \}$ for the GR clusters in the whole population. (d) Fraction of spiking groups. Distributions of conjunction indices $\{ {\cal{C}}^{(I)} \}$ for the (e1) in-phase and (e2) out-of-phase spiking groups. Bin size for the histograms in (c) and (e1)-(e2) is 0.1. (f) Ranges of $\{ {\cal{C}}^{(I)} \}$ in the in-phase and the out-of-phase spiking groups.
}
\label{fig:HC1}
\end{figure}

\subsection{Dependence of Diversity Degree $\cal{D}$ and Learning Gain Degree ${\cal{L}}_g$ on $p_c$ (Connection
            Probability from GO to GR Cells)}
\label{subsec:PC}
In the above subsections, we consider only the optimal case of $p_c^*=0.06$ (i.e., $6\%$) where the spiking patterns of the GR clusters are the most diverse.
From now on, we vary the connection probability $p_c$ from GO to GR cells, and investigate the dependence of the diversity degree $\cal D$ for the spiking patterns of the GR clusters and the learning gain degree ${\cal L}_g$ on $p_c$.

We first consider the highly-connected case of $p_c=0.6$ (i.e., $60\%$). Figure \ref{fig:HC1}(a) shows the raster plot of spikes of $10^3$ randomly chosen GR cells, and the population-averaged firing activity in the whole population of GR cells may be well seen in the instantaneous whole-population spike rate $R_{\rm GR}(t)$ in Fig.~\ref{fig:HC1}(b). As shown in Fig.~\ref{fig:RN}(b), each GR cluster is bounded by two glomeruli (corresponding to the terminals of the MFs) at both ends. Each glomerulus receives inhibitory inputs from nearby 81 GO cells with the connection probability $p_c=0.6$. Hence, on average, about 49 GO cell axons innervate each glomerulus. Then, each GR cell in a GR cluster receives about 97 inhibitory inputs via two dendrites which contact the two glomeruli at both ends.
Due to the increased inhibitory inputs from the pre-synaptic GO cells, spike density in the raster plot is decreased, and the top part of $R_{\rm GR}(t)$ becomes lowered and broadly flattened, in comparison with the optimal case in Fig.~\ref{fig:WP}(b). Thus, $R_{\rm GR}(t)$ becomes more different from the firing rate $f_{\rm MF}$ for the MF signal in Fig.~\ref{fig:OKR}(b1).

GR cells in each GR cluster shares the same inhibitory and the excitatory inputs through their dendrites which synaptically contact the two glomeruli
at both ends. Thus, GR cells in each GR cluster exhibit similar firing activity. Then, similar to the case of $R_{\rm GR}(t)$, the cluster-averaged firing activity in the $I$th GR cluster ($I=1,\cdots,2^{10}$) may be well described in terms of its instantaneous cluster spike rate $R_{\rm GR}^{(I)}(t)$ of Eq.~(\ref{eq:ISPSR}).
In this case, the conjunction index ${\cal C}^{(I)}$ of the $I$th GR cluster
(representing the similarity degree between the spiking behavior [$R_{\rm GR}^{(I)}(t)$] of the $I$th GR cluster and that of the whole population
[$R_{\rm GR}(t)$]) is given by the cross-correlation at the zero time lag between $R_{\rm GR}^{(I)}(t)$ and $R_{\rm GR}(t)$ [see Eq.~(\ref{eq:CI})].

Figure \ref{fig:HC1}(c) shows the distribution of conjunction indices $\{ C^{(I)} \}$ with a peak at 0.65.
When compared with the optimal case in Fig.~\ref{fig:Char}(a) with a peak at 0.55, the whole distribution is moved to the right, and
the values of $ C^{(I)} $ for all the GR clusters are positive.  Thus, all the anti-phase and complex out-of-phase spiking patterns with negative
values of $ C^{(I)} $ disappear. Only the in-phase and out-of-phase spiking patterns with positive values of $C^{(I)}$ persist.
Consequently, the mean of the distribution $\{ C^{(I)} \}$ is increased to 0.613, while the standard deviation is decreased to 0.125, in comparison
to the optimal case where the mean and the standard deviation are 0.320 and 0.516, respectively.
Then, the diversity degree ${\cal D}$ of the spiking patterns $\{ R_{\rm GR}^{(I)}(t) \}$ in all the GR clusters, representing a quantitative measure for diverse recoding in the granular layer, is given by the relative standard deviation for the distribution $\{ C^{(I)} \}$ [see Eq.~(\ref{eq:DD})].
In the highly-connected case of $p_c=0.6$, its diversity degree is ${\cal D} \simeq 0.204$  which is much smaller than ${\cal D}^*~(\simeq 1.613)$
in the optimal case.

The reduction in $\cal D$ for the spiking patterns (corresponding to the firing outputs) of the GR clusters arises due to decrease in differences between the total synaptic inputs into each GR clusters. As the connection probability $p_c$ from the GO to GR cells is increased, differences between the total inhibitory synaptic inputs from the pre-synaptic GO cells into each GR clusters are decreased due to increase in the number of pre-synaptic GO cells. On the other hand, the excitatory inputs into each GR clusters via MFs are Poisson spike trains with the same firing rates, and hence they are essentially the same. Thus, differences between the total synaptic inputs (including both the inhibitory and the excitatory inputs) into each GR clusters become reduced. These less different inputs into the GR clusters produce less different outputs (i.e. spiking patterns) in the GR clusters, which leads to decreases in the diversity degree $\cal D$ in the highly-connected case.

We decompose the whole GR clusters into the in-phase and the out-of-phase spiking groups. Unlike the optimal case of $p_c^*=0.06$,
no anti-phase spiking group appears. Figure \ref{fig:HC1}(d) shows the fraction of spiking groups. The in-phase spiking group is a major one
with fraction 81.5$\%$, while the out-of-phase spiking group is a minor one with fraction 18.5$\%$. In comparison with the optimal case
where the fraction of in-phase spiking group is 50.2$\%$, the fraction of in-phase spiking group for $p_c=0.6$ is so much increased.
In this highly-connected case, the spiking-group ratio, given by the ratio of the fraction of the in-phase spiking group to that
of the out-of-phase spiking group, is ${\cal{R}} \simeq 4.405$ which is much larger than that (${\cal R}^* \simeq 1.008$) in the optimal case.
Thus, good balance between the in-phase and the out-of-phase spiking groups in the optimal case becomes broken up because the in-phase spiking group is a dominant one. In this unbalanced case, the diversity degree $\cal D$ for the spiking patterns in the GR clusters is decreased so much to ${\cal D} \simeq 0.204$, in comparison with ${\cal D}^*~(\simeq 1.613)$ in the optimal case of $p_c^*=0.06$.

Figures \ref{fig:HC1}(e1) and \ref{fig:HC1}(e2) also show plots of the fractions of conjunction indices ${\cal{C}}^{(I)}$
of the GR clusters in the in-phase and the out-of-phase spiking groups, respectively. The ranges for the distributions $\{ {\cal{C}}^{(I)} \}$ in the two
spiking groups are also given in the bar diagram in Fig.~\ref{fig:HC1}(f).
As in the optimal case of $p_c^*=0.06$ in Figs.~\ref{fig:DSP} and \ref{fig:Char}, we determine the threshold ${\cal C}_{th}~(\simeq 0.41)$ between the in-phase and
the out-of-phase spiking patterns by making intensive examination of phase difference of $R_{\rm GR}^{(I)}(t)$ of the GR clusters relative to $R_{\rm GR}(t)$.
For ${\cal C}^{(I)} > {\cal C}_{th},$ in-phase spiking patterns with one or more peaks in the middle part of cycle exist.
On the other hand, when passing the threshold ${\cal C}_{th}$ from the above (i.e., for ${\cal C}^{(I)} < {\cal C}_{th})$ out-phase spiking patterns appear.
These out-of-phase spiking patterns have left-skewed (right-skewed) peaks near the 1st (3rd) quartile of cycle (i.e., near $t=500$ (1500) msec).

\begin{figure}
\includegraphics[width=0.9\columnwidth]{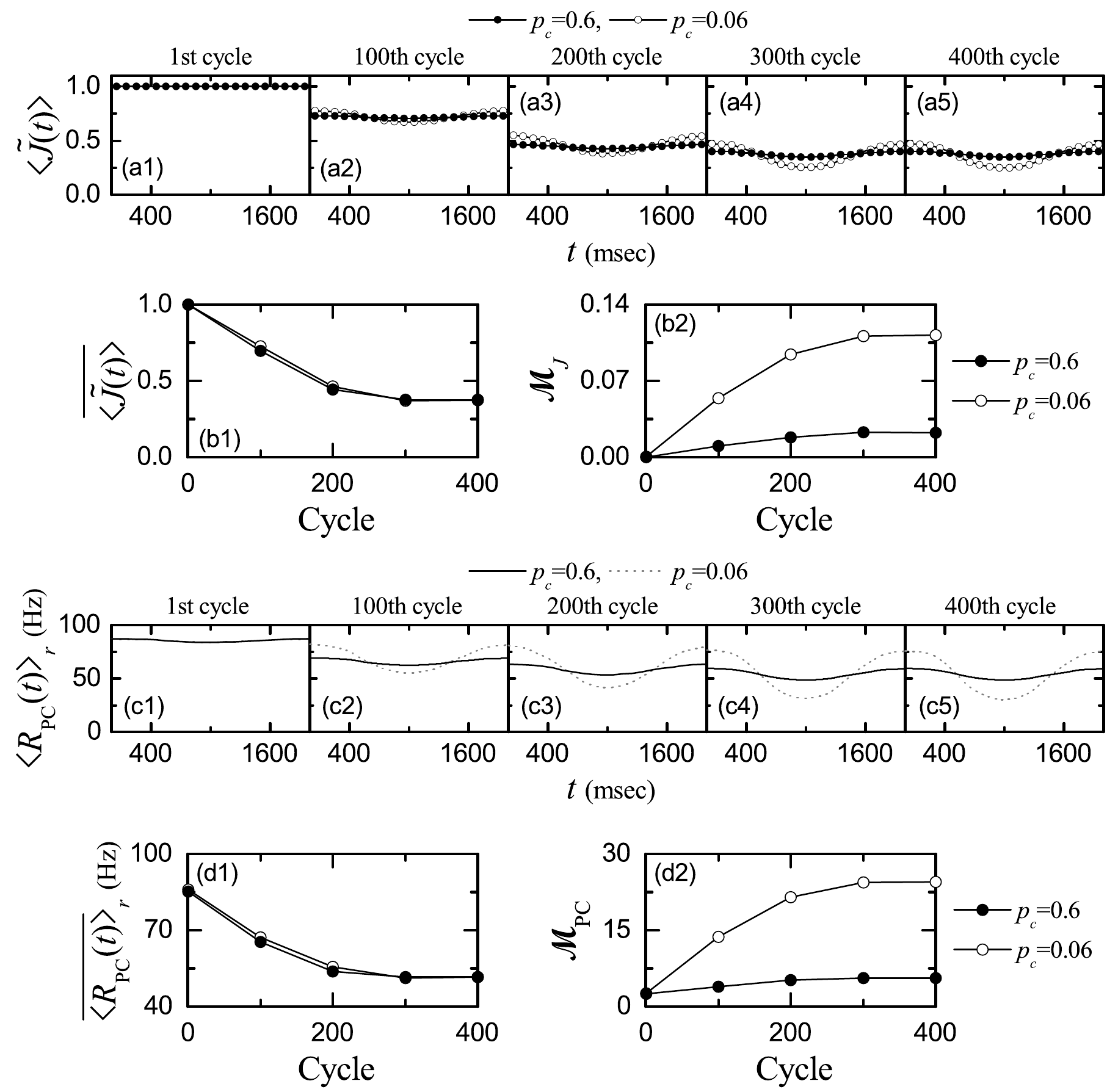}
\caption{Change in synaptic weights of active PF-PC synapses and firing activity of PCs during learning in the highly-connected case of $p_c = 0.6$;
for comparison, data in the optimal case of $p_c^*=0.06$ are also given. (a1)-(a5) Cycle-evolution of bin-averaged (normalized) synaptic weights $\langle {\tilde J}(t) \rangle$ of active PF signals (bin size: $\Delta t=$ 100 msec). Plots of (b1) cycle-averaged mean $\overline{ \langle {\tilde J}(t) \rangle }$ and (b2) modulation ${\cal M}_J$ for $\langle {\tilde J}(t) \rangle$ versus cycle. (c1)-(c5) Realization-averaged instantaneous population spike rate $\langle R_{PC}(t) \rangle_r$; the number of realizations is 100. Plots of (d1) cycle-averaged mean
$\overline{ \langle R_{PC}(t) \rangle_r }$ and (d2) modulations ${\cal M}_{PC}$ for $\langle R_{PC}(t) \rangle_r$ versus cycle.
In (a1)-(a5), (b1)-(b2), and (d1)-(d2), solid (open) circles represent data in the case of $p_c=$ 0.6 (0.06).
In (c1)-(c5), solid (dotted) lines denote data in the case of $p_c=$ 0.6 (0.06).
}
\label{fig:HC2}
\end{figure}

In the case of in-phase spiking group, the distribution $\{ {\cal{C}}^{(I)} \}$ with a peak at 0.65 has positive values in the range of $(0.41, 0.92)$.
When compared with the optimal case where the range is (0.39, 0.85), the range in the highly-connected case of $p_c=0.6$ is shifted to the right
and a little widened. Also, the mean and standard deviation for $p_c=0.6$ are 0.677 and 0.094, respectively. In comparison with the optimal case where
the mean is 0.538 and the standard deviation is 0.181, the relative standard deviation for $p_c=0.6$ is much reduced.

Also, in the case of the out-of-phase spiking group, the distribution $\{ {\cal{C}}^{(I)} \}$ with a peak at 0.35 has only positive values in the range
of $(0.09,0.41)$, and its mean and standard deviations are 0.325 and 0.067, respectively.
Unlike the optimal case, there are no out-of-phase spiking patterns with negative values of ${\cal{C}}^{(I)}$.
Only out-of-phase spiking patterns which are developed from the in-phase spiking patterns appear, and hence they have just positive values
of ${\cal{C}}^{(I)}$. Thus, the range for $p_c=0.6$ is moved to the positive region, and it becomes narrowed, in comparison to the optimal case
where the range is (-0.20,0.39). Also, the relative standard deviation for $p_c=0.6$ is much decreased, in comparison to the optimal case where
the mean and the standard deviation are 0.102 and 0.328, respectively.

In the optimal case of $p_c^*=0.06$, the in-phase and the out-of-phase spiking groups are shown to play their own roles for synaptic plasticity at the PF-PC synapses, respectively. As a result of cooperation in their good-balanced state, effective depression (strong/weak LTD) at the PF-PC synapses occurs, which
eventually leads to effective motor learning for the OKR adaptation in the PC-VN-IO system. On the other hand, in the highly-connected case of $p_c=0.6$, the
in-phase spiking group becomes a dominant one, and hence good balance between the in-phase and the out-of-phase spiking groups is broken up. In such an unbalanced
state, contribution of the out-of-phase spiking group to the synaptic plasticity at the PF-PC synapses is decreased so much, which is clearly shown below.

Figure \ref{fig:HC2} shows change in synaptic weights of active PF-PC synapses and firing activity of PCs during learning in the highly-connected case of $p_c = 0.6$. Cycle-evolution of  bin-averaged synaptic weights $\langle {\tilde J}(t) \rangle$ (solid circles) of active PF signals is shown in
Figs.~\ref{fig:HC2}(a1)-\ref{fig:HC2}(a5). For comparison, data for $\langle {\tilde J}(t) \rangle$ (open circles) in the optimal case of $p_c^*=0.06$ are also given. In comparison with the optimal case, the bin-averaged synaptic weights $\langle {\tilde J}(t) \rangle$ at the middle stage are less depressed, while at the initial and the final stages, they are more depressed. Thus, the modulation of $\langle {\tilde J}(t) \rangle$ is reduced so much. This small modulation is
distinctly in contrast to the big modulation in the optimal case. $\langle {\tilde J}(t) \rangle$ also becomes saturated at about the 300th cycle, as in the optimal case.

In this highly-connected case of $p_c=0.6,$ less depression (in the bin-averaged synaptic weights $\langle {\tilde J}(t) \rangle$) at the middle stage may be easily understood in the following way. We note that, at the middle stage, the in-phase spiking group makes a dominant contribution to $\langle {\tilde J}(t) \rangle$
(i.e., the contribution of the out-of-phase spiking group to $\langle {\tilde J}(t) \rangle$ may be negligible), as discussed in Fig.~\ref{fig:SWFF} in the optimal case of $p_c=0.06.$ However, the degree of depression changes depending on the relative phase difference between the in-phase PF (student) signals and the error-teaching (instructor) CF signals. The CF signals are in-phase ones with respect to the firing rate $f_{\rm DS}(t)$ of the Poisson spike trains for the IO desired signal (for a desired eye-movement) in Fig.~\ref{fig:OKR}(b2), and the in-phase PF signals are in-phase ones relative to the instantaneous whole-population spike rate $R_{\rm GR}(t)$. We note that, depending on $p_c,$ the reference signal $R_{\rm GR}(t)$ for the in-phase PF signals has a varying ``matching'' degree relative to the sinusoidally-modulating IO desired signal $f_{\rm DS}(t)$. Here, the matching degree ${\cal M}_d$ is quantitatively given by the cross-correlation at the zero-time lag [$Corr_{\rm M}(0)$] between $R_{\rm GR}(t)$ and $f_{\rm DS}(t)$:
\begin{equation}
Corr_{\rm M} (\tau) = \frac{\overline{\Delta f_{\rm DS}(t+\tau) \Delta R_{\rm GR}(t)}} {\sqrt{\overline{\Delta f_{\rm DS}^{2}(t)}} \sqrt{\overline{{\Delta R_{\rm GR}^2(t)}}} },
\label{eq:MD}
\end{equation}
where $\Delta f_{\rm DS}(t) = f_{\rm DS}(t)-\overline{f_{\rm DS}(t)}$, $\Delta R_{\rm GR}(t) = R_{\rm GR}(t)-\overline{R_{\rm GR}(t)}$, and the overline denotes the time average over a cycle.

With increasing $p_c$, the top part of $R_{\rm GR}(t)$ becomes more broadly flattened, and hence its matching degree with respect to $f_{\rm DS}(t)$ is decreased.
For $p_c=0.6,$ ${\cal M}_d=0.625,$ which is smaller than that (${\cal M}_d=0.857$) in the optimal case of $p_c^* =0.06$.
Hence, due to decrease in the matching degree between $R_{\rm GR}(t)$ and $f_{\rm DS}(t)$, on average, in-phase PF signals for $p_c=0.6$ are less depressed by
the CF signals than those in the optimal case. Thus, in the highly-connected case of $p_c=0.6$, the bin-averaged synaptic weights $\langle {\tilde J}(t) \rangle$
(solid circles) at the middle stage (with a dominant in-phase spiking group) are less depressed than those (open circles) in the optimal case.
In this way. at the middle stage where the contribution of the in-phase spiking group is dominant, the depression degree for $\langle {\tilde J}(t) \rangle$
is determined by the matching degree ${\cal M}_d$ between $R_{\rm GR}(t)$ and $f_{\rm DS}(t)$.

More depression (in the bin-averaged synaptic weights $\langle {\tilde J}(t) \rangle$) at the initial and the final stages for $p_c=0.6$ can also be clearly understood as follows. In the optimal case, at the initial and the final stages, the firing fractions $F^{(G)}(t)$ of the in-phase ($51\%$) and the out-of-phase ($49\%$) spiking groups are nearly the same (see Eq.~(\ref{eq:FF}) and Fig.~\ref{fig:SWFF}). Hence, at the initial and the final stages, somewhat less LTD occurs at the PF-PC synapses, in contrast to strong LTD at the middle stage, because both the out-of-phase spiking group (with weak LTD) and the in-phase spiking group make contributions together. However, in the highly-connected case of $p_c=0.6$, the fraction of the in-phase spiking group is so much increased to $81.5\%$,
in comparison to the optimal case ($50.2\%$). Hence, the firing fraction of the in-phase spiking group becomes much larger even at the initial and the final stages
($F^{(G)}(t) = 0.82$). Thus, the bin-averaged synaptic weights $\langle {\tilde J}(t) \rangle$ (solid circles) at the initial and the final stages are more depressed than those (open circles) in the optimal case. In this way. at the initial and the final stages, the depression degree for $\langle {\tilde J}(t) \rangle$ is determined by the relative fractions of the in-phase and the out-of-phase spiking groups; as the fraction of the in-phase spiking group is increased, $\langle {\tilde J}(t) \rangle$ is more depressed.

Figures \ref{fig:HC2}(b1) and \ref{fig:HC2}(b2) show plots of cycle-averaged mean $\overline{ \langle {\tilde J}(t) \rangle }$ and modulation ${\cal M}_J$ for $\langle {\tilde J}(t) \rangle$ versus cycle, respectively; $p_c=0.6$ (solid circles) and $p_c^*=0.06$ (open circles). Both the cycle-averaged mean $\overline{ \langle {\tilde J}(t) \rangle }$ and the modulation ${\cal M}_J$ for $\langle {\tilde J}(t) \rangle$ become saturated at about the 300th cycle.
With increasing the cycle, the cycle-averaged mean $\overline{ \langle {\tilde J}(t) \rangle }$ decreases from 1 to 0.374 due to LTD at the PF-PC synapses,
which is similar to the optimal case of $p_c^*=0.06$ where $\overline{ \langle {\tilde J}(t) \rangle }$ decreases from 1 to 0.372.
On the other hand, the modulation ${\cal M}_J$ increases very slowly from 0 to 0.023, in contrast to the optimal case with a big modulation where it increases quickly from 0 to 0.112. When compared with the optimal case, bin-averaged synaptic weights $\langle {\tilde J}(t) \rangle$ at the initial and the final stages come down more rapidly (i.e., they are more depressed), while at the middle stage, they come down relatively slowly (i.e., they are less depressed).
This kind of less-effective synaptic plasticity at the PF-PC synapses arises due to the decreased matching degree of Eq.~(\ref{eq:MD}) between $R_{\rm GR}(t)$ and $f_{\rm DS}(t)$ (leading to less depression at the middle stage) and the much-increased fraction of in-phase spiking group (resulting in more depression at the initial and the final stages). As a result, the modulation ${\cal M}_J$ makes a slow increase to its saturated value (= 0.023), which is markedly in contrast to the optimal case with a big modulation.

We next consider the effect of PF-PC synaptic plasticity with a reduced small modulation on the subsequent learning process in the PC-VN system.
Figures \ref{fig:HC2}(c1)-\ref{fig:HC2}(c5) show cycle-evolution of realization-averaged instantaneous population spike rate $\langle R_{\rm PC}(t) \rangle_r$
of the PCs (number of realizations: 100); $p_c=0.6$ (solid line) and $p_c^*=0.06$ (dotted line). As a result of PF-PC synaptic plasticity,
$\langle R_{\rm PC}(t) \rangle_r$ becomes lower at the middle stage than at the initial and the final stages. Thus, like the case of $\langle {\tilde J}(t) \rangle,$ $\langle R_{\rm PC}(t) \rangle_r$ forms a well-shaped curve, and it becomes saturated at about the 300th cycle. At the middle stage, $\langle R_{\rm PC}(t) \rangle_r$ (solid line) for $p_c=0.6$ is larger than that (dotted line) in the optimal case of $p_c^*=0.06$ due to less-depressed $\langle {\tilde J}(t) \rangle$. On the other hand, at the initial and the final stages $\langle R_{\rm PC}(t) \rangle_r$ (solid line) for $p_c=0.6$ is smaller than that (dotted line) in the optimal case of $p_c^*=0.06$ because of more-depressed $\langle {\tilde J}(t) \rangle$. As a result of such less-effective PF-PC synaptic plasticity, the modulation of $\langle R_{\rm PC}(t) \rangle_r$ for $p_c=0.6$ becomes so small, in contrast to the optimal case with a big modulation (which arises from the effective PF-PC synaptic plasticity).

Figures \ref{fig:HC2}(d1) and \ref{fig:HC2}(d2) show plots of cycle-averaged mean $\overline{ \langle R_{\rm PC}(t) \rangle_r }$ and modulation ${\cal M}_{\rm PC}$
for $\langle R_{\rm PC}(t) \rangle_r$ versus cycle, respectively; $p_c=0.6$ (solid circles) and $p_c^*=0.06$ (open circles).
Both $\overline{ \langle R_{\rm PC}(t) \rangle_r }$ and ${\cal M}_{\rm PC}$ become saturated at about the 300th cycle.
With increasing the cycle, the cycle-averaged mean $\overline{ \langle R_{\rm PC}(t) \rangle_r }$ decreases from 86.1 Hz to 51.8 Hz due to LTD at the PF-PC synapses,
which is similar to the optimal case where $\overline{ \langle R_{\rm PC}(t) \rangle_r }$ decreases from 86.1 Hz to 51.7 Hz.
On the other hand, the modulations ${\cal M}_{\rm PC}$ increases slowly from 2.6 Hz to 5.6 Hz, which is distinctly in contrast to
the optimal case where it increases rapidly from 2.6 Hz to 24.1 Hz. Such a small modulation in $\langle R_{\rm PC}(t) \rangle_r$ for $p_c=0.6$ arises due to the
less-effective depression at the PF-PC synapses. These principal PCs of the cerebellar cortex also exert less-effective inhibitory coordination on the VN neuron which evokes OKR eye-movement.

\begin{figure}
\includegraphics[width=0.95\columnwidth]{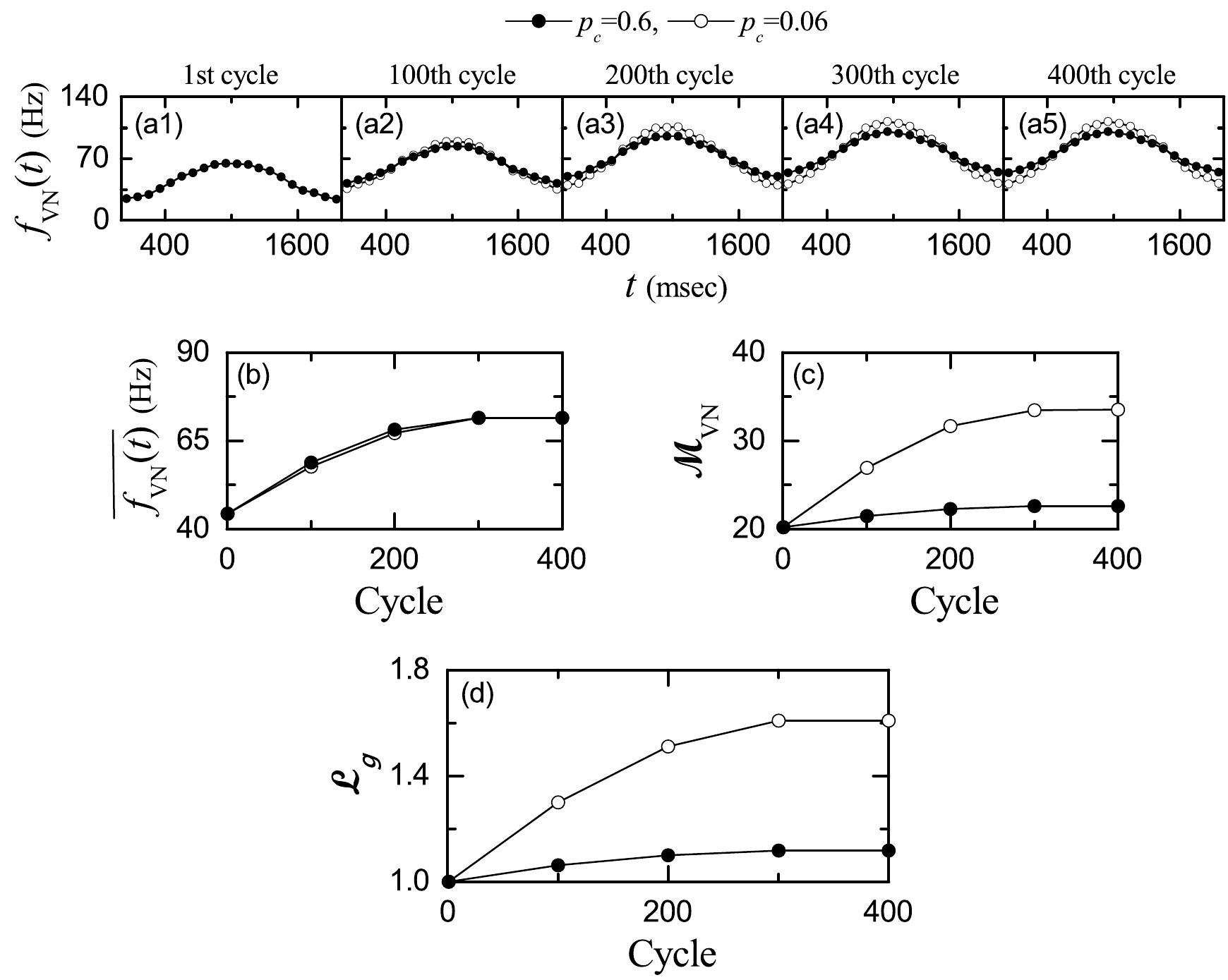}
\caption{Change in firing activity of the VN neuron during learning in the highly-connected case of $p_c = 0.6$; for comparison, data in the optimal case of $p_c^*=0.06$ are also given. (a1)-(a5) Cycle-evolution of bin-averaged instantaneous individual firing rates $f_{\rm VN}(t)$; the bin size is $\Delta t=$ 100 msec. Plots of (b) cycle-averaged mean $\overline{f_{\rm VN}(t)}$ and (c) modulation ${\cal M}_{\rm VN}$ for $f_{\rm VN}(t)$, and (d) learning gain degree ${\cal L}_g$ versus cycle.
In (a1)-(a5) and (b)-(d), solid (open) circles represent data for $p_c=$ 0.6 (0.06).
}
\label{fig:HC3}
\end{figure}

Change in firing activity of the VN neuron during learning in the highly-connected case of $p_c=0.6$ (solid circles) is shown in Fig.~\ref{fig:HC3};
for comparison, data in the optimal case of $p_c^*=0.06$ (open circles) are also given. Figures \ref{fig:HC3}(a1)-\ref{fig:HC3}(a5) show
cycle-evolution of bin-averaged instantaneous individual firing rate $f_{\rm VN}(t)$. It seems to be saturated at about the 300th cycle.
Due to the inhibitory coordination of PCs on the VN neuron, the maximum of $f_{\rm VN}(t)$ occurs at the middle stage, while the minima appear at the initial and the final stages. Thus, $f_{\rm VN}(t)$ forms a bell-shaped curve, in contrast to the well-shaped curves of $\langle R_{\rm PC}(t) \rangle_r$.
$f_{\rm VN}(t)$ (solid circles) at the middle stage is smaller than that (open circles) in the optimal case of $p_c^*=0.06$
due to more inhibition of the PCs on the VN neuron, while at the initial and the final stages, $f_{\rm VN}(t)$ (solid circles) is larger than
that (open circles) in the optimal case because of less inhibition of PCs on the VN neuron.
As a result of such less-effective inhibitory coordination of the PCs, the modulation of $f_{\rm VN}(t)$ becomes smaller than that in the optimal case.

\begin{figure}
\includegraphics[width=0.9\columnwidth]{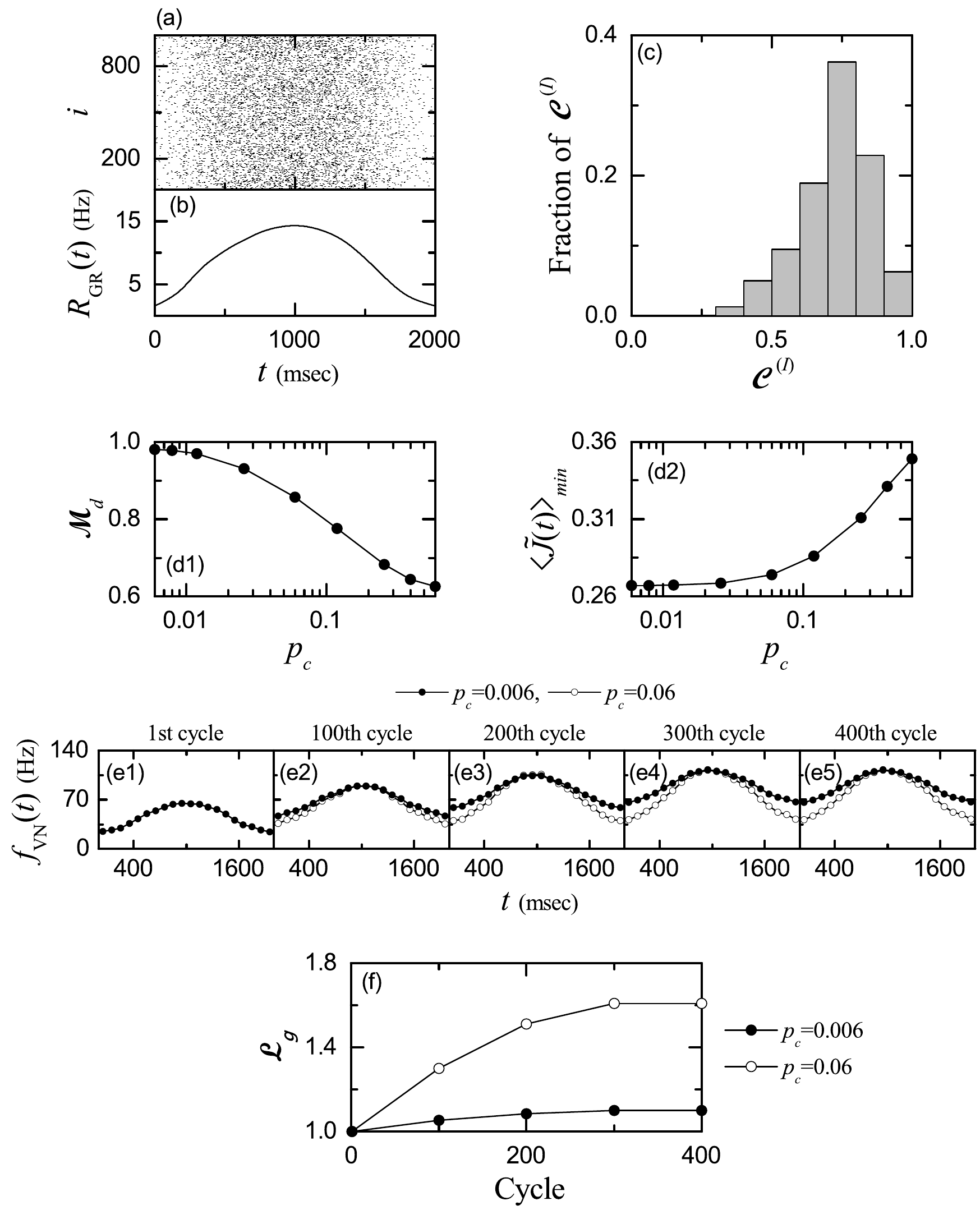}
\caption{Lowly-connected case of $p_c=0.006$.
(a) Raster plot of spikes of $10^3$ randomly chosen GR cells. (b) Instantaneous whole-population spike rate $R_{\rm GR}(t)$ in the whole population of GR cells.
Band width for $R_{\rm GR}(t)$: $h=10$ msec. (c) Distribution of conjunction indices $\{ C^{(I)} \}$ for the GR clusters in the whole population.
(d1) Plot of matching degree ${\cal M}_d$ versus $p_c$. (d2) Plot of the minimum $\langle {\tilde J}(t) \rangle_{min}$ versus $p_c$.
(e1)-(e5) Cycle-evolution of bin-averaged instantaneous individual firing rates $f_{\rm VN}(t)$; the bin size is $\Delta t=$ 100 msec.
(f) Plot of learning gain degree ${\cal L}_g$ (solid circles) versus cycle; for comparison, data (open circles) in the optimal case of $p_c^*=0.06$ are also given.
}
\label{fig:LC}
\end{figure}

Figures \ref{fig:HC3}(b) and \ref{fig:HC3}(c) show plots of cycle-averaged mean $\overline{ f_{\rm VN}(t) }$ and modulation ${\cal M}_{\rm VN}$ for $f_{\rm VN}(t)$.
Both $\overline{ f_{\rm VN}(t) }$ and ${\cal M}_{\rm VN}$ become saturated at about the 300th cycle.
$\overline{ f_{\rm VN}(t) }$ increases from 44.3 Hz to 71.4 Hz, which is nearly the same as that in the optimal case where it increases
from 44.3 Hz to 71.5 Hz. ${\cal M}_{\rm VN}$ also increases slowly from 20.2 Hz to 22.6 Hz, which is in contrast to the optimal case
where ${\cal M}_{\rm VN}$ increases quickly from 20.2 Hz to 32.5 Hz.
Then, the learning gain degree ${\cal L}_g$, given by the normalized gain ratio, is shown in Fig.~\ref{fig:HC3}(d).
${\cal L}_g$ increases from 1 and becomes saturated at ${\cal L}_g^* \simeq 1.118$ at about the 300th cycle which is smaller than
the saturated learning gain degree ${\cal L}_g^*~(\simeq 1.608)$ in the optimal case.
Consequently, due to the less-effective coordination of PCs, the saturated learning gain degree becomes smaller.

To sum up the highly-connected case of $p_c=0.6$, good balance between the in-phase and the out-of-phase spiking group in the optimal case
(${\cal R}^* \simeq 1.008$) is broken up, because the fraction of in-phase spiking group becomes dominant (${\cal R} \simeq 4.405$).
Thus, the diversity degree for the spiking patterns of the GR clusters is so much decreased (${\cal D} \simeq 0.204$), in comparison with
${\cal D}^*~(\simeq 1.613)$ in the optimal case. Due to increase in the fraction of the in-phase spiking group and decrease in the matching degree
between $R_{\rm GR}(t)$ and $f_{\rm DS}(t)$, synaptic plasticity at the PF-PC synapses becomes less effective, which also results in less-effective
motor learning for the OKR eye-movement.

We next consider the lowly-connected cases of $p_c=0.006$ (i.e. 0.6 $\%$).
Figures \ref{fig:LC}(a) and \ref{fig:LC}(b) show the raster plot of spikes of $10^3$ randomly-chosen GR cells and the
instantaneous whole-population spike rate $R_{\rm GR}(t)$ in the whole population of GR cells for $p_c = 0.006$, respectively.
In this lowly-connected case, the inhibitory inputs from GO cells into the GR clusters is so much reduced, and
the excitatory MF signals into the GR clusters become dominant inputs. Hence, the raster plot of spikes becomes more dense, and
$R_{\rm GR}(t)$ becomes more similar to the firing rate $f_{\rm MF}(t)$ of the Poisson spike train for the MF signal in Fig.~\ref{fig:OKR}(b1),
in contrast to the highly-connected case of $p_c=0.6$ with broadly flattened top part in $R_{\rm GR}(t)$ [see Fig.~\ref{fig:HC1}(b)].

Due to so much decrease in inhibitory inputs from the GO cells into the GR clusters, only the in-phase spiking group [where spiking patterns are similar to $f_{\rm MF}(t)$] appears (i.e., all the out-of-phase spiking group disappears), which is distinctly in contrast to the optimal case of $p_c^*=0.06$ where diverse spiking groups such as the in-phase, anti-phase, and complex out-of-phase spiking groups coexist.
The distribution of conjunction indices $\{ {\cal C}^{(I)} \}$ of the GR clusters (${\cal C}^{(I)}$: representing the similarity degree
between the spiking behavior [$R_{\rm GR}^{(I)}(t)$: instantaneous cluster spike rate] of the $I$th GR cluster and that [$R_{\rm GR}(t)$] in the whole population) is shown in Fig.~\ref{fig:LC}(c). In comparison to the optimal case in Fig.~\ref{fig:Char}(a), the distribution is shifted to the positive region, due to existence of only the in-phase GR clusters with positive values of ${\cal C}^{(I)}$. Thus, it has a peak at 0.75, and its range is $(0.32, 0.98)$.

In this lowly-connected case of $p_c=0.006$, the mean and the standard deviation for the distribution $\{ {\cal C}^{(I)} \}$ are 0.737 and 0.129, respectively, which is in contrast to the optimal case with the smaller mean (= 0.320) and the larger standard deviation (= 0.516). Then, the diversity degree ${\cal D}$ of the spiking patterns in all the GR clusters, given by the relative standard deviation for the distribution $\{ C^{(I)} \}$ [see Eq.~(\ref{eq:DD})], is ${\cal D} \simeq 0.175$  which is much smaller than ${\cal D}^*~(\simeq 1.613)$ in the optimal case. Consequently, the degree in diverse recoding in the granular layer is so much reduced in the lowly-connected case of $p_c=0.006$, due to existence of only the in-phase spiking group without the out-of-phase spiking group.

We also compare the lowly-connected case of $p_c=0.006$ with the highly-connected case of $p_c=0.6$ in Fig.~\ref{fig:HC1}.
The diversity degree ${\cal D}~(\simeq 0.204)$ for $p_c=0.6$ is also much reduced in comparison with ${\cal D}^*~(\simeq 1.613)$ in the optimal case.
However, it is a little larger than ${\cal D} \simeq 0.175$ for $p_c=0.006$, because for $p_c=0.6$ a minor out-of-phase spiking group with positive values of ${\cal C}^{(I)}$ exists, along with the major in-phase spiking group. We also  note that, the in-phase spiking patterns in both the lowly- and the highly-connected cases have completely different waveforms. For $p_c=0.006$ the in-phase spiking patterns are more similar to the sinusoidally-modulating MF signal $f_{\rm MF}(t)$, while those for $p_c=0.6$ are more different from $f_{\rm MF}(t)$ due to broad flatness in their top part. Thus, the matching degree ${\cal M}_d~(=0.981)$ of Eq.~(\ref{eq:MD}) for $p_c=0.006$ is larger than that ($=0.625$) for $p_c=0.6.$ In this way, there are two independent ways via increase or decrease in $p_c$ from the optimal value $p_c^*~(=0.06)$ to break up the good balance between the in-phase and the out-of-phase spiking groups, which results in decrease in the diversity degree in recoding of the GR cells.

As a result of reduced diversity in recoding of the GR cells, the modulation for the bin-averaged synaptic weights $\langle {\tilde J}(t) \rangle$
is much decreased, in contrast to the big modulation in the optimal case of $p_c^*=0.06$. Figure \ref{fig:LC}(d1) shows the plot of the matching degree ${\cal M}_d$ between $R_{\rm GR}(t)$ and $f_{\rm DS}(t)$ versus $p_c$. With decreasing $p_c$, ${\cal M}_d$ increases monotonically. Hence, ${\cal M}_d~(=0.981)$ for $p_c=0.006$ is larger than that (=0.857) in the optimal case of $p_c=0.06$. Due to increase in ${\cal M}_d$, at the middle stage in the lowly-connected case of $p_c=0.006$, more depression in $\langle {\tilde J}(t) \rangle$ is expected to occur, in comparison to the optimal case. Figure \ref{fig:LC}(d2) shows the plot of the minimum $\langle {\tilde J}(t) \rangle_{min}$ of the well-shaped curve for $\langle {\tilde J}(t) \rangle$ (appearing at $t=1000$ msec) versus $p_c$. We note that, as $p_c$ is decreased from $p_c^*~(=0.06)$, $\langle {\tilde J}(t) \rangle_{min}$ decreases so much slowly. Hence, $\langle {\tilde J}(t) \rangle_{min}~(=0.267)$ for $p_c=0.006$ becomes just a little smaller than that (=0.274) in the optimal case. Thus, for $p_c=0.006$ $\langle {\tilde J}(t) \rangle$ is only a little more depressed than that in the optimal case of $p_c=0.06$. At the initial and the final stages, only in-phase spiking group exists, unlike the optimal case where both the in-phase and the out-of-phase spiking groups coexist. Hence, much more depression in $\langle {\tilde J}(t) \rangle$ occurs in comparison to the optimal case.
Thus, modulation in $\langle {\tilde J}(t) \rangle$ for $p_c=0.006$ becomes much reduced, in comparison to the optimal case.

To make this point more clear, we divide the distribution of conjunction indices $\{ {\cal C}^{(I)} \}$ with a range (0.32, 0.98) in Fig.~\ref{fig:LC}(c) into the two parts by taking the mean (=0.737) as a reference. Thus, the 1st in-phase spiking sub-group has its conjunction indices higher than the mean (=0.737)
[i.e., the range is (0.737, 0.98)], while the 2nd in-phase spiking sub-group has its conjunction indices lower than than the mean [i.e., the range is (0.32, 0.737)]. In this way, the whole in-phase spiking group is decomposed into the two sub-groups.
Then, we obtain the firing fraction $F_i^{(G)}$ of active PF signals in the $G$ spiking group [see Eq.~(\ref{eq:FF})], where $G$ corresponds to the 1st or the 2nd in-phase spiking sub-group. Similar to Figs.~\ref{fig:SWFF}(d1)-\ref{fig:SWFF}(d5), the firing fraction $F^{(G)}(t)$ for the 1st (2nd) in-phase spiking sub-group
is found to form a bell-shaped (well-shaped) curve. At the initial and the final stages, the firing fractions $F^{(G)}(t)$ for the 1st and the 2nd in-phase spiking sub-groups are equal (i.e., $50 \%$). On the other hand, at the middle stage, the values of $F^{(G)}(t)$ for the 1st and the 2nd in-phase spiking sub-groups are 0.61 ($61 \%$) and 0.39 ($ 39 \%$), respectively. We note that more depression in $\langle {\tilde J}(t) \rangle$ occurs for the 1st in-phase spiking group because
their conjunction indices are larger than those in the 2nd in-phase spiking sub-group. Judging from the ratio of their firing fractions, at the middle stage
($0.61:0.39$),  $\langle {\tilde J}(t) \rangle$ is a little more depressed, in comparison to the initial and the final stages ($0.5:0.5$).
Thus, in the lowly-connected case of $p_c=0.006,$ a small modulation appears in $\langle {\tilde J}(t) \rangle$, in contrast to the big modulation in the optimal case.
The less-effective synaptic plasticity at the PF-PC synapses leads to reduced modulation in the realization-averaged instantaneous population spike rate $\langle R_{\rm PC}(t) \rangle_r$ of the principal PCs in the cerebellar cortex which also exert less-effective inhibitory coordination on the VN neuron which produces the final output of the cerebellum (i.e., OKR eye-movement). Figures \ref{fig:LC}(e1)-\ref{fig:LC}(e5) show cycle-evolution of the bin-averaged instantaneous firing rate $f_{\rm VN}(t)$ (solid circles) of the VN neuron for $p_c=0.006$; for comparison, data (open circles) in the optimal case of $p_c^*=0.06$ are also given. With increasing the cycle, $f_{\rm VN}(t)$ seems to be saturated at about the 300th cycle. $f_{\rm VN}(t)$ (solid circles) at the middle stage is a little larger than that (open circles) in the optimal case due to a little less inhibition of the PCs on the VN neuron for $p_c=0.006$.
On the other hand, at the initial and the final stages, $f_{\rm VN}(t)$ (solid circles) is much larger than that (open circles) in the optimal case because of
much less inhibition of PCs on the VN neuron in comparison to the optimal case.

In this way, as a result of less-effective inhibitory coordination of the PCs, the modulation of $f_{\rm VN}(t)$ becomes much smaller than that in the optimal case.
Plots of the learning gain degree ${\cal L}_g$ [corresponding to the modulation gain ratio for $f_{\rm VN}(t)$] are shown in Fig.~\ref{fig:LC}(f)
in both the lowly-connected case of $p_c=0.006$ (solid circles) and the optimal case of $p_c^*=0.06$ (open circles).
For $p_c=0.006$ ${\cal L}_g$ increases very slowly and becomes saturated at ${\cal L}_g^*~\simeq 1.099$ at about the 300th cycle.
This saturated learning gain degree ${\cal L}_g^*~(\simeq 1.099)$ is much smaller than that (${\cal L}_g^*~\simeq 1.608$) in the optimal case.
Consequently, low diversity in recoding of the GR cells for $p_c=0.006$ results in less-effective motor learning for the OKR adaptation.

Before proceeding to the general variation of $p_c$, we now summarize the main points obtained in the above highly-connected ($p_c=0.6$) and lowly-connected ($p_c=0.006$) cases. In both the highly- and the lowly-connected cases, their diversity degrees $\cal D$ [=0.204 ($P_c=0.6)$ and 0.175 ($p_c=0.006$)] for spiking patterns of the GR clusters are so much decreased in comparison to that (${\cal D} = 1.613$) in the optimal case of $p_c^*=0.06$, because the fraction of the in-phase spiking group is increased from 50.2$\%$ to 81.5$\%$ ($p_c=0.6$) and 100$\%$ ($p_c=0.006$).
At the initial and the final stages, the degree of depression in the bin-averaged (normalized) synaptic weight $\langle {\tilde J}(t) \rangle$ is determined by the firing fractions $F^{(G)}$ of Eq.~(\ref{eq:FF}) of the in-phase and the out-of-phase spiking groups. For $p_c=0.6$ and 0.006, $F^{(i)}$ of the in-phase spiking group is 0.82 (82$\%$) and 1.0 (100$\%$), respectively, in comparison to 0.51 (51$\%$) in the optimal case of $p_c^*=0.06$. Hence, for $p_c=0.6$ and 0.006, due to increased firing fraction of the in-phase spiking group (with strong LTD), more depression in $\langle {\tilde J}(t) \rangle$ occurs at the initial and the final stages of cycle, which is in contrast to the optimal case (e.g., see Fig.~\ref{fig:HC2}).

In the middle stage, the in-phase spiking group make a dominant contribution, independently of $p_c$; $F^{(i)}=$ 0.94 (94$\%$ for $p_c^*=0.06$),
0.98 (98$\%$ for $p_c=0.6$), and 1.0 (100$\%$ for $p_c=0.006$). However, the effect of the dominant in-phase spiking group on depression in $\langle {\tilde J}(t) \rangle$ varies depending on $p_c$. The degree of depression in $\langle {\tilde J}(t) \rangle$ is determined by the matching degree ${\cal M}_d$ of Eq.~(\ref{eq:MD}) between the instantaneous whole-population spike rate $R_{\rm GR}(t)$ and the IO desired signal $f_{\rm DS}(t)$; $R_{\rm GR}(t)$ plays a ``reference'' signal for the in-phase ``student'' PF signals and the error-teaching instructor CF signals are in-phase with $f_{\rm DS}(t)$.
Depending on $p_c$, the reference signal $R_{\rm GR}(t)$ has a changing matching degree ${\cal M}_d$ with respect to $f_{\rm DS}(t)$.
With decreasing $p_c$, ${\cal M}_d$ is monotonically increased [see Fig.~\ref{fig:LC}(d1)] [i.e, $R_{\rm GR}(t)$ becomes more similar to the sinusoidal IO desired signal $f_{\rm DS}(t)$].

In the highly-connected case of $p_c=0.6,$ ${\cal M}_d=0.625$, which is smaller in comparison with ${\cal M}_d=0.857$ in the optimal case of $p_c^*=0.06$.
Due to decrease in ${\cal M}_d$ for $p_=0.6$, in the middle stage (where the in-phase spiking group is dominant), less depression in $\langle {\tilde J}(t) \rangle$ occurs (see Fig.~\ref{fig:HC2}). Consequently, for $p_c=0.6$ the modulation in $\langle {\tilde J}(t) \rangle$ decreases, in comparison to that in the optimal case, due to more depression at the initial and the final stages and less depression at the middle stage.

In the lowly-connected-connected case of $p_c=0.006$, ${\cal M}_d=0.981$, which is larger than that (${\cal M}_d=0.857$) in the optimal case of $p_c^*=0.06$.
However, as shown in Fig.~\ref{fig:LC}(d2), the minimum $\langle {\tilde J}(t) \rangle_{min}$ of the well-shaped curve for $\langle {\tilde J}(t) \rangle$
decreases so much slowly with decreasing $p_c$. Thus, for $p_c=0.006$ only a little more depression in $\langle {\tilde J}(t) \rangle$ in comparison with the optimal case. As explained in the above, at the initial and the final stages, only the in-phase spiking group (with strong LTD) exists, in contrast to the optimal case where both the out-of-phase (with weak LTD) and the in-phase spiking groups. Hence, much more depression takes place in comparison to the optimal case.
As a result, for $p_c=0.006$ the modulation in $\langle {\tilde J}(t) \rangle$ is decreased, in comparison with that in the optimal case, because of much more depression at the initial and the final stages and a little more depression at the middle stage.

Such decrease in modulation in $\langle {\tilde J}(t) \rangle$ for $p_c=0.6$ and 0.006 causes a small modulation in firing activity
$R_{\rm PC}(t)$ of the PCs, which then exerts less-effective inhibitory coordination on the VN neuron (which evokes the OKR).
Consequently, a small modulation in the firing activity $f_{\rm VN}(t)$ of the VN neuron arises [see Fig.~\ref{fig:HC3} for $p_c=0.6$ and Fig.~\ref{fig:LC}
for $p_c=0.006$]. Consequently, for $p_c=0.6$ and 0.006 where the diversity degree in recoding of GR cells is decreased,
less-effective motor learning for the OKR adaptation occurs, in comparison to the optimal case of $p_c=0.06$.

\begin{figure}
\includegraphics[width=0.9\columnwidth]{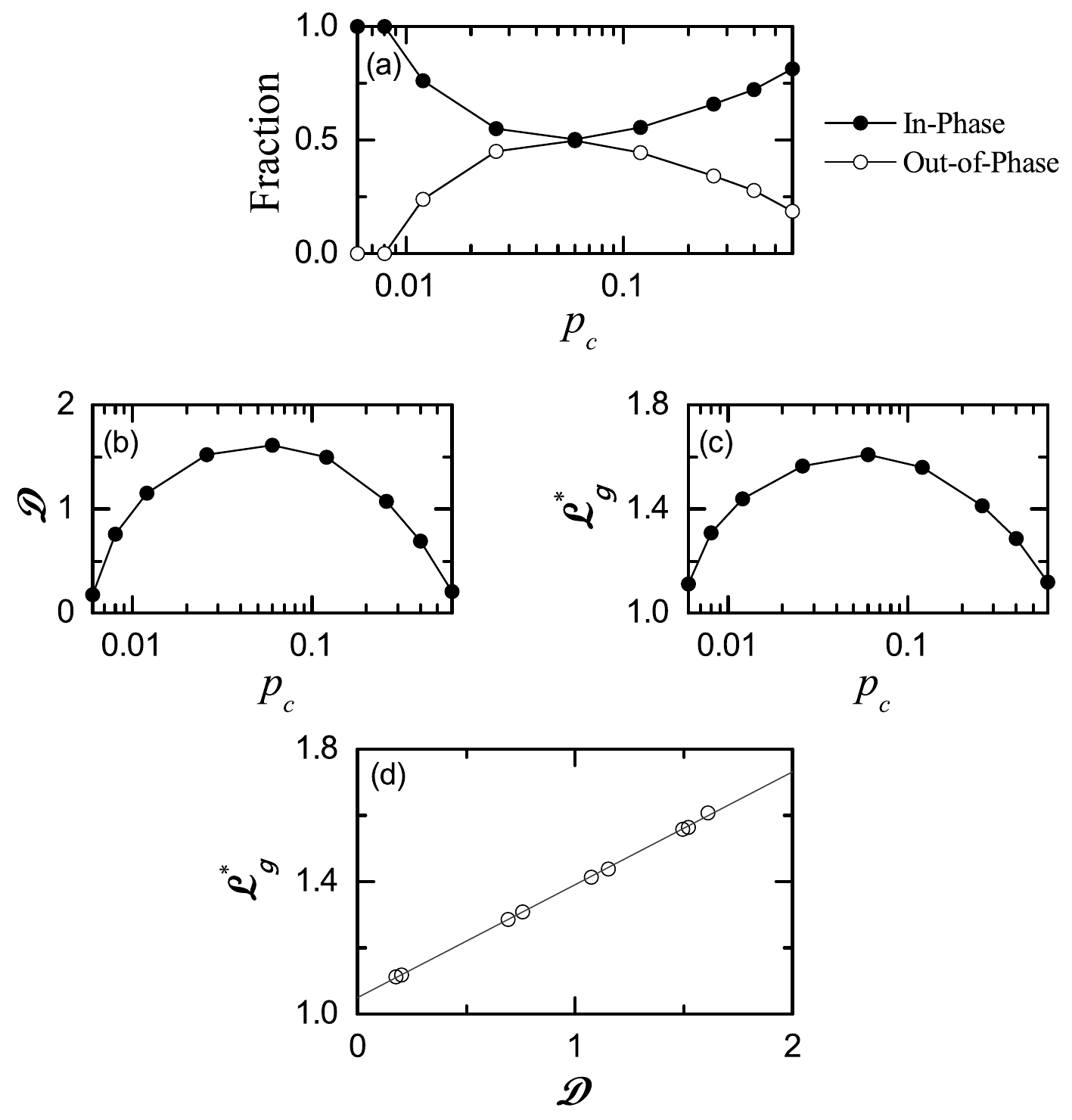}
\caption{Strong correlation between the diversity degree $\cal {D}$ and the saturated learning gain degree ${\cal L}_g^*$. (a) Fraction of in-phase (solid circles) and out-of-phase (open circles) spiking groups. (b) Plot of diversity degree $\cal {D}$ for the spiking patterns of all the GR clusters versus $p_c$.  (c) Saturated learning gain degree ${\cal L}_g^*$ versus $p_c$. (d) Plot of ${\cal L}_g^*$ versus $\cal {D}$.
}
\label{fig:Final}
\end{figure}

Finally, based on the above two examples for the highly- and the lowly-connected cases, we investigate dependence of the diversity degree $\cal D$ for the spiking patterns of the GR clusters and the saturated learning gain degree ${\cal L}_g^*$ on $p_c$ by varying it from the optimal value ($p_c^*=0.06$).
Figure \ref{fig:Final}(a) shows plots of fractions of the in-phase and the out-of-phase spiking groups versus $p_c$.
The fraction of the in-phase spiking group (solid circles) forms a well-shaped curve with a minimum at $p_c = p_c^*~(=0.06)$, while
the fraction of the out-of-phase spiking group (open circles) forms a bell-shaped curve with a maximum at the optimal value of $p_c^*=0.06$.
For sufficiently small $p_c$, we have two sample cases where the fraction of in-phase spiking group is 1 (i.e., the fraction of out-of-phase spiking group is 0).
We note that, in the optimal case of $p_c^*=0.06$, fractions of the in-phase (50.2$\%$) and the out-of-phase spiking (49.8$\%$) groups are well balanced
(i.e., good balance between the in-phase and the out-of-phase spiking groups).  As $p_c$ is changed (i.e., increased or decreased) from $p_c^*$, the fraction of the in-phase spiking group increases, and then the spiking-group ratio $\cal R$ (i.e., the ratio of the fraction of the in-phase spiking group to that of the
out-of-phase spiking group) increases from the golden spiking-group ratio ${\cal R}^*~(\simeq 1.008)$ in the optimal case.

Figures \ref{fig:Final}(b) and \ref{fig:Final}(c) show plots of the diversity degree $\cal D$ for the spiking patterns of the GR clusters and the saturated
learning gain degree ${\cal L}_g^*$ versus $p_c$, respectively. The diversity degrees $\cal D$ forms a bell-shaped curve with a maximum
${\cal D}^*~(\simeq 1.613)$ in the optimal case of $p_c^*=0.06$ with the golden spiking-group ratio ${\cal R}^*\simeq 1.008$ (i.e., good balance between the in- and the out-of-phase spiking group). We note that, in this optimal case where the recoding of the GR cells is the most diverse, the saturated learning gain degree ${\cal L}^*$ also has its maximum (${\cal L}_g^*~\simeq 1.608$).
As $p_c$ is changed (i.e., increased or decreased) from the optimal value (= 0.06), the spiking-group ratio $\cal R$ is increased, because of increase in
the fraction of in-phase spiking group. Then, the diversity degree $\cal D$ in recodings of the GR cells becomes decreased, which also results
in decrease in the saturated learning gain degree ${\cal L}_g^*$ from the maximum. Thus, ${\cal L}_g^*$ also forms a bell-shaped curve, as in the case of $\cal D$, and they have their maxima at the optimal values ($p_c^*=0.06$).

Figure \ref{fig:Final}(d) shows a plot of ${\cal L}_g^*$ versus $\cal D$. As shown clearly in Fig.~\ref{fig:Final}(d), both
${\cal L}_g^*$ and $\cal D$ have a strong correlation with the Pearson's correlation coefficient $r\simeq 0.9998$.
Consequently, the more diverse in recoding of the GR cells, the more effective in motor learning for the OKR adaptation.

\section{Summary and Discussion}
\label{sec:SUM}
We are interested in the gain adaptation of OKR for the eye movement. Various experimental works on the OKR have been done in rabbits, mice, and zebrafishes  \cite{OKRExp1,OKRExp2,OKRExp3,OKRExp4,OKRExp5,OKRExp6,OKRExp7,OKRExp8}. Moreover, some features of the OKR adaptation were successfully reproduced through
computational simulations in the adaptive filter model \cite{OKRCom1} and the spiking network model \cite{Yama1}. However, effects of diverse recoding of GR cells
on the OKR adaptation in previous computational works are necessary to be more clarified in several dynamical aspects. Particularly, the previous works lacked complete dynamical classification of diverse spiking patterns of GR cells and their association with the error-teaching CF signals. We note that such dynamical classification of diverse recoding of GR cells may be a basis for clear understanding of the synaptic plasticity at the PF-PC synapses and the subsequent learning progress in the PC-VN-IO system.

For the effective study of OKR, we first introduced a cerebellar ring network. This ring network with a simple architecture has advantage for computational and analytical efficiency, and its visual representation may also been easily made. Moreover, we also employed a refined rule for the synaptic plasticity, based on the experimental result in \cite{Safo}. We note that this one-dimensional cerebellar ring network with a refined synaptic plasticity is in contrast to the two-dimensional square-lattice network \cite{Yama1}.

For the first time, we made complete quantitative classification of diverse spiking patterns in the GR clusters via introduction of the conjunction index $\{ {\cal C}^{(I)} \}$ and the diversity degree $\cal D$. Each spiking pattern in the $I$th GR cluster may be characterized in terms of its conjunction index ${\cal C}^{(I)},$ denoting the degree for the association of the spiking behavior of each $I$th GR cluster [characterized by the instantaneous cluster spike rate $R_{\rm GR}^{(I)}(t)$] with the population-averaged firing activity in the whole population [given by the instantaneous whole-population spike rate $R_{\rm GR}(t)$].
Then, the whole spiking patterns are decomposed into the two types of in-phase and out-of-phase spiking groups which are in-phase and out-of-phase with respect to the ``reference'' signal $R_{\rm GR}(t)$, respectively. Furthermore, the degree of diverse recoding of the GR cells may be quantified in terms of the diversity degree $\cal D$, given by the relative standard deviation in the distribution of $\{ {\cal C}^{(I)} \}$. Thus, $\cal D$ gives a quantitative measure for diverse recoding of GR cells. Dynamical origin for the appearance of diverse spiking patterns (i.e., outputs of the GR clusters) has also been investigated. It has thus been found that diverse total synaptic inputs (including both the excitatory inputs via MFs and the inhibitory inputs from the pre-synaptic GO cells) into the GR clusters
lead to production of diverse spiking patterns in the GR clusters.

Based on the above dynamical classification of diverse spiking patterns in the GR clusters, we made intensive investigations on the effect of diverse recoding
of GR cells on the OKR adaptation (i.e., its effect on the synaptic plasticity at the PF-PC synapses and the subsequent learning process in the PC-VN-IO system).
To the best of our knowledge, this type of approach, based on the in-phase and the out-of-phase spiking groups, is unique for the study of OKR.
Diversely-recoded student PF signals from GR cells and the instructor error-teaching CF signals from the IO neuron are fed into the PCs.
We note that the CF signals are always in-phase with respect to the reference signal $R_{\rm GR}(t)$.
During the whole learning process, these in-phase and out-of-phase spiking groups have been found to play their own roles, respectively.

The in-phase student PF signals have been found to be strongly depressed (i.e., strong LTD) by the instructor CF signals, because they are well-matched with the
in-phase CF signals. On the other hand, the out-of-phase student PF signals have been found to be weakly depressed (i.e., weak LTD) by the instructor CF signals, because they are ill-matched with the in-phase CF signals. However, contributions of these in-phase and out-of-phase spiking groups vary depending on the stage of cycle. In the middle stage of each cycle, strong LTD takes place through dominant contributions of the in-phase spiking group, which results in emergence of a minimum of the bin-averaged (normalized) synaptic weight $\langle {\tilde J} \rangle$ of active PF signals. In contrast, at the initial and the final stages of each cycle, less LTD occurs (i.e., maxima of $\langle {\tilde J} \rangle$ appear) because both the out-of-phase spiking group with weak LTD and the in-phase spiking group with strong LTD make contributions together.
Hence, the bin-averaged synaptic weights $\langle {\tilde J} \rangle$ have been found to form a well-shaped curve (i.e., appearance of a minimum in the middle stage and maxima at both the initial and final stages). In this way, a big modulation in $\langle {\tilde J} \rangle$ emerges through interplay of the in-phase and the out-of-phase spiking groups.

Due to this kind of effective depression (i.e., strong/weak LTD) at the PF-PC synapses, the (realization-averaged) instantaneous population spike rate $\langle R_{\rm PC}(t) \rangle_r$ of PCs (corresponding to the principal outputs of the cerebellar cortex) has been found to form a well-shaped curve with a minimum in the middle stage. Consequently, a big modulation occurs in $\langle R_{\rm PC}(t) \rangle_r$. These PCs exert effective inhibitory coordination on the VN neuron (which evokes OKR eye-movement). Thus, the (realization-averaged) instantaneous individual firing rate $f_{\rm VN}(t)$ of the VN neuron has been found to form a bell-shaped curve with a maximum in the middle stage. The maximum in the middle stage of cycle is formed due to dominant contribution of the in-phase spiking group with strong LTD, while appearance of minima at the initial and the final stages is made via comparable contributions of the in-phase and the out-of-phase spiking groups.
In this case, the learning gain degree ${\cal{L}}_g$, corresponding to the modulation gain ratio (i.e., normalized modulation divided by that at the 1st cycle for $f_{\rm VN}$), has been found to increase with learning cycle and to be saturated at about the 300th cycle.

By varying $p_c,$ we investigated dependence of the diversity degree $\cal D$ of the spiking patterns and the saturated learning gain degree ${\cal L}_g^*$  on $p_c.$  Both $\cal D$ and ${\cal L}_g^*$ have been found to form bell-shaped curves with peaks (${\cal{D}}^* \simeq 1.613$ and ${{\cal{L}}_g}^* \simeq 1.608$)
at the same optimal value of $p_c^*~=0.06$. Also, in this optimal case, each GR cell receives inhibitory inputs from about 10 nearby GO cells.
In the references \cite{Yama2,Yama1} where the parameter values were taken from physiological data, the average number of nearby GO cell axons innervating each GR cell is about 8, which is close to that in the optimal case. Hence, we hypothesize that the granular layer in the cerebellar cortex has evolved toward the goal
of the most diverse recoding. Moreover, Both $\cal D$ and ${\cal L}_g^*$ have also been found to have a strong correlation with the Pearson's correlation coefficient $r \simeq 0.9998$. Consequently, the more diverse in the spiking patterns of GR cells, the more effective in motor learning for the OKR adaptation, which is the main result in our work.

For examination of our main result, we propose a real experiment for the OKR. To control $p_c$ in a given species of animals (e.g., a species of
rabbit, mouse, or zebrafish) in an experiment seems to be practically difficult, unlike the case of computational neuroscience where $p_c$ can be easily changed.
Instead, we consider an experiment for several species of animals (e.g., 3 species of rabbit, mouse, and zebrafish).
In each species, we consider a large number of randomly chosen GR cells ($i=1, \cdots, L$).
Then, through many learning cycles, one can obtain the peristimulus time histogram (PSTH) for each $i$th GR cell [i.e., (bin-averaged) instantaneous individual firing rate $f_{\rm GR}^{(i)}(t)$ of the $i$th GR cell]. GR cells are expected to exhibit diverse PSTHs.
Then, in the case of each $i$th GR cell, we obtain its conjunction index ${\cal C}_i$ between its PSTH $f_{\rm GR}^{(i)}(t)$ and the CF signal from the IO neuron
[i.e., the PSTH of the IO neuron $f_{\rm IO}(t)$]. In this case, the conjunction index ${\cal C}_i$ is given by the cross-correlation at the zero-time lag between $f_{\rm GR}^{(i)}(t)$ and $f_{\rm IO}(t)$. Thus, we get the diversity degree ${\cal D}$ of PSTHs of GR cells, given by the relative standard deviation in the distribution of $\{ {\cal C}_i \}$, for the species.

Besides the PSTHs of GR cells, under the many learning cycles, we can also get a bell-shaped PSTH of a VN neuron [i.e., (bin-averaged) instantaneous individual firing rate $f_{\rm VN}(t)$ of the VN neuron]. The normalized modulation of $f_{\rm VN}(t)$ (divided by that at the 1st cycle) corresponds to
the learning gain degree ${\cal L}_g$. Thus, a set of $({\cal D}, {\cal L}_g)$ can be experimentally obtained for each species, and the set of $({\cal D}, {\cal L}_g)$ may vary depending on the species. Then, for example in the case of 3 different species of rabbit, mouse, and zebrafish, with the three different data sets for $({\cal D}, {\cal L}_g)$, one can examine our main result (i.e., whether more diversity in PSTHs of GR cells results in more effective motor learning for the OKR).

We also discuss our results briefly in comparison with other previous works \cite{Gao,Cayco,Inagaki}.
It was discussed in \cite{Gao} that various forms of synaptic plasticity occur at different sites in the cerebellum, and they work synergistically to create optimal outputs for behavior. In the case of synaptic plasticity at the PF-PC synapse, LTD occurs when firing of the PF signal is in-phase (in conjunction) with that of
the CF signal. On the other hand, in the absence of the CF signal, the PF signal becomes out-of-phase, and LTP takes place. This mechanism for the synaptic plasticity at the PF-PC synapse is essentially similar to that in our work. However, in our case, we employed a refined rule for the synaptic plasticity
with a time window for the LTD (see Fig.~\ref{fig:TW}), based on the experimental result in \cite{Safo}, which has more quantitative advantages. In the presence
of a CF firing, a major LTD ($\Delta {\rm LTD}^{(1)}$) occurs in conjunction with earlier PF firings, while a minor LTD ($\Delta {\rm LTD}^{(2)}$) takes place in association with later PF firings. Outside the effective range of LTD, PF firings alone result in LTP. The GR cells receive both the excitatory synaptic inputs via the MFs and the inhibitory synaptic inputs from the GO cells. It was shown in \cite{Cayco} that sparse excitatory synaptic connectivity via $N_{syn}~(=3 \sim 5)$ MFs is crucial for pattern separation of the MF inputs. In contrast to the work in \cite{Cayco}, we controlled the inhibitory synaptic inputs into the GR cells by changing the connection probability $p_c$ from the GO cells to the GR cells, and investigated the effect of diverse recoding of GR cells on the motor learning. It was thus found that the saturated learning gain degree ${\cal{L}}_g^*$ for the OKR is maximum for an optimal value of $p_c~(=0.06)$.

In \cite{Inagaki}, Inagaki and Hirata studied the motor learning for the vestibulo-ocular reflex in the case that each GR cell receives 6 excitatory MF inputs and 3 inhibitory GO-cell inputs. There are 3 types of vestibular, efference-copy, and retinal-slip input signals from the MF. Thus, 9 diverse types of GR cells were prepared depending on the fractions of the vestibular, the efference-copy, and the retinal-slip signals (conveyed via MF inputs).
In this case, motor learning for the vestibular ocular reflex was induced by a combination of LTD and LTP at the PF-PC synapses. Moreover, different
types of signal processing were required for high- and low-gain motor learning. In contrast to this work, we considered a single type of sinusoidal MF input.
Then, diverse spiking patterns of GR cells were shown to arise due to inhibitory coordination of the GO cells. Based on the in-phase and the out-of-phase spiking groups, we investigated the effect of diverse recoding of the GR cells on the motor learning for the OKR.

Finally, we discuss limitations of our present work and future works.
In the present work, we did not take into consideration intra-population synaptic connections, and hence no motor rhythms appear
in the presence of just inter-population synaptic connections. When we add intra-population couplings between inhibitory GO cells,
a granular motor rhythm of 7 {\small $\sim$} 25 Hz may appear in the granular layer (i.e., GR-GO feedback system) \cite{Dangelo}.
Ultrafast motor rhythm of {\small $\sim$} 200 Hz may also appear in the Purkinje (molecular) layer by adding intra-population synaptic connections between
PCs (BCs) \cite{Solages}. In the system of IO neurons, {\small $\sim$} 10 Hz motor rhythm appears in the presence of electric gap junctions between IO neurons \cite{Llinas}. Hence, in a future work, it would be interesting to investigate the effect of motor rhythms on diverse recoding of GR cells and learning process
in the PC-VN-IO system by adding intra-population couplings. Beyond the synaptic plasticity at PF-PC synapses (considered in this work), diverse synaptic plasticity occurs at other synapses in the cerebellum \cite{Gao,Hansel} such as synaptic plasticity at PF-BC and BC-PC synapses \cite{Lennon}, at MF-cerebellar nucleus and PC-cerebellar nucleus synapse \cite{Zheng}, and at MF-GR cells synapses \cite{Dangelo2}. Therefore, as a future work, it would be interesting to study the effect of diverse synaptic plasticity at other synapses on cerebellar motor learning. In addition to variation in $p_c$, another possibility to change synaptic inputs into the GR cells is to vary NMDA receptor-mediated maximum conductances $\bar{g}_{\rm NMDA}^{\rm (GR)}$ and $\bar{g}_{\rm NMDA}^{\rm (GO)}$, associated with persistent long-lasting firing activities. It would also be interesting to investigate the effect of NMDA receptor-mediated synaptic inputs on diverse recoding of GR cells and motor learning in the OKR adaptation by changing $\bar{g}_{\rm NMDA}^{\rm (GR)}$ and $\bar{g}_{\rm NMDA}^{\rm (GO)}$.  This work is beyond the scope of the present work, and hence it is left as a future work.

We also discuss another interesting future work. In the present work, we varied the connection probability $p_c$ (from the GO cells to the GR cells) while the synaptic weight $J^{\rm (GR,GO)}~(=10.0)$ per synapse is unchanged. Hence, with increasing (decreasing) $p_c,$ the total inhibitory synaptic strength $K^{\rm (GR,GO)}$ on each GR cell also increases (decreases). Instead, one may consider another case where $p_c$ is changed while maintaining a constant $K^{\rm (GR,GO)}$.
For keeping the value of $K^{\rm (GR,GO)}$  to be a constant, the synaptic weight $J^{\rm (GR,GO)}$ must also change depending on the variation of $p_c$;
when $p_c$ is increased (decreased), $J^{\rm (GR,GO)}$ should decrease (increase) such that $K^{\rm (GR,GO)}$ is a constant.
We made a preliminary work on this issue by fixing the constant value of $K^{\rm (GR,GO)}$ at ${K^{\rm (GR,GO)}}^*$ in the optimal case of $p_c^*=0.06$.
Particularly, we were interested in how the spiking patterns in the GR clusters, characterized by $R_{\rm GR}^{(I)}(t)$, and the population-averaged
firing activity, given by the instantaneous whole-population spike rate $R_{\rm GR}(t)$, change when varying $p_c$.
We first considered the highly-connected case of $p_c=0.6$ (with $K^{\rm (GR,GO)}= {K^{\rm (GR,GO)}}^*$).
In comparison with the highly-connected case in Fig.~\ref{fig:HC1} where $K^{\rm (GR,GO)} > {K^{\rm (GR,GO)}}^*$,
shapes of $R_{\rm GR}(t)$ and $R_{\rm GR}^{(I)}(t)$ in the case of $K^{\rm (GR,GO)} = {K^{\rm (GR,GO)}}^*$ were nearly unchanged,
while their amplitudes were increased because $K^{\rm (GR,GO)}$ was decreased to ${K^{\rm (GR,GO)}}^*$.
We note that the conjunction index ${\cal C}^{(I)}$ between $R_{\rm GR}^{(I)}(t)$ and $R_{\rm GR}(t)$ represents the degree for similarity between their shapes.
Hence, the distribution of $\{ {\cal C}^{(I)} \}$ in the case of $K^{\rm (GR,GO)} = {K^{\rm (GR,GO)}}^*$ was found to be nearly the same as that in Fig.~\ref{fig:HC1}(c).
We also get the diversity degree ${\cal D}~(\simeq 0.195)$ for the spiking patterns, which is close to that (${\cal D}~\simeq 0.204)$ in Fig.~\ref{fig:HC1}(c).
In this highly-connected case, decrease in the synaptic weight $J$ for maintaining the value of $K^{\rm (GR,GO)}$ seems to have effect mainly on
the amplitudes of spiking patterns without much alteration of their shapes, and hence the diversity for the spiking patterns seems to be determined
mainly by the high connectivity from the GO to the GR cells. Consequently, in both cases with and without maintaining the constant ${K^{\rm (GR,GO)}}^*$,
the diversity degree $\cal D$ for the spiking patterns seems to be nearly the same.

Next, we considered the lowly-connected case of $p_c=0.006$ where $K^{\rm (GR,GO)}= {K^{\rm (GR,GO)}}^*$. In comparison to the lowly-connected case in Fig.~\ref{fig:LC} with $K^{\rm (GR,GO)} < {K^{\rm (GR,GO)}}^*,$ both the shapes and the amplitudes of $R_{\rm GR}^{(I)}(t)$ and $R_{\rm GR}(t)$ were changed.
Because $K^{\rm (GR,GO)}$ was increased to ${K^{\rm (GR,GO)}}^*$, amplitudes of $R_{\rm GR}^{(I)}(t)$ and $R_{\rm GR}(t)$ were lowered  and
their top shapes became flattened. Thus, the spiking patterns in the case of $K^{\rm (GR,GO)}= {K^{\rm (GR,GO)}}^*$ tended to be somewhat similar to those in the optimal case of $p_c^*=0.06$. We note that, in this lowly-connected case, both the low-connectivity $p_c$ and the increase in synaptic weight $J$
seem to determine the diversity for the spiking patterns, unlike the highly-connected case.
Mainly due to increased synaptic weight $J$, conjunction indices for the spiking patterns were broadly distributed in a range of (-0.17,0.75), in contrast to that in
Fig.~\ref{fig:LC}(c). Out-of-phase spiking patterns with negative conjunction indices also appear, unlike the case in Fig.~\ref{fig:LC} where
only in-phase spiking patterns exist. In this case, we obtain the diversity degree is ${\cal D}~(\simeq 1.192)$, which is much larger than that
(${\cal D} \simeq 0.175)$ in the case of Fig.~\ref{fig:LC}. However, it is still smaller than that (${\cal D}^* \simeq 1.613)$ in the optimal case.
To get the effect of low connectivity, we start from the optimal case of $p_c^*=0.06$ in Fig.~\ref{fig:Char}, and decrease $p_c$ from 0.06 to 0.006
(lowly-connected case) while maintaining the constant ${K^{\rm (GR,GO)}}^*$. The range in the distribution of conjunction indices in the optimal case is (-0.57,0.85). We note that this range became narrowed in the case of lowly-connected case of $p_c=0.006$. Through decrease from $p_c^*$ to 0.006, in-phase and anti-phase spiking patterns with higher magnitudes of conjunction indices seems to be ``degraded,'' and accordingly their magnitude of conjunction indices
seem to be lowered. We conjecture that the connection probability of $p_c=0.006$ would be too small to warrant the high-degree conjunction for the in-phase and the anti-phase spiking patterns. Base on the results of this preliminary work, we hypothesize a possibility that the diversity degree $\cal D$ might have its maximum
at the same optimal value of $p_c^*=0.06$ even in the case of maintaining the constant ${K^{\rm (GR,GO)}}^*$; in the lowly-connected case of $p_c < p_c^*,$ $\cal D$ would decrease in a much slow way in comparison with the case without maintaining the constant ${K^{\rm (GR,GO)}}^*$. To confirm this hypothesis, more intensive future work is necessary by decreasing $p_c$ from $p_c^*$ in the lowly-connected case with the constant ${K^{\rm (GR,GO)}}^*$ where changes in both $p_c$ and $J$ have effects on spiking patterns.

\begin{table}
\caption{Parameter values for LIF neuron models with AHP currents for the granule (GR) cell and the Golgi (GO) cell in the granular layer, the Purkinje cell (PC) and the basket cell (BC) in the Purkinje-molecular layer, and the vestibular nucleus (VN) and the inferior olive (IO) neurons.
}
\label{tab:SingleParm}
\begin{tabular}{|c|c|c|c|c|c|c|c|}
\hline
\multicolumn{2}{|c|}{} & \multicolumn{2}{c|}{\multirow{2}{*}{Granular}} & \multicolumn{2}{c|}{Purkinje} &  &  \\
\multicolumn{2}{|c|}{\multirow{3}{*}{$X$-population}} & \multicolumn{2}{c|}{\multirow{2}{*}{Layer}} & \multicolumn{2}{c|}{-Molecular} & \multirow{2}{*}{VN} &  \multirow{2}{*}{IO} \\
\multicolumn{2}{|c|}{} & \multicolumn{2}{c|}{} & \multicolumn{2}{c|}{Layer} & \multirow{2}{*}{neuron} &  \multirow{2}{*}{neuron} \\ \cline{3-6}
\multicolumn{2}{|c|}{} & GR & GO & \multirow{2}{*}{PC} & \multirow{2}{*}{BC} & & \\
\multicolumn{2}{|c|}{} & cell & cell & & & & \\
\hline
\multicolumn{2}{|c|}{$C_X$} & 3.1 & 28.0 & 107.0 & 107.0 & 122.3 & 10.0 \\ \hline
\multirow{2}{*}{$I_L^{(X)}$} & $g_L^{(X)}$ & 0.43 & 2.3 & 2.32 & 2.32 & 1.63 & 0.67 \\ \cline{2-8}
& $V_L^{(X)}$ & -58.0 & -55.0 & -68.0 & -68.0 & -56.0 & -60.0 \\ \cline{2-8}
\hline
\multirow{4}{*}{$I_{AHP}^{(X)}$} & $\bar{g}_{AHP}^{(X)}$ & 1.0 & 20.0 & 100.0 & 100.0 & 50.0 & 1.0 \\ \cline{2-8}
& $\tau_{AHP}^{(X)}$ & 5.0 & 5.0 & 5.0 & 2.5 & 2.5 & 10.0 \\ \cline{2-8}
& $V_{AHP}^{(X)}$ & -82.0 & -72.7 & -70.0 & -70.0 & -70.0 & -75.0 \\ \cline{2-8}
& $v_{th}^{(X)}$ & -35.0 & -52.0 & -55.0 & -55.0 & -38.8 & -50.0 \\ \cline{2-8}
\hline
\multicolumn{2}{|c|}{$I_{ext}^{(X)}$} & 0.0 & 0.0 & 250.0 & 0.0 & 700.0 & 0.0 \\
\hline
\end{tabular}
\end{table}

\section*{Acknowledgments}
This research was supported by the Basic Science Research Program through the National Research Foundation of Korea (NRF) funded by the Ministry of Education (Grant No. 20162007688).

\begin{table*}
\caption{Parameter values for synaptic currents $I_R^{(T,S)}(t)$ into the granule (GR) and the Golgi (GO) cells in the granular layer, the Purkinje cells (PCs) and the basket cells (BCs) in the Purkinje-molecular layer, and  the vestibular nucleus (VN) and the inferior olive (IO) neurons in the other parts. In the granular layer, the GR cells receive excitatory inputs via mossy fibers (MFs) and inhibitory inputs from GO cells, and the GO cells receive excitatory inputs via parallel fibers (PFs) from GR cells. In the Purkinje-molecular layer, the PCs receive two types of excitatory inputs via PFs from GR cells and through climbing fibers (CFs) from the IO and one type of inhibitory inputs from the BCs. The BCs receive excitatory inputs via PFs from GR cells. In the other parts, the VN neuron receives excitatory inputs via MFs and inhibitory inputs from PCs, and the IO neuron receives excitatory input via the desired signal (DS) and inhibitory input from the VN neuron.
}
\label{tab:SynParm}
\begin{tabular}{|c|c|c|c|c|c|c|c|c|c|c|c|c|c|c|}
\hline
& \multicolumn{5}{c|}{Granular Layer} & \multicolumn{4}{c|}{Purkinje-Molecular Layer} & \multicolumn{5}{c|}{Other Parts} \\
\hline
Target & \multicolumn{3}{c|}{\multirow{2}{*}{GR}} & \multicolumn{2}{c|}{\multirow{2}{*}{GO}} & \multicolumn{3}{c|}{\multirow{2}{*}{PC}} & \multirow{2}{*}{BC} & \multicolumn{3}{c|}{\multirow{2}{*}{VN}} & \multicolumn{2}{c|}{\multirow{2}{*}{IO}} \\
Cells ($T$) & \multicolumn{3}{c|}{} & \multicolumn{2}{c|}{} & \multicolumn{3}{c|}{} & \multirow{2}{*}{} & \multicolumn{3}{c|}{} & \multicolumn{2}{c|}{}  \\
\hline
Source & \multirow{2}{*}{MF} & \multirow{2}{*}{MF} & \multirow{2}{*}{GO} & \multirow{2}{*}{PF} & \multirow{2}{*}{PF} & \multirow{2}{*}{PF} & \multirow{2}{*}{CF} & \multirow{2}{*}{BC} & \multirow{2}{*}{PF} & \multirow{2}{*}{MF} & \multirow{2}{*}{MF} & \multirow{2}{*}{PC} & \multirow{2}{*}{DS} & \multirow{2}{*}{VN} \\
Cells ($S$) &  &  &  &  &  & & & & & & & & & \\
\hline
Receptor & \multirow{2}{*}{AMPA} & \multirow{2}{*}{NMDA} & \multirow{2}{*}{GABA} & \multirow{2}{*}{AMPA} & \multirow{2}{*}{NMDA} & \multirow{2}{*}{AMPA} & \multirow{2}{*}{AMPA} & \multirow{2}{*}{GABA} & \multirow{2}{*}{AMPA} & \multirow{2}{*}{AMPA} & \multirow{2}{*}{NMDA} & \multirow{2}{*}{GABA} & \multirow{2}{*}{AMPA} & \multirow{2}{*}{GABA} \\
$(R)$ & & & & & & & & & & & & & &  \\
\hline
$\bar{g}_{R}^{(T)}$ & 0.18 & 0.025 & 0.028 & 45.5 & 30.0 & 0.7 & 0.7 & 1.0 & 0.7 & 50.0 & 25.8 & 30.0 & 1.0 & 0.18\\
\hline
$J_{ij}^{(T,S)}$ & 8.0 & 8.0 & 10.0 & 0.00004 & 0.00004 & 0.006 & 1.0 & 5.3 & 0.006 & 0.002 & 0.002 & 0.008 & 1.0 & 5.0 \\
\hline
$V_{R}^{(S)}$ & 0.0 & 0.0 & -82.0 & 0.0 & 0.0 & 0.0 & 0.0 & -75.0 & 0.0 & 0.0 & 0.0 & -88.0 & 0.0 & -75.0\\
\hline
$\tau_{R}^{(T)}$ & 1.2 & 52.0 & 7.0, 59.0 & 1.5 & 31.0, 170.0 & 8.3 & 8.3 & 10.0 & 8.3 & 9.9 & 30.6 & 42.3 & 10.0 & 10.0\\
\hline
$A_1$, $A_2$ & & & 0.43, 0.57 & & 0.33, 0.67 & & & & & & & & & \\
\hline
\end{tabular}
\end{table*}

\begin{table*}
\caption{Glossary for various terms which characterize the cerebellar model.
}
\label{tab:Glossary}
\begin{tabular}{|l|l|}
\hline
\multicolumn{1}{|c|}{Terms} & \multicolumn{1}{|c|}{Description} \\
\hline
$A_{i}(t)$: activation degree & Fraction of active GR cells \\
\hline
$A_{i}^{(G)}(t)$: activation degree & \multirow{2}{*}{Fraction of active GR cells in the $G$ spiking group} \\
\multicolumn{1}{|l|}{\hspace{1cm} in the $G$ spiking group} &  \\
\hline
\multirow{2}{*}{${\cal C}^{(I)}$: output conjunction index} & The degree for the conjunction of the spiking behavior in each $I$th GR cluster \\
 & with the population-averaged spiking behavior in the whole population \\
\hline
\multirow{2}{*}{${\cal C}_{\rm in}^{(I)}$: input conjunction index} & The degree for the conjunction of the total synaptic input into each $I$th GR cluster \\
&  with the cluster-averaged total synaptic input\\
\hline
${\cal D}:$ output diversity degree & Relative standard deviation for the distribution of $\{ {\cal C}^{(I)} \}$ \\
\hline
${\cal D}_{\rm in}$: input diversity degree & Relative standard deviation for the distribution of $\{ {\cal C}_{\rm in}^{(I)} \}$ \\
\hline
$f_{\rm GR}^{(i)} (t)$: instantaneous individual & \multirow{2}{*}{Individual firing activity of active GR cells} \\
\multicolumn{1}{|l|}{\hspace{1cm} firing rate} & \\
\hline
$f_{\rm GR}^{(p)} (t)$: instantaneous population & \multirow{2}{*}{Population firing activity of the whole GR cells} \\
\multicolumn{1}{|l|}{\hspace{1cm} firing rate} & \\
\hline
$f_{X} (t)$: firing rate & Firing activity of the $X$ cell ($X={\rm VN}$ and ${\rm IO}$) \\
\hline
\multirow{2}{*}{${\cal L}_{g}$: learning gain degree} & Modulation gain ratio. Normalized modulation of the individual firing rate $f_{\rm VN} (t)$ of \\
& the VN neuron divided by that at the 1st cycle \\
\hline
\multirow{2}{*}{${\cal L}_{p}$: learning progress degree} & Ratio of the cycle-averaged inhibitory input from the VN neuron to the cycle- \\
& averaged excitatory input via the IO desired signal \\
\hline
\multirow{2}{*}{${\cal M}_{d}$: matching degree} & Matching degree between $R_{\rm GR} (t)$ (instantaneous whole-population spike rate in the \\
& whole population of GR cells) and $f_{\rm DS} (t)$ (IO desired signal) \\
\hline
$R_{\rm GR}^{(I)} (t)$: instantaneous cluster spike rate & Spiking behavior in each $I$th GR cluster \\
\hline
$R_{X} (t)$: instantaneous whole- & \multirow{2}{*}{Population behavior of $X$ cells} \\
\multicolumn{1}{|l|}{\hspace{1cm} population spike rate} & \\
\hline
$\langle W_{J}^{(G)} (t) \rangle$: weighted synaptic weight & Contribution of the $G$ spiking group to $\langle \tilde{J}(t) \rangle$ of active PF signals in the whole\\
\multicolumn{1}{|l|}{\hspace{1cm} in the $G$ spiking group} &  population \\
\hline
\end{tabular}
\end{table*}

\appendix
\section{Parameter Values for The LIF Neuron Models and The Synaptic Currents}
\label{app:A}
In this appendix, we list two tables which show parameter values for the LIF neuron models in Subsec.~\ref{subsec:LIF} and the synaptic currents in   Subsec.~\ref{subsec:SC}. These values are adopted from physiological data \citep{Yama2,Yama1}.

For the LIF neuron models,  the parameter values for the capacitance $C_X$, the leakage current $I_L^{(X)}$, the AHP current $I_{AHP}^{(X)}$, and the external constant current $I_{ext}^{(X)}$ are given in Table \ref{tab:SingleParm}.

For the synaptic currents, the parameter values for the maximum conductance  $\bar{g}_{R}^{(T)}$, the synaptic weight $J_{ij}^{(T,S)}$, the synaptic reversal potential $V_{R}^{(S)}$, the synaptic decay time constant $\tau_{R}^{(T)}$, and the amplitudes $A_1$ and $A_2$ for the type-2 exponential-decay function in the granular layer, the Purkinje-molecular layer, and the other parts for the VN and IO are given in Table \ref{tab:SynParm}.

\section{Glossary}
\label{app:B}
In this appendix, glossary for various terms characterizing the cerebellar model is given in Table \ref{tab:Glossary}.

\end{document}